\documentclass[12pt]{report}
\usepackage{amssymb}

\usepackage{amsmath}
\usepackage{jeep}

\begin{document}

\author{Axel Gelfert \\
%EndAName
University of Cambridge\thanks{%
Current address: Axel Gelfert, Department of History and Philosophy of
Science, University of Cambridge, Free School Lane, Cambridge CB2 3RH,
England; email: axel@gelfert.net.}\\
and\\
Humboldt University of Berlin}
\title{On the Role of Dimensionality in Many-Body Theories of Magnetic
Long-Range Order}
\date{1 June 2001}
\maketitle

\begin{abstract}
Starting from general considerations concerning phase transitions and the
many-body problem, a pedagogical introduction to the Mermin-Wagner theorem
is given. Particular attention is paid to how the Mermin-Wagner theorem fits
in with more general methods in many-particle physics. An attempt at an
assessment of its validity for approximative methods is made by contrasting
the results obtained within Onsager reaction-field theory and those obtained
by the well-known Tyablikov method.
\end{abstract}

\tableofcontents

\chapter{Introduction}

\section{Outline}

In this paper, we shall investigate the role of spatial dimensionality in
the context of many-body theories for magnetic long-range order. After a
rather general introduction to the problem (this chapter), and an exposition
of the basic notions of phase transitions and spontaneous symmetry-breaking
(section 2.1), the Bogoliubov inequality is discussed (section 2.2). This is
a rigorous relation between observables of a physical system and the
system's Hamiltonian, and it shall be used at length in Chapter III in the
course of an extended proof of the absence of a magnetic phase transition in
films described by the Heisenberg, Hubbard, s-f (Kondo-lattice) and Periodic
Anderson model. Chapter IV investigates the connection of the Bogoliubov
inequality and related correlation inequalities with the Green's function
technique widely used in many-body physics. Finally (Chapter V), two
examples of approximative theories are discussed to illustrate the
theoretical findings of the previous chapters.

\section{The many-body problem}

Loosely speaking, the success of physics in describing properties and
processes of the physical world is, to a large part, the result of taking a
reductionist viewpoint, in the sense that macroscopic phenomena are regarded
as being fully determined, at least in principle, by underlying microscopic
processes which behave according to some (more or less well-understood)
fundamental law of nature. In the case of electrons in solids, one would
expect the non-relativistic Schr\"{o}dinger equation (plus the Pauli
principle) to provide an accurate description of the behaviour of the
physical system, at least in the low-energy regime. For non-interacting
electrons in a completely periodic potential of ions located at the lattice
sites of a perfectly ordered crystal, this can indeed be accomplished.
However, it turns out that, for the many-body problem of \textit{strongly} 
\textit{interacting} electrons, such an approach is highly impracticable,
and attempting to calculate macroscopic properties from a set of 10$^{23}$
one-particle equations simultaneously, would be fighting a losing battle.

One solution to this problem is to propose reduced many-body models which
retain the basic features of the microscopic dynamics, e.g. Coulomb
repulsion or the Pauli exclusion principle, but at the same time leave out
less important features, e.g. by neglecting interactions between electrons
and phonons or by limiting Coulomb repulsion between electrons to those at
the same lattice site. In most instances, such models still cannot be solved
exactly; methodologically, a ``computational gap''\footnote{%
This expression was coined by \cite{Red 1980}.} remains which one can only
hope to close, to some extent, by use of approximative methods, whose
validity must be constantly reassessed.

\subsection{Summary of many-body Hamiltonians}

The main focus of this paper is on magnetic phenomena. It can be rigorously
shown \textit{(Bohr-van Leeuwen theorem)} that the existence of magnetic
phenomena (diamagnetism, paramagnetism, \textit{and} collective magnetism)
of an ensemble of particles in a crystal is inconsistent with classical
statistical mechanics.\footnote{%
For a concise discussion of this fact see \cite{QdM1 1986}.} This striking
fact indicates that a satisfactory description of spontaneous magnetic order
will have to include quantum effects from the very beginning. In the
following, we shall discuss almost exclusively quantum many-body models.
Nevertheless, classical models such as the\textbf{\ Ising model}\footnote{%
Also, and more appropriately, known as \textit{Lenz-Ising model}, to give
credit to the true initiator of the model, W. Lenz (for a broad account of
the history of the subject, see \cite{Dom 1996}).}\ defined by the Hamilton
function 
\begin{equation}
H=-\sum_{ij}J_{ij}S_{i}S_{j},
\end{equation}%
where $S_{i}$ is a classical (c-number) ``spin'' at lattice site $i,$ have
played, and continue to play, an important role in the conceptual
development of the theory of phase transitions. They serve as a test bed for
new mathematical approaches, such as the renormalization group approach, and
as limiting cases of more sophisticated models.

The Ising model, for example, can be derived from the more general\textbf{\
Heisenberg model} as defined by the Hamiltonian 
\begin{equation}
H=-\sum_{ij}J_{ij}\left( S_i^xS_j^x+S_i^yS_j^y+S_i^zS_j^z\right)
\end{equation}
by singling out one of the components $(x,y,$ or $z)$. The Heisenberg model
is a quantum many-body model and the spin $\vec{S}_i$ must now be
interpreted as a quantum-mechanical operator. From the $x$ and $y$%
-components of $\vec{S}_i$ one constructs \textit{spin} \textit{raising} and 
\textit{lowering operators} 
\begin{equation}
S_i^{\pm }=S_i^x\pm iS_i^y.
\end{equation}
The operators $\vec{S}_i$ and $S_i^{\pm }$ obey the commutation relations
for angular momentum operators so that, among others, the following
identities hold: 
\begin{eqnarray}
\left[ S_i^z,S_j^{\pm }\right] _{-} &=&\pm \hbar \delta _{ij}S_i^{\pm }
\label{zplus} \\
\left[ S_i^{+},S_j^{-}\right] _{-} &=&2\hbar \delta _{ij}S_i^z
\label{plusminus}
\end{eqnarray}
and 
\begin{equation}
\vec{S}_i\cdot \vec{S}_j=\frac 12\left( S_i^{+}S_j^{-}+S_i^{-}S_j^{+}\right)
+S_i^zS_j^z.
\end{equation}

Note that the indices $i$ and $j$ are taken to identify uniquely the lattice
sites of spins $\vec{S}_i$ and $\vec{S}_j.$ Instead of numbering all sites
consecutively, one could identify a lattice site by more than one index,
e.g. in film systems where it will turn out to be useful to identify lattice
sites by numbering layers $\alpha ,\beta ...$ and positions $i,j,...$ within
the layers separately: $i\mapsto (i,\alpha ).$ The Kronecker symbol $\delta
_{ij}$ in (\ref{zplus}) and (\ref{plusminus}) must then be replaced by the
product $\delta _{ij}\delta _{\alpha \beta }.$

At times it will be helpful to make use of Fourier transformation into the
reciprocal lattice: 
\begin{equation}
S^{(+,-,x,y,z)}(\vec{k})=\sum_ie^{i\vec{k}\cdot \vec{R}_i}S_i^{(+,-,x,y,z)}
\label{fouriernachk}
\end{equation}
with the inverse transformation 
\begin{equation}
S_i^{(+,-,x,y,z)}=\frac 1N\sum_{\vec{k}}e^{-i\vec{k}\cdot \vec{R}%
_i}S^{(+,-,x,y,z)}(\vec{k}).  \label{fouriernachi}
\end{equation}

The Heisenberg model was proposed some 75 years ago \cite{Hei 1926}, in
order to describe the behaviour of magnetic insulators such as EuO
(ferromagnetic), EuTe (antiferromagnetic) and EuSe (ferrimagnetic). Whether
a Heisenberg-type magnet will order ferromagnetically or
antiferromagnetically depends on the sign of the coupling constants $J_{ij}$%
, also known as \textit{exchange integrals.} When $J_{ij}>0$ $(J_{ij}<0)$
for neighbouring spins $S_i,$ $S_j$, ferromagnetic (antiferromagnetic) order
will be favoured, as this will minimize their contribution to the total
energy of the system. Different exchange mechanisms exist; the feature they
have in common is that by virtue of quantum mechanical symmetry
requirements, electrostatic correlations lead to an effective spin-spin
interaction. The numerical values of $J_{ij}$ depend on details of the
substance and the exchange mechanism involved.

The \textbf{Hubbard model,} in contrast to the Heisenberg model, describes
the case of itinerant electrons in narrow energy bands, as found in the
prominent ferromagnetic metals Co, Ni, Fe. It was proposed in 1963 by
Hubbard \cite{Hub 1963} (and simultaneously by Gutzwiller \cite{Gut 1963}
and Kanamori \cite{Kan 1963}) and takes into account that in these systems
the same group of electrons is responsible for magnetic effects and
electrical conduction. Further empirical findings that are taken into
account in the Hubbard model are, firstly, that narrow $3d$-bands are
responsible for the appearance of collective magnetism; secondly, the
screening of the Coulomb interaction between $d$-electrons through electrons
of broad $s$- and $p$-bands; and thirdly, the importance of the lattice
structure (for a detailed derivation see \cite{Hub 1963},\cite{QdM2 1986}).
The individual points can be discussed by looking at the structure of the
Hubbard Hamiltonian in its simplest (one-band) form 
\begin{equation}
H=\sum_{ij\sigma }T_{ij}c_{i\sigma }^{+}c_{j\sigma }+\frac U2\sum_{i\sigma
}n_{i\sigma }n_{i-\sigma },
\end{equation}
where $c_{i\sigma }^{+}$ and $c_{j\sigma }$ are creation and annihilation
operators for an electron with spin $\sigma $ at lattice sites $i$ and $j,$
respectively; and $n_{i\pm \sigma }=c_{i\pm \sigma }^{+}c_{i\pm \sigma }$ is
the number operator which tests whether or not an electron with spin $\pm
\sigma $ is present at lattice site $i.$\footnote{%
For a detailed account of second quantization in many-body physics, see \cite%
{Bd7 1997}.} The Hubbard Hamiltonian consists of a kinetic part and an
interaction part reflecting the Coulomb repulsion between electrons which
becomes relevant only if two electrons happen to be at the same lattice site
(which, because of the Pauli principle, must then have different spin
quantum numbers). The hopping integrals $T_{ij}$, i.e. the probability
amplitude for an electron with spin $\sigma $ at site $j$ to move to another
site $i$, are related to the one-particle Bloch energy dispersion $%
\varepsilon (\vec{k})$ by Fourier transformation: 
\begin{eqnarray}
T_{ij} &=&\frac 1N\sum_{\vec{k}}\varepsilon (\vec{k})e^{i\vec{k}\cdot (\vec{R%
}_i-\vec{R}_j)} \\
\varepsilon (\vec{k}) &=&\frac 1N\sum_{ij}T_{ij}e^{-i\vec{k}\cdot (\vec{R}_i-%
\vec{R}_j)}.
\end{eqnarray}

Spin-like operators, i.e. pseudo-spins $\sigma $, are built from the
electron creation and annihilation operators, in the following way: 
\begin{eqnarray}
\sigma _i^{+} &=&\hbar c_{i\uparrow }^{+}c_{i\downarrow }  \label{sigmpl} \\
\sigma _i^{-} &=&\hbar c_{i\downarrow }^{+}c_{i\uparrow }  \label{sigmmi} \\
\sigma _i^z &=&\frac \hbar 2(n_{i\uparrow }-n_{i\downarrow })  \notag \\
&=&\frac \hbar 2(c_{i\uparrow }^{+}c_{i\uparrow }-c_{i\downarrow
}^{+}c_{i\downarrow }).  \label{pseudosz}
\end{eqnarray}

Fourier transforms into reciprocal lattice are defined in exactly the same
way as in (\ref{fouriernachk}) and (\ref{fouriernachi}), e.g. 
\begin{equation}
\sigma ^{+}(\vec{k})=\hbar \sum_ie^{i\vec{k}\cdot \vec{R}_i}c_{i\uparrow
}^{+}c_{i\downarrow }.
\end{equation}

In the s-f (Kondo-lattice) model and the Periodic Anderson Model (PAM), the
situation is somewhat different from the Heisenberg and Hubbard models, as
one must now distinguish between two spin systems. Both models deal with a
situation that can be found in the rare earths, where the electrons of the
inner $4f$ ($5f$) shells, which are incompletely filled and which correspond
to strictly localized states, are screened by the broader $6s$ ($7s$)
states. Thus, the magnetic moments associated with the $4f$ ($5f$) electrons
couple according to Hund's rules and form a local magnetic moment. The s-f
model describes the coupling of such a local $f$-spin $S$ to the electrons
of the conduction band: 
\begin{equation*}
H=\sum_{ij\sigma }T_{ij}c_{i\sigma }^{+}c_{j\sigma }-\frac J2\sum_{i\sigma
}(z_\sigma S_i^zn_{i\sigma }+S_i^\sigma c_{i-\sigma }^{+}c_{i\sigma }),
\end{equation*}
where $z_{\sigma =_{\downarrow }^{\uparrow }}=\pm 1.$ In the PAM the
Hamiltonian includes the hopping $H_s$ of the electrons of the conduction
band (as in the s-f case), a term $H_f$ defining the energy level of the $f$%
-electrons, the Coulomb repulsion $H_{ff}$ between electrons in the strongly
localized $f$-states, and the hybridization $H_{hyb}$ of the $f$-level with
the conduction band: 
\begin{eqnarray}
H &\equiv &H_s+H_f+H_{ff}+H_{hyb} \\
&=&\sum_{ij\sigma }T_{ij}c_{i\sigma }^{+}c_{j\sigma }+\varepsilon
_f\sum_{i\sigma }f_{i\sigma }^{+}f_{i\sigma }+\frac U2\sum_in_{i\uparrow
}^fn_{i\downarrow }^f  \notag \\
&&+V\sum_{i\sigma }\left( c_{i\sigma }^{+}f_{i\sigma }+f_{i\sigma
}^{+}c_{i\sigma }\right) .
\end{eqnarray}

\chapter{Phase Transitions and the Bogoliubov Inequality}

Among the general features of physical systems at a macroscopic scale, the
display of different phases plays a fundamental role. By this we mean that
the overall characteristics of a physical system, for example a large bulk
volume of a certain substance, can be described by a limited number of
system-specific parameters (in many cases of interest by a single one) which
themselves depend on a set of certain external conditions, usually an
external field, and the temperature. When a system changes from one phase to
another, e.g. as a result of increasing temperature, this transition will
often be marked by a rapid change in one of the system's characteristic
quantities, such as the density in the liquid-vapour transition. From early
on, it has been considered a challenging task to describe the experimental
findings through theoretical models. On a phenomenological level as well as
by starting from first principles, substantial progress in the description
of classical systems has been achieved since the pioneering work by, for
example, \textit{van der Waals} and \textit{Wei\ss }\ in the liquid-vapour
and the ferromagnetic phase transition. With the advent of quantum theory
and progress in the understanding of the laws governing the microscopic
behaviour of physical systems, it became possible to ``design''\
mathematical models in such a way as to reflect certain aspects of the
microscopic dynamics of real systems while omitting others. This has led to
a remarkable wealth of models, some of which have been described in some
detail in the previous chapter.

\section{Phase transitions and spontaneous symmetry-breaking}

In the theory of condensed many-body systems one is often interested in
describing phases that fail to exhibit certain symmetry properties of the
underlying Hamiltonians. This is, in fact, a common feature and one can
classify such systems according to their symmetry properties. Thus,
crystals, by their very lattice structure, break the translational symmetry
encountered in the continuum description of fluids; ferromagnets, in
addition to the spatial symmetry-breaking due to their crystal structure,
are not invariant towards rotations in spin-space, even though the
underlying Hamiltonians describing the system may well be. Less obvious
types of symmetry-breaking occur in other quantum systems, such as
superfluids and superconductors, where gauge invariance is broken.

\subsection{Bogoliubov quasi-averages}

Bogoliubov \cite{Bog 1960},\cite{Bog 1962} has devised a method to describe
the occurrence of spontaneous symmetry-breaking in terms of\textit{\
quasi-averages.\footnote{%
This section follows \cite{Bog 1960} and \cite{Akh Pel 1981}.}} Normally,
for systems in statistical equilibrium, we have defined the average value of 
$A$ as the trace over the equilibrium \textit{density} (or\textit{\
statistical) operator} $\rho =\exp (-\beta \mathcal{H})$ times the
observable $A,$ so that for the infinite system $(V\rightarrow \infty )$ one
has 
\begin{eqnarray}
\left\langle A\right\rangle &\equiv &\lim_{V\rightarrow \infty }tr\left(
\rho A\right)  \notag \\
&=&\lim_{V\rightarrow \infty }tr\left( e^{-\beta \mathcal{H}}A\right) ,
\label{averag}
\end{eqnarray}
where $\mathcal{H}$ is the grand-canonical Hamiltonian, $\mathcal{H}=H-\mu 
\hat{N}$ ($\hat{N}$: number operator). However, it turns out that under
certain conditions such averages may be unstable with respect to an
infinitesimal perturbation of the Hamiltonian. If a corresponding additive
contribution $H_\nu \equiv \nu H^{\prime }$ of the order of $\nu $ is added
(where $\nu $ is a small positive number which will eventually be taken to
zero: $\nu \rightarrow 0)$, i.e. 
\begin{equation}
\mathcal{H}_\nu =H+H_\nu -\mu \hat{N},  \label{calhanu}
\end{equation}
one can define the \textit{quasi-average} of $A$ in the following way: 
\begin{equation}
\left\langle A\right\rangle _q\equiv \lim_{\nu \rightarrow
0}\lim_{V\rightarrow \infty }tr(e^{-\beta \mathcal{H}_\nu }A).
\label{quaverag}
\end{equation}
The average (\ref{averag}) and the quasi-average (\ref{quaverag}) need not
coincide, since the two limits in (\ref{quaverag}) may fail to commute
within some parameter region (i.e. for some combination of $\mu $ and $\beta
)$.

Quasi-averages are appropriate for cases in which spontaneous
symmetry-breaking occurs, as can be shown by a simple argument. Suppose the
Hamiltonian $\mathcal{H}$ displays a continuous symmetry $\mathcal{S}$, i.e.
it commutes with the generators $\Gamma _{\mathcal{S}}^i$ of the
corresponding symmetry group, 
\begin{equation}
\left[ \mathcal{H},\Gamma _{\mathcal{S}}^i\right] _{-}=0.
\end{equation}
If some operator $B$ is not invariant under the transformations of $\mathcal{%
S}$, 
\begin{equation}
\left[ B,\Gamma _{\mathcal{S}}^i\right] _{-}\equiv C^i\neq 0,
\end{equation}
the (normal) average of the commutator $C^i$ vanishes, 
\begin{equation}
\left\langle C^i\right\rangle =0,  \label{vanichi}
\end{equation}
as can be readily seen from eqn. (\ref{averag}) by use of cyclic invariance
of the trace. In those instances, however, where the perturbative part $%
H_\nu $ of (\ref{calhanu}) does not commute with $\Gamma _{\mathcal{S}}^i,$
this will give a \textit{non-vanishing} \textit{quasi-average:} 
\begin{eqnarray}
\left\langle C^i\right\rangle _q &=&\lim_{\nu \rightarrow 0}tr\left(
e^{-\beta \mathcal{H}_\nu }\left[ B,\Gamma _{\mathcal{S}}^i\right]
_{-}\right)  \notag \\
&\neq &0.
\end{eqnarray}
Thus, even though one might na\"{\i}vely expect the quasi-average to
coincide with the ``normal'' average in the limit $\nu \rightarrow 0,$ quite
generally this will not be the case. Note that the quasi-average depends on
the nature of the perturbation added to the ``original'' Hamiltonian.

In the following, we will make extensive use of the concept of
quasi-averages as in (\ref{quaverag}).\footnote{%
The symbol $\left\langle ...\right\rangle $ will be used for averages and
quasi-averages alike, the distinction between the two being obvious from
context in all cases under consideration.} As a first example, let us
examine the Heisenberg model 
\begin{equation}
H=-\sum_{ij}J_{ij}\left( \vec{S}_i\cdot \vec{S}_j\right) ,
\end{equation}
which is invariant under the continuous rotation group generated by the
total spin vector $\vec{S}=\sum_i\vec{S}_i$ because of $\left[ H,\vec{S}%
\right] _{-}=0.$ Thus, we can take $\vec{S}$ as the operator $B$, and from (%
\ref{vanichi}) one finds $\left\langle \left[ S^\alpha ,S^\beta \right]
_{-}\right\rangle =0,$ where $S^\alpha ,S^\beta $ are the components of the
total spin vector $\vec{S}.$ Together with the commutation relations for
spin operators, e.g. $\left[ S^x,S^y\right] _{-}=i\hbar S^z,$ one sees that
the conventional average of the magnetization vanishes. This is just a
manifestation of the fact, that on the macroscopic level, for an ideal,
infinitely extended system, there is no preferred direction in space. One
can say this is a situation of ``degeneracy'' with respect to spatial
orientation. Adding an external field, $\vec{B}_0=B_0\vec{e}_z$ along the $z$%
-axis for example, lifts this degeneracy and via the contribution $H_b\sim
B_0M(T,B_0)$ to the Hamiltonian $(M$: magnetization) one can, thus,
construct appropriate quasi-averages for arbitrary operators according to
equation (\ref{quaverag}).

\subsection{Order parameters}

In the theory of phase transitions, the first step is to identify a quantity
whose (quasi-)average vanishes on one side of the transition, but takes a
finite value on the other side. This quantity is called the \textit{order
parameter. }In a continuous phase transition, the order parameter may
gradually evolve from zero at the critical point to a finite value on one
side (usually the low-temperature side) of the transition. For different
kinds of phases, different order parameters must be \textit{chosen.} From a
phenomenological point of view, one must consider each physical system anew.
We have already mentioned the liquid-vapour and the
ferromagnetic-paramagnetic transition as prototypes for phase transitions.
In the former, the obvious choice for the order parameter would be the
difference between the mean densities, i.e. $\rho -\rho _{vapour};$ in the
latter the relevant order parameter is the magnetization $M.$ Within a given
many-body model, the magnetization can be defined in microscopic terms, as
we shall see shortly.

Sometimes, as in the transition to the superconducting state, it may even be
possible to characterize the \textit{same} type of phase transition by use
of different order parameters. According to the standard theories of
superconductivity, at low temperatures electrons with opposite spins form
Cooper pairs. Thus, a possible order parameter would be the average
probability amplitude to find a Cooper pair at a given lattice site in the
crystal. Alternatively, one could characterize the phase transition through
the gap parameter whose modulus is the difference in energy per electron of
the Cooper pair condensate and the energy at the Fermi level. We shall
briefly return to the problem of superconductivity when we consider the
possibility of pairing at finite temperatures in film systems.

\section{Bogoliubov inequality}

\subsection{Proof of the Bogoliubov inequality}

The Bogoliubov inequality is a rigorous relation between two essentially
arbitrary operators $A$ and $B$ and a valid Hamiltonian $H$ of a physical
system. In its original form, proposed in \cite{Bog 1962}, it is given by 
\begin{equation}
\left| \left\langle \left[ C,A\right] _{-}\right\rangle \right| ^2\leq \frac
\beta 2\left\langle \left[ A,A^{+}\right] _{+}\right\rangle \left\langle %
\left[ C^{+},[H,C]_{-}\right] _{-}\right\rangle ,  \label{bogsimple}
\end{equation}
where $\beta =1/k_BT$ is the inverse temperature and $\left\langle
....\right\rangle $ denotes the thermodynamic expectation value. $A$ and $B$
do not necessarily have an obvious physical interpretation from the very
beginning, so the physical significance of (\ref{bogsimple}) will depend on
the suitable choice for the operators involved.

The inequality can be proved by introducing a scalar product which is based
on the energy eigenvalues $E_{n}$ and the corresponding energy eigenstates $%
\left| n\right\rangle $ of the Hamiltonian $H:$%
\begin{eqnarray}
\left\langle n|m\right\rangle &=&\delta _{nm}  \notag \\
H\left| n\right\rangle &=&E_{n}\left| n\right\rangle \\
\Rightarrow \left\langle n\left| H\right| n\right\rangle &=&E_{n}  \notag
\end{eqnarray}%
with the \textit{completeness relation }(\textsf{1}$:$ identity operator) 
\begin{equation}
\mathsf{1}=\sum_{n}\left| n\right\rangle \left\langle n\right| .
\label{completeness}
\end{equation}%
By further introducing the \textit{statistical weight} 
\begin{equation}
w_{n}=\frac{\exp (-\beta E_{n})}{tr\left( \exp (-\beta H)\right) },
\end{equation}%
the Bogoliubov scalar product $\mathfrak{B}$ of two arbitrary operators $A,B$
is defined as 
\begin{equation}
\mathfrak{B}(A;B)=\sum_{\substack{ n,m  \\ E_{n}\neq E_{m}}}
\label{bogscalprod}
\end{equation}%
Note that the sum is restricted to terms with $E_{n}\neq E_{m}$ only. Since $%
\mathfrak{B}$ is positive semidefinite (see \cite{QdM2 1986} or appendix A),
the Schwarz inequality holds: 
\begin{equation}
\left| \mathfrak{B}(A;B)\right| ^{2}\leq \mathfrak{B}(A;A)\mathfrak{B}(B;B).
\label{schwarzenegger}
\end{equation}

With the specific choice 
\begin{equation}
B=\left[ C^{+},H\right] _{-}
\end{equation}%
the mixed product $\mathfrak{B}(A;B)$ on the LHS of (\ref{schwarzenegger})
is 
\begin{eqnarray}
\mathfrak{B}(A;B) &=&\sum_{\substack{ n,m  \\ E_{n}\neq E_{m}}}\left\langle
n\left| A^{+}\right| m\right\rangle \left\langle m\left| \left[ C^{+},H%
\right] _{-}\right| n\right\rangle \frac{w_{m}-w_{n}}{E_{n}-E_{m}}  \notag
\label{wequot} \\
&=&\sum_{n,m}\left\langle n\left| A^{+}\right| m\right\rangle \left\langle
m\left| C^{+}\right| n\right\rangle (w_{m}-w_{n})  \notag \\
&=&\sum_{m}w_{m}\left\langle m\left| C^{+}A^{+}\right| m\right\rangle
-\sum_{n}w_{n}\left\langle n\left| A^{+}C^{+}\right| n\right\rangle  \notag
\\
&=&\left\langle \left[ C^{+},A^{+}\right] _{-}\right\rangle  \label{mixd}
\end{eqnarray}%
by use of the completeness relation (\ref{completeness}) and the definition
of $w_{n}.$

The norm $\mathfrak{B}(B;B)=\left| \left| B\right| \right| ^2\geq 0$ can be
calculated from eqn. (\ref{mixd}) by specializing $A=B=\left[ C^{+},H\right]
_{-},$ thus arriving at 
\begin{equation}
\mathfrak{B}(B;B)=\left\langle \left[ C^{+},\left[ H,C\right] _{-}\right]
_{-}\right\rangle \geq 0.  \label{dblcm}
\end{equation}

For the remaining norm $\mathfrak{B}(A;A)=\left| \left| A\right| \right|
^{2} $ we start from (\ref{bogscalprod}) and note that the quotient $%
(w_{m}-w_{n})/(E_{n}-E_{m})>0$ $(E_{m}\neq E_{n})$ can be bounded from
above: 
\begin{eqnarray}
\frac{w_{m}-w_{n}}{E_{n}-E_{m}} &=&\frac{1}{tr(\exp (-\beta H))}\frac{\left(
e^{-\beta E_{m}}+e^{-\beta E_{n}}\right) }{E_{n}-E_{m}}\frac{e^{-\beta
E_{m}}-e^{-\beta E_{n}}}{e^{-\beta E_{m}}+e^{-\beta E_{n}}}  \notag \\
&=&\frac{\left( w_{m}+w_{n}\right) }{E_{n}-E_{m}}\tanh \left( \frac{\beta }{2%
}\left( E_{n}-E_{m}\right) \right) <\frac{\beta }{2}\left(
w_{m}+w_{n}\right) ,
\end{eqnarray}%
where the condition $E_{m}\neq E_{n}$ has again been used. It is now
straightforward to show that the norm $\left| \left| A\right| \right| ^{2}$
is bounded from above: 
\begin{eqnarray}
\mathfrak{B}(A;A) &<&\frac{\beta }{2}\sum_{\substack{ n,m  \\ E_{n}\neq
E_{m} }}\left\langle n\left| A^{+}\right| m\right\rangle \left\langle
m\left| A\right| n\right\rangle \left( w_{m}+w_{m}\right)  \notag \\
&\leq &\frac{\beta }{2}\sum_{n,m}\left\langle n\left| A^{+}\right|
m\right\rangle \left\langle m\left| A\right| n\right\rangle \left(
w_{m}+w_{m}\right)  \notag \\
&=&\frac{\beta }{2}\sum_{m}w_{m}\left\langle m\left| \left[ A,A^{+}\right]
_{+}\right| m\right\rangle =\frac{\beta }{2}\left\langle \left[ A,A^{+}%
\right] _{+}\right\rangle .  \label{nrmvaa}
\end{eqnarray}%
This is all that is needed for deriving the Bogoliubov inequality; inserting
(\ref{mixd}), (\ref{dblcm}), and (\ref{nrmvaa}) into the Schwarz inequality (%
\ref{schwarzenegger}) immediately recovers (after minor rearrangement of the
LHS) the Bogoliubov inequality in the form proposed in eqn. (\ref{bogsimple}%
). From the derivation it is noteworthy that the two factors on the RHS each
are \textit{upper bounds to a norm} and, thus, greater than or equal to
zero. In particular, if, for example, the double commutator depends on some
parameter $k,$ one will always have 
\begin{equation}
\left\langle \left[ \lbrack C,H]_{-},C^{+}\right] _{-}\right\rangle
(k)+\left\langle \left[ [C,H]_{-},C^{+}\right] _{-}\right\rangle (k^{\prime
})\geq \left\langle \left[ \lbrack C,H]_{-},C^{+}\right] _{-}\right\rangle
(k).
\end{equation}

In later calculations a slightly modified version of the Bogoliubov
inequality will turn out to be useful. Dividing both sides of the Bogoliubov
inequality (\ref{bogsimple}) by the double commutator and summing over all
wave vectors $\vec{k}$ associated with the first Brillouin zone in the
reciprocal lattice, one arrives at 
\begin{equation}
\sum_{\vec{k}}\frac{\left| \left\langle \left[ C,A\right] _{-}\right\rangle
\right| ^2}{\left\langle \left[ [C,H]_{-},C^{+}\right] _{-}\right\rangle (%
\vec{k})}\leq \frac \beta 2\sum_{\vec{k}}\left\langle \left[ A,A^{+}\right]
_{+}\right\rangle (\vec{k}).  \label{bog}
\end{equation}

\subsection{Applications of the Bogoliubov inequality}

Hohenberg \cite{Hoh 1967} was the first to note that the Bogoliubov
inequality could be used to exclude phase transitions, showing that there
could be no finite-temperature phase transition in one- and two-dimensional
superfluid systems. At roughly the same time Mermin and Wagner, following
Hohenberg, considered the case of spontaneous magnetization in the
Heisenberg model. Incidentally, the \textit{Mermin-Wagner theorem, }as it
became known later, was published shortly before Hohenberg's proof, and it
has since become a landmark paper in the theory of phase transitions \cite%
{Mer Wag 1966}.\footnote{%
Note that Mermin and Wagner never claimed priority, instead explicitly
mentioning Hohenberg's suggestion in their paper.} A variety of extensions
of the theorem as to the possibility of spontaneous magnetization have been
given (see introductory remark in the next section) for a number of
many-body models. Here, we shall briefly mention further applications to
other order parameters.\footnote{%
A survey of all papers concerning extensions of the Mermin-Wagner theorem is
forthcoming in a review article by this author, see ref. \cite{Gel Nol 2001}.%
}

Mermin \cite{Mer 1968} applied the Bogoliubov inequality to the possibility
of crystalline order in two dimensions, confirming earlier suggestions by
Peierls and Landau that there could be no two-dimensional crystalline
ordered state. The proof is carried out in detail for classical crystals,
giving only an outline of the quantum case. For the latter case, the
derivation was given by Fern\'{a}ndez \cite{Fer 1970b}.

Kishore and Sherrington \cite{Kis She 1972} considered a quite general
Hamiltonian of electrons interacting non-relativistically among themselves,
and with spatially ordered or disordered scatterers, excluding spontaneous
low-dimensional magnetic order. The restriction to non-relativistic
interactions means that spin-orbit effects are excluded from the problem so
that the results keep the backdoor open for possible Ising-type, or other
suitable anisotropies.

The problem of partially restricted geometries was considered by Chester et
al. \cite{Che et al 1969} for three-dimensional Bose systems of finite cross
section or thickness. The results of the generically one or two-dimensional
case are, in principle, reproduced. Their main line of argument rests on the
operators in the Bogoliubov inequality being defined as Fourier transforms
in a restricted space $D\times D\times L$ or $D\times L\times L,$
respectively, where $D$ remains finite in the thermodynamic limit while $L$
goes to infinity. This way, the operators lack the physical meaning of the
usual real-space or $k$-space operators, which means that an intuitive
physical interpretation is not readily obvious.

\chapter{Absence of Finite-Temperature Long-Range Order in Partially
Restricted Geometries}

Following the well-known 1966 paper by Mermin and Wagner \cite{Mer Wag 1966}%
, in which the authors prove the absence of spontaneous magnetization in the
one- and two-dimensional isotropic Heisenberg ferromagnet (and the
two-sublattice antiferromagnet), much research has been carried out in order
to extend the statement to other spin systems. Wegner considered a model
describing a system with locally interacting itinerant electrons \cite{Weg
1967}. Walker and Ruijgrok discussed a band model for interacting electrons
in a metal \cite{Wal Rui 1968}; Ghosh \cite{Gho 1971}, more specifically,
recovered the Mermin-Wagner theorem for the Hubbard model. A proof for the
s-d interaction model is given by van den Bergh and Vertogen \cite{Ber Ver
1974}. A paper by Robaszkiewicz and Micnas \cite{Rob Mic 1976} extends the
Mermin-Wagner result to a general model with localized and itinerant
electrons, covering the modified Zener model, the extended Hubbard model and
s-d models as particular cases. In addition to extending the Mermin-Wagner
theorem to several microscopic many-body models, generalizations to more
complicated geometries were also found. Baryakhtar and Yablonskii \cite{Bar
Yab 1975}, for example, proved the Mermin-Wagner theorem for systems with an
arbitrary number of magnetic sublattices, also excluding non-collinear
magnetic order when an external field is applied. Thorpe \cite{Tho 1971}
considers the case of ferromagnetism in phenomenological models with double
and higher-order exchange terms. Similar results were obtained in the
multi-sublattice case by Krzemi\'{n}ski \cite{Krz 1976}. Recently, Matayoshi
and Matayoshi \cite{Mat Mat 1997} have discussed models with $n$-th nearest
neighbour exchange interactions and tried to extend the Mermin-Wagner
theorem to anisotropic exchange interactions; their proof, however, requires
extremely special conditions on the parameters of the model which appear to
be of no physical significance. Also, attempts to extend the Mermin-Wagner
theorem to special cases in three dimensions remain somewhat inconclusive %
\cite{Ras Tas 1989}. More recent papers will be discussed later in this
chapter.\footnote{%
Refs. \cite{Pro Lop 1983}, \cite{Cas 1992}, \cite{Uhr 1992}, \cite{Noc Cuo
1999}.}

The general procedure is essentially the same for most of the papers above:
The Bogoliubov inequality is used with suitable operators defined in such a
way as to give an upper bound for the desired order parameter, i.e. the
(bulk or sublattice) magnetization. In general, one arrives at an upper
bound that, in the limit $B_0\rightarrow 0$, behaves in one or two
dimensions as $\sim (MB_0)^{1/3}$ or $\sim 1/(\ln MB_0)^{1/2},$ thus
excluding a finite value of the order parameter ($M$ denotes the bulk
magnetization and $B_0$ is the external magnetic field). In three
dimensions, no such behaviour is found, i.e. the usual line of reasoning
fails and spontaneous magnetization cannot be ruled out. Within the usual
scheme following Mermin and Wagner, the dimensionality of a system enters
into the calculation only through the volume element when integrating the
final version of the Bogoliubov inequality over $k$-space. Thus, a
microscopic picture of the transition from two to three dimensions does not
emerge.

\section{Films with finite thickness}

As a result of the rapid progress of thin-film technology in recent years,
one can now prepare and study systems with restricted geometries, such as
films, in great detail. Thus, important parameters, such as the Curie
temperature, magnetization and susceptibility, can be measured and discussed
as functions of the number of layers $d$ in a magnetic film. Experiments
indicate \cite{Li Bab 1991} that, for real systems, the transition from 2D
to 3D behaviour of, for example, the critical exponent $\beta $, occurs
within a narrow crossover region of $d.$ The critical temperature $T_c$ also
shows a strong $d$-dependence, with $T_c(d)$ quickly approaching the bulk
value $T_c(\infty )$ as $d$ increases \cite{Li Bab 1991}, \cite{Far 1993}.

With these observations taken into account, it appears desirable to improve
one's theoretical understanding of how the transition from the monolayer to
the bulk occurs. By referring only to the truly microscopic properties of
the respective many-body model, it should be possible to study in detail the
transition from two to three dimensions. Work in this direction has been
done by studying film systems with symmetry-breaking contributions to the
Hamiltonian, thus making possible the study of the Curie temperature as a
function of $d$ and the anisotropy parameter (for analytic approaches to
Heisenberg films see \cite{Hau Bro Cor Cos 1972}, for an extensive review of
the magnetic properties of thin itinerant-electron films see \cite{Wu Nol
2000}).

At a more fundamental level, the validity of exact results, such as the
Mermin-Wagner theorem in two dimensions, to thin films may be tested. In
this chapter, a proof of the Mermin-Wagner theorem for film systems within
the main many-body models (Heisenberg, Hubbard, s-f (Kondo-lattice) model,
and Periodic Anderson Model) will be given.

\subsection{Film Hamiltonians}

For reasons of simplicity, we shall restrict our attention to film systems
composed of $d$ identical layers stacked on top of each other. The
calculations can easily be generalized to account for more complicated
geometries. One can think of the film geometry as consisting of a
two-dimensional Bravais lattice, i.e. the first layer, with $N$ lattice
sites and a $d$-atomic basis that corresponds to the $d$ layers being
stacked up. Each lattice vector then decomposes into 
\begin{equation}
\vec{R}_{i\alpha }=\vec{R}_i+\vec{r}_\alpha ,
\end{equation}
where $\vec{R}_i$ is a vector of the Bravais lattice and $\vec{r}_\alpha $
is the basis vector pointing to the $\alpha $-th layer. In all of the
following calculations, Greek indices label layers and Roman indices refer
to sites of the Bravais lattice. It should be noted that translational
invariance can only be assumed within each layer. E.g., the notion of a 
\textit{reciprocal lattice} only makes sense when referring to the
two-dimensional Bravais lattice. A similar statement applies to Fourier
transforms being thought of as connecting real-space quantities with those
defined in wave-vector space.

In the following we shall discuss the Mermin-Wagner theorem for the
Heisenberg model, the Hubbard model, s-f model (Kondo-lattice model), and
the Periodic Anderson Model (PAM).\footnote{%
Parts of this proof have been published in \cite{Gel Nol 2000}.} For film
systems one has to distinguish between Bravais lattice indices and layer
indices for all site-dependent quantities, such as spin operators $%
S_{i\alpha }^{(+,-,x,y,z)},$ annihilation and creation operators $c_{i\alpha
}^{(+)}$ or coupling constants $J_{ij}^{\alpha \beta }$ which depend on two
lattice sites. In this notation, the Hamiltonian for \textit{Heisenberg films%
} is given by

\begin{equation}
H=-\sum_{ij\alpha \beta }J_{ij}^{\alpha \beta }(S_{i\alpha }^{+}S_{j\beta
}^{-}+S_{i\alpha }^zS_{j\beta }^z)-b\sum_{i\alpha }e^{-i\vec{K}\cdot \vec{R}%
_i}S_{i\alpha }^z,
\end{equation}
where the term $b\sum_{i\alpha }e^{-i\vec{K}\cdot \vec{R}_i}S_{i\alpha }^z$
is due to the interaction with an external magnetic field $b=\frac{g_J\mu
_BB_0}\hbar .$ The interaction with an external magnetic field leads to the
magnetization 
\begin{equation}
M=\frac 1{Nd}\frac{g_J\mu _B}\hbar \sum_{i\alpha }e^{-i\vec{K}\cdot \vec{R}%
_i}\left\langle S_{i\alpha }^z\right\rangle \equiv \frac 1d\sum_\alpha
M_\alpha ,
\end{equation}
where the phase factor $e^{-i\vec{K}\cdot \vec{R}_i}$ accounts for both
ferromagnetic and antiferromagnetic ordering, depending on the choice of $%
\vec{K}.$ We also assume the coupling constants $J_{ij}^{\alpha \beta }$ as
satisfying some general properties, i.e. 
\begin{equation}
J_{ij}^{\alpha \beta }=J_{ji}^{\beta \alpha }
\end{equation}
\begin{equation*}
J_{ll}^{\varepsilon \varepsilon }=0
\end{equation*}
and 
\begin{equation}
\frac 1{Nd}\sum_{\gamma \varepsilon }\sum_{mp}\left| J_{pm}^{\varepsilon
\gamma }\right| \frac{\left( \vec{R}_m-\vec{R}_p\right) ^2}4\equiv \tilde{Q}%
<\infty .  \label{exchconv}
\end{equation}
The last condition (\ref{exchconv}) is very weak considering that the
exchange integrals $J$ will usually decay exponentially with distance.

As an example for itinerant-electron systems, we shall discuss the \textit{%
Hubbard model}

\begin{equation}
H=\sum_{ij\alpha \beta \sigma }T_{ij}^{\alpha \beta }c_{i\alpha \sigma
}^{+}c_{j\beta \sigma }+\frac U2\sum_{i\alpha \sigma }n_{i\alpha \sigma
}n_{i\alpha -\sigma }-b\sum_{i\alpha }e^{-i\vec{K}\cdot \vec{R}_i}\sigma
_{i\alpha }^z,  \label{Hubbard real-space}
\end{equation}
where $T_{ij}^{\alpha \beta }$ describes the hopping of an electron from
lattice site $j$ of the $\beta $-th layer to site $i$ of the $\alpha $-th
layer and $U$ is the energy associated with having two electrons at the same
lattice site. Again, a term corresponding to an external magnetic field is
included. The $z$-component of the pseudo-spin, $\sigma _{i\alpha }^z,$ is
defined according to (\ref{pseudosz}). Similarly to the Heisenberg case, we
require the hopping constants $T_{ij}^{\alpha \beta }$ to satisfy the
isotropy conditions $T_{ij}^{\alpha \beta }=T_{ji}^{\beta \alpha }$ as well
as to converge upon summation over all lattice sites: 
\begin{equation}
\frac 1{Nd}\sum_{\gamma \beta }\sum_{nk}\left| T_{nk}^{\gamma \beta }\right| 
\frac{\left( \vec{R}_n-\vec{R}_k\right) ^2}4\equiv \tilde{q}<\infty .
\end{equation}

In the \textit{s-f model} one deals with two spin sub-systems, $\left\{ \vec{%
S}_{i\alpha }\right\} $ and $\left\{ \vec{\sigma}_{i\alpha }\right\} $, the
former consisting of localized $f$-electrons, the latter being associated
with itinerant $s$-electrons:

\begin{equation}
H=\sum_{ij\alpha \beta \sigma }T_{ij}^{\alpha \beta }c_{i\alpha \sigma
}^{+}c_{j\beta \sigma }-\frac J2\sum_{i\alpha \sigma }(z_\sigma S_{i\alpha
}^zn_{i\alpha \sigma }+S_{i\alpha }^\sigma c_{i\alpha -\sigma
}^{+}c_{i\alpha \sigma })-b\sum_{i\alpha }e^{-i\vec{K}\cdot \vec{R}%
_i}(S_{i\alpha }^z+\sigma _{i\alpha }^z).
\end{equation}

A similar situation occurs in the \textit{Periodic Anderson Model}, where
one must distinguish between the electrons of the conduction band (described
by fermionic creation and annihilation operators $c_{i\alpha \sigma }^{(+)})$
and the $f$-electrons (whose creation and annihilation operators we shall
denote by $f_{i\alpha \sigma }^{(+)})$: 
\begin{eqnarray}
H &=&\sum_{ij\alpha \beta \sigma }t_{ij}^{\alpha \beta }c_{i\alpha \sigma
}^{+}c_{j\beta \sigma }+\varepsilon _f\sum_{i\alpha \sigma }f_{i\alpha
\sigma }^{+}f_{i\alpha \sigma }+\frac U2\sum_{i\alpha }n_{i\alpha \uparrow
}^fn_{i\alpha \downarrow }^f  \notag \\
&&+V\sum_{i\alpha \sigma }\left( c_{i\alpha \sigma }^{+}f_{i\alpha \sigma
}+f_{i\alpha \sigma }^{+}c_{i\alpha \sigma }\right) -b\sum_{i\alpha }e^{-i%
\vec{K}\cdot \vec{R}_i}(\sigma _{c_{i\alpha }}^z+\sigma _{f_{i\alpha }}^z) \\
&=&H_s+H_f+H_{ff}+H_{hyb}+H_b.
\end{eqnarray}
Here, $\sigma _{c_{i\alpha }}^z$ and $\sigma _{f_{i\alpha }}^z$ are the $z$%
-components associated with the pseudo-spins of the conduction and the $f$%
-electrons, respectively: 
\begin{eqnarray}
\sigma _{c_{i\alpha }}^z &=&\frac \hbar 2(n_{i\alpha \uparrow }^c-n_{i\alpha
\downarrow }^c)=\frac \hbar 2(c_{i\alpha \uparrow }^{+}c_{i\alpha \uparrow
}-c_{i\alpha \downarrow }^{+}c_{i\alpha \downarrow }) \\
\sigma _{f_{i\alpha }}^z &=&\frac \hbar 2(n_{i\alpha \uparrow }^f-n_{i\alpha
\downarrow }^f)=\frac \hbar 2(f_{i\alpha \uparrow }^{+}f_{i\alpha \uparrow
}-f_{i\alpha \downarrow }^{+}f_{i\alpha \downarrow }).
\end{eqnarray}

\section{Absence of a magnetic phase transition in film systems}

\subsection{Choice of operators and evaluation of the Bogoliubov inequality}

By inspecting eqn. (\ref{bogsimple}) one can easily conclude that the choice
of suitable operators $A$ and $C$ is crucial; it determines whether the
inequality will be physically meaningful or not. In film systems, long-range
magnetic order\textit{\ within a given layer} is conceivable where the bulk
(or even sub-lattice) magnetization of the whole system vanishes, e.g. 
\begin{eqnarray*}
\ldots {} &\uparrow \uparrow \uparrow \uparrow \uparrow \uparrow \uparrow
\uparrow \uparrow \uparrow \uparrow \uparrow \uparrow \uparrow \uparrow
\uparrow \ldots \\
\ldots {} &\uparrow \downarrow \uparrow \downarrow \uparrow \downarrow
\uparrow \downarrow \uparrow \downarrow \uparrow \downarrow \uparrow
\downarrow \uparrow \downarrow \ldots \\
\ldots {} &\uparrow \downarrow \uparrow \downarrow \uparrow \downarrow
\uparrow \downarrow \uparrow \downarrow \uparrow \downarrow \uparrow
\downarrow \uparrow \downarrow \ldots \\
\ldots {} &\downarrow \downarrow \downarrow \downarrow \downarrow \downarrow
\downarrow \downarrow \downarrow \downarrow \downarrow \downarrow \downarrow
\downarrow \downarrow \downarrow \ldots
\end{eqnarray*}

Excluding long-range magnetic order for every layer within a film would,
therefore, be a considerably stronger statement. The general idea of our
proof is to use the Bogoliubov inequality to find an upper bound for the
layer magnetization $M_\beta ,$ i.e. 
\begin{equation*}
M_\beta \leq f(B_0,M),
\end{equation*}
where $f$ is a function that approaches zero as $B_0\rightarrow 0$ and does
not depend on any layer-specific quantities.

This is best achieved by choosing $C$ as a sum $\sum_{\alpha }(...)$ of spin
operators and $A$ as a spin operator associated with a specific layer $\beta 
$. Hence, the numerator of the LHS of the Bogoliubov inequality will be a
layer-dependent quantity, while the double commutator in the denominator
will be summed over all lattice sites and thus be layer-independent. One may
then expect to be able to replace the RHS of the inequality by a
(layer-independent) upper bound. More specifically, for the \textit{%
Heisenberg model} we set 
\begin{equation}
A_{(\alpha )}\equiv S_{\alpha }^{-}(-\vec{k}-\vec{K})
\end{equation}%
and 
\begin{equation}
C\equiv \sum_{\beta }C_{\beta }\equiv \sum_{\beta }S_{\beta }^{+}(\vec{k}),
\end{equation}%
where the Fourier transforms $S_{\beta }^{\pm }(\vec{k})$ of the
layer-dependent operators are defined within the two-dimensional reciprocal
lattice according to eqn. (\ref{fouriernachk}). Note that in $\vec{k}$-space
we have $\left( S_{\beta }^{+}(\vec{k})\right) ^{+}=S_{\beta }^{-}(-\vec{k})$
as is obvious from the definition of the Fourier transform. The operators $S$
in the context of the Heisenberg model are spin operators for which the
usual commutation relations (\ref{zplus}) and (\ref{plusminus}) hold true,
which, after Fourier transformation into $\vec{k}$-space, run 
\begin{equation}
\left[ S_{\alpha }^{+}(\vec{k}_{1}),S_{\beta }^{-}(\vec{k}_{2})\right]
_{-}=2\hbar \delta _{\alpha \beta }S_{\alpha }^{z}(\vec{k}_{1}+\vec{k}_{2})
\end{equation}%
and 
\begin{equation}
\left[ S_{\alpha }^{z}(\vec{k}_{1}),S_{\beta }^{\pm }(\vec{k}_{2})\right]
_{-}=\pm \hbar \delta _{\alpha \beta }S_{\alpha }^{\pm }(\vec{k}_{1}+\vec{k}%
_{2}).
\end{equation}

In the\textit{\ Hubbard model }a similar definition for the operators $A$
and $C$ is used. However, as mentioned above, the operators $S^{(...)}$ are
now built up from fermionic creation and annihilation operators, e.g. 
\begin{equation}
\sigma _\alpha ^{-}(\vec{k}_1)=\hbar c_{\vec{k}_1\alpha \downarrow }^{+}c_{%
\vec{k}_1\alpha \uparrow }  \label{sigma minus}
\end{equation}
and 
\begin{equation}
\sigma _\beta ^{+}(\vec{k}_2)=\hbar c_{\vec{k}_2\uparrow \beta }^{+}c_{\vec{k%
}_2\beta \downarrow }.  \label{sigma plus}
\end{equation}
The operators $A$ and $C$ are, then, 
\begin{equation}
A_{(\alpha )}\equiv \sigma _\alpha ^{-}(-\vec{k}-\vec{K})=\hbar c_{-\vec{k}-%
\vec{K},\alpha \downarrow }^{+}c_{-\vec{k}-\vec{K},\alpha \uparrow }
\end{equation}
and 
\begin{equation}
C\equiv \sum_\beta \sigma _\beta ^{+}(\vec{k})=\hbar \sum_\beta c_{\vec{k}%
\beta \uparrow }^{+}c_{\vec{k}\beta \downarrow }\left( =\sum_\beta C_\beta
\right) ,
\end{equation}
respectively. The commutation relations for spin operators may, in a purely
formal sense, be used in this case as well.

In the \textit{s-f model }and the \textit{Periodic Anderson Model}, one must
not forget that one is dealing with two distinct spin sub-systems. In the
former, one of the subsystems can be described by ordinary spin operators,
while the other is associated with itinerant electrons; in the PAM both
subsystems must be described in terms of fermionic creation and annihilation
operators. In the s-f model, the two systems are independent from one
another in the sense that spin operators and creation or annihilation
operators commute: 
\begin{equation}
\left[ S_{i\alpha }^{(+,-,x,y,z)},c_{j\beta \sigma }^{(+)}\right] _{-}=0
\end{equation}
and, thus, 
\begin{equation}
\left[ S_\alpha ^{(+,-,x,y,z)}(\vec{k}_1),c_{\vec{k}_2\beta \sigma }^{(+)}%
\right] _{-}=0,
\end{equation}
while in the PAM similar anticommutation relations hold between creation and
annihilation operators of the conduction electrons and the $f$-electrons: 
\begin{eqnarray}
\left[ c_{i\alpha \sigma }^{(+)},f_{j\beta \sigma ^{\prime }}^{(+)}\right]
_{+} &=&0  \label{pamacr} \\
\left[ c_{\vec{k}_1\alpha \sigma }^{(+)},f_{\vec{k}_2\beta \sigma ^{\prime
}}^{(+)}\right] _{+} &=&0.
\end{eqnarray}

With (\ref{sigma plus}),(\ref{sigma minus}) we define as operators $A$ and $%
C $ in the s-f model

\begin{equation}
A_{(\gamma )}=S_\gamma ^{-}(-\vec{k}-\vec{K})+\sigma _\gamma ^{-}(-\vec{k}-%
\vec{K})
\end{equation}
and 
\begin{equation}
C\equiv \sum_\beta C_\beta =\sum_\beta \left( S_\beta ^{+}(\vec{k})+\sigma
_\beta ^{+}(\vec{k})\right) ,
\end{equation}
and in the Periodic Anderson Model 
\begin{eqnarray}
A_{(\gamma )} &=&\sigma _{c_\gamma }^{-}(-\vec{k}-\vec{K})+\sigma _{f_\gamma
}(-\vec{k}-\vec{K}) \\
C &\equiv &\sum_\beta C_\beta =\sum_\beta \left( \sigma _{c_\beta }^{+}(\vec{%
k})+\sigma _{f_\beta }^{+}(\vec{k})\right) .
\end{eqnarray}

\subsection{Hamiltonian-independent quantities}

It is obvious from the structure of the Bogoliubov inequality (\ref{bog})
that the numerators on both sides of the inequality are determined entirely
by the choice of operators $A$ and $C$ as mentioned earlier. The Hamiltonian 
$H$ of the many-body model enters the calculation only via the double
commutator $\left\langle \left[ [C,H]_{-},C^{+}\right] _{-}\right\rangle .$
For simplicity, we shall start with the Hamiltonian-independent expectation
values appearing in the Bogoliubov inequality, more specificially with the $%
[C,A]_{-}$\textbf{-commutator} 
\begin{eqnarray}
&&\left\langle \left[ C,A_{(\gamma )}\right] _{-}\right\rangle  \notag \\
&=&\sum_\beta \left\langle \left[ S_\beta ^{+}(\vec{k})+\sigma _\beta ^{+}(%
\vec{k}),S_\gamma ^{-}(-\vec{k}-\vec{K})+\sigma _\gamma ^{-}(-\vec{k}-\vec{K}%
)\right] _{-}\right\rangle \\
&=&\sum_{mn\beta }e^{i\vec{k}\cdot \vec{R}_m}e^{-i(\vec{k}+\vec{K})\cdot 
\vec{R}_n}\left\langle \left[ S_{m\beta }^{+}+\hbar c_{m\beta \uparrow
}^{+}c_{m\beta \downarrow },S_{n\gamma }^{-}+\hbar c_{n\gamma \downarrow
}^{+}c_{n\gamma \uparrow }\right] _{-}\right\rangle  \notag
\end{eqnarray}
for the s-f model. For reasons described above, $S$ and $\sigma $ operators
commute, so the commutator can be evaluated directly by use of the
fundamental commutation relations. This leads to

\begin{eqnarray}
\left\langle \left[ C,A_{(\gamma )}\right] _{-}\right\rangle &=&2\hbar
\sum_{mn\beta }\delta _{mn}\delta _{\beta \gamma }e^{i\vec{k}\vec{R}_m}e^{-i(%
\vec{k}+\vec{K})\cdot \vec{R}_n}\left\langle S_{m\beta }^z+\sigma _{m\beta
}^z\right\rangle  \notag \\
&=&\frac{2\hbar ^2N}{g_J\mu _B}M_\gamma (T,B_0),  \label{sfca}
\end{eqnarray}
where we have introduced the layer magnetization 
\begin{equation}
M_\gamma (T,B_0)=\frac 1N\frac{g_J\mu _B}\hbar \sum_ne^{-i\vec{K}\cdot \vec{R%
}_n}\left\langle S_{n\gamma }^z+\sigma _{n\gamma }^z\right\rangle .
\end{equation}

An analogous result is found for the Periodic Anderson Model where one also
has to take into account two spin systems, associated with the conduction
and the $f$-electrons, respectively. Thus, with a slightly modified
definition for the magnetization (where, for reasons of simplicity, we
choose identical prefactors $g_J\mu_B / \hbar $ for both spin systems), 
\begin{equation}
M_\gamma (T,B_0)=\frac 1N\frac{g_J\mu _B}\hbar \sum_ne^{-i\vec{K}\cdot \vec{R%
}_n}\left\langle \sigma _{c_{n\gamma }}^z+\sigma _{f_{n\gamma
}}^z\right\rangle ,
\end{equation}
eqn. (\ref{sfca}) holds also for the PAM. The same formula (\ref{sfca}),
obviously, is true for the Heisenberg and the Hubbard model, the only
difference being the fact that in these cases only one spin system
contributes to the magnetization. Eqn. (\ref{sfca}) is therefore valid for
all the four models discussed in this section, as indeed one would expect
from the very general considerations that led to our choice for the
operators $A$ and $C.$

The RHS of the Bogoliubov inequality (\ref{bog}) is proportional to the 
\textbf{anticommutator sum} 
\begin{equation}
\sum_{\vec{k}}\left\langle \left[ A,A^{+}\right] _{+}\right\rangle
\end{equation}
which, for the s-f model, is given by 
\begin{equation}
\sum_{\vec{k}}\left[ A,A^{+}\right] _{+}=\sum_{\vec{k}mn}e^{-i(\vec{k}+\vec{K%
})\cdot (\vec{R}_m-\vec{R}_n)}\left( \left[ \left( S_{n\gamma }^{-}+\hbar
c_{n\gamma \downarrow }^{+}c_{n\gamma \uparrow }\right) ,\left( S_{m\gamma
}^{+}+\hbar c_{m\gamma \uparrow }^{+}c_{m\gamma \downarrow }\right) \right]
_{+}\right) .
\end{equation}
Summing the exponential over all $\vec{k}$ gives the delta function $N\delta
_{mn}.$ The mixed commutators involving both spin and creation/annihilation
operators in this case do not vanish, since we have the anticommutator
rather than the commutator. Thus, we arrive at the expression

\begin{eqnarray}
\left\langle \sum_{\vec{k}}\left[ A,A^{+}\right] _{+}\right\rangle
&=&N\sum_n\left( \left\langle \left[ S_{n\gamma }^{-},S_{n\gamma }^{+}\right]
_{+}\right\rangle +2\hbar \left\langle c_{n\gamma \downarrow }^{+}c_{n\gamma
\uparrow }S_{n\gamma }^{+}+c_{n\gamma \uparrow }^{+}c_{n\gamma \downarrow
}S_{n\gamma }^{-}\right\rangle \right.  \notag \\
&&\left. +{\hbar }^2\left\langle c_{n\gamma \downarrow }^{+}c_{n\gamma
\uparrow }c_{n\gamma \uparrow }^{+}c_{n\gamma \downarrow }+c_{n\gamma
\uparrow }^{+}c_{n\gamma \downarrow }c_{n\gamma \downarrow }^{+}c_{n\gamma
\uparrow }\right\rangle \right) .
\end{eqnarray}
\linebreak Since we are interested in the quantity $\left\langle \sum_{\vec{k%
}}\left[ A,A^{+}\right] _{+}\right\rangle $ as an upper bound in the
Bogoliubov inequality, it suffices to find upper bounds for the individual
expectation values on the RHS. For the\textit{\ Heisenberg term} we find

\begin{equation}
\sum_n\left\langle \left[ S_{n\gamma }^{-},S_{n\gamma }^{+}\right]
_{+}\right\rangle =2\sum_n\left\langle (S_{n\gamma }^x)^2+(S_{n\gamma
}^y)^2\right\rangle \leq 2\sum_n\left\langle \vec{S}_{n\gamma
}^2\right\rangle =2\hbar ^2S(S+1)N.
\end{equation}
For the \textit{Hubbard contribution} we have 
\begin{eqnarray}
\sum_n\left\langle c_{n\gamma \downarrow }^{+}c_{n\gamma \uparrow
}c_{n\gamma \uparrow }^{+}c_{n\gamma \downarrow }+c_{n\gamma \uparrow
}^{+}c_{n\gamma \downarrow }c_{n\gamma \downarrow }^{+}c_{n\gamma \uparrow
}\right\rangle &\leq &\sum_n\left\langle c_{n\gamma \downarrow
}^{+}c_{n\gamma \downarrow }\left( 1-c_{n\gamma \uparrow }^{+}c_{n\gamma
\uparrow }\right) \right.  \notag \\
&&\left. +c_{n\gamma \uparrow }^{+}c_{n\gamma \uparrow }\left( 1-c_{n\gamma
\downarrow }^{+}c_{n\gamma \downarrow }\right) \right\rangle  \notag \\
&\leq &\sum_n\left( \left\langle n_{n\gamma \downarrow }\right\rangle
+\left\langle n_{n\gamma \uparrow }\right\rangle \right)  \notag \\
&\leq &2N.
\end{eqnarray}

For the \textit{mixed terms} appearing in the s-f model, an upper bound is
given by

\begin{equation}
\sum_{n\sigma }\left\langle c_{n\gamma \sigma }^{+}c_{n\gamma -\sigma
}S_{n\gamma }^{-\sigma }\right\rangle \leq 2\left( 4+2S(S+1)\right) \hbar ^2N
\label{sfacsum}
\end{equation}
as will be shown in the following mathematical digression (see also the
derivation of (\ref{gemischte c}) in appendix B).

\subsubsection{Mathematical digression: Constructing an upper bound for $%
\sum_{\protect\sigma }\left\langle \hbar c_{n\protect\gamma \protect\sigma %
}^{+}c_{n\protect\gamma -\protect\sigma }S_{n\protect\gamma }^{-\protect%
\sigma }\right\rangle $}

The problem in evaluating this expectation value arises from the fact that
we are dealing with seemingly unrelated operators. It is, therefore,
convenient to use the identity 
\begin{eqnarray}
\hbar c_{n\gamma \sigma }^{+}c_{n\gamma -\sigma }S_{n\gamma }^{-\sigma }
&=&\frac 14\left\{ \left( \hbar c_{n\gamma \sigma }^{+}c_{n\gamma -\sigma
}+S_{n\gamma }^\sigma \right) \left( \hbar c_{n\gamma -\sigma
}^{+}c_{n\gamma \sigma }+S_{n\gamma }^{-\sigma }\right) \right.  \notag \\
&&-\left( \hbar c_{n\gamma \sigma }^{+}c_{n\gamma -\sigma }-S_{n\gamma
}^\sigma \right) \left( \hbar c_{n\gamma -\sigma }^{+}c_{n\gamma \sigma
}-S_{n\gamma }^{-\sigma }\right)  \notag \\
&&+i\left( \hbar c_{n\gamma \sigma }^{+}c_{n\gamma -\sigma }+iS_{n\gamma
}^\sigma \right) \left( \hbar c_{n\gamma -\sigma }^{+}c_{n\gamma \sigma
}-iS_{n\gamma }^{-\sigma }\right)  \notag \\
&&\left. -i\left( \hbar c_{n\gamma \sigma }^{+}c_{n\gamma -\sigma
}-iS_{n\gamma }^\sigma \right) \left( \hbar c_{n\gamma -\sigma
}^{+}c_{n\gamma \sigma }+iS_{n\gamma }^{-\sigma }\right) \right\}  \notag \\
&\equiv &\frac 14\sum_{j=1}^4\phi (j)B_{j\sigma }B_{j\sigma }^{+},
\label{symccs}
\end{eqnarray}
where $j$ labels the individual terms and $\phi $ is a phase factor ($\phi
(1)=+1,$ $\phi (2)=-1,$ $\phi (3)=i,$ $\phi (4)=-i).$ The original
expectation value has thus been decomposed into the sum of expectation
values of pairs of adjunct operators $B_{j\sigma },B_{j\sigma }^{+}.$ Using
the spectral theorem, we find 
\begin{equation}
\sum_\sigma \left\langle \hbar c_{n\gamma \sigma }^{+}c_{n\gamma -\sigma
}S_{n\gamma }^{-\sigma }\right\rangle =\frac 1{4\hbar
}\sum_{j=1}^4\int\limits_{-\infty }^\infty dE\frac 1{e^{\beta E}+1}\phi
(j)\sum_\sigma S_{B_{j\sigma }^{+}B_{j\sigma }}^{(-)}(E),
\end{equation}
where $S_{B_{j\sigma }^{+}B_{j\sigma }}^{(-)}(E)=\frac 1{2\pi }\left\langle %
\left[ B_{j\sigma }^{+},B_{j\sigma }\right] _{+}\right\rangle (E)$ is the
spectral density in its energy representation. The LHS is the expectation
value of the sum of two adjunct operators and is therefore real. We may now
make use of the fact that for pairs of adjunct operators, the spectral
density is positive definite, which, together with the triangle inequality
and $\left| \phi (j)\right| =1$ gives an upper bound for the RHS: 
\begin{equation}
\frac 1{4\hbar }\sum_{j=1}^4\int\limits_{-\infty }^\infty dE\frac 1{e^{\beta
E}+1}\phi (j)\sum_\sigma S_{B_{j\sigma }^{+}B_{j\sigma }}^{(-)}(E)\leq \frac
14\sum_\sigma \sum_{j=1}^4\frac 1\hbar \int\limits_{-\infty }^\infty
dES_{B_{j\sigma }^{+}B_{j\sigma }}^{(-)}(E).
\end{equation}
The sum on the RHS now consists of the $0$-th spectral moments associated
with the operator pairs $B_{j\sigma }^{+},B_{j\sigma }.$ Each of the
spectral moments is given by the expectation value of the anticommutator 
\begin{equation}
\frac 1\hbar \int\limits_{-\infty }^\infty dES_{B_{j\sigma }^{+}B_{j\sigma
}}^{(-)}(E)\equiv M_{B_{j\sigma }^{+}B_{j\sigma }}^{(0)}=\left\langle \left[
B_{j\sigma }^{+},B_{j\sigma }\right] _{+}\right\rangle .
\end{equation}
The anticommutators can be easily evaluated: 
\begin{eqnarray*}
M_{B_{1\sigma }^{+}B_{1\sigma }}^{(0)} &=&\left\langle \left[ \hbar
c_{n\gamma -\sigma }^{+}c_{n\gamma \sigma }+S_{n\gamma }^{-\sigma },\hbar
c_{n\gamma \sigma }^{+}c_{n\gamma -\sigma }+S_{n\gamma }^\sigma \right]
_{+}\right\rangle \\
&\leq &\left\langle n_{n\gamma \sigma }-2n_{n\gamma \sigma }n_{n\gamma
-\sigma }+n_{n\gamma -\sigma }\right\rangle \hbar ^2+2S(S+1)\hbar ^2 \\
&&+2\hbar \left\langle S_{n\gamma }^\sigma c_{n\gamma -\sigma
}^{+}c_{n\gamma \sigma }+S_{n\gamma }^{-\sigma }c_{n\gamma \sigma
}^{+}c_{n\gamma -\sigma }\right\rangle \\
&\leq &4\hbar ^2+2S(S+1)\hbar ^2+2\hbar \left\langle S_{n\gamma }^\sigma
c_{n\gamma -\sigma }^{+}c_{n\gamma \sigma }+S_{n\gamma }^{-\sigma
}c_{n\gamma \sigma }^{+}c_{n\gamma -\sigma }\right\rangle \\
M_{B_{2\sigma }^{+}B_{2\sigma }}^{(0)} &\leq &4\hbar ^2+2S(S+1)\hbar
^2-2\hbar \left\langle S_{n\gamma }^\sigma c_{n\gamma -\sigma
}^{+}c_{n\gamma \sigma }+S_{n\gamma }^{-\sigma }c_{n\gamma \sigma
}^{+}c_{n\gamma -\sigma }\right\rangle \\
M_{B_{3\sigma }^{+}B_{3\sigma }}^{(0)} &\leq &4\hbar ^2+2S(S+1)\hbar
^2+2\hbar i\left\langle S_{n\gamma }^\sigma c_{n\gamma -\sigma
}^{+}c_{n\gamma \sigma }-S_{n\gamma }^{-\sigma }c_{n\gamma \sigma
}^{+}c_{n\gamma -\sigma }\right\rangle \\
M_{B_{4\sigma }^{+}B_{4\sigma }}^{(0)} &\leq &4\hbar ^2+2S(S+1)\hbar
^2-2\hbar i\left\langle S_{n\gamma }^\sigma c_{n\gamma -\sigma
}^{+}c_{n\gamma \sigma }-S_{n\gamma }^{-\sigma }c_{n\gamma \sigma
}^{+}c_{n\gamma -\sigma }\right\rangle
\end{eqnarray*}
and hence 
\begin{equation}
\sum_\sigma \left\langle c_{n\gamma \sigma }^{+}c_{n\gamma -\sigma
}S_{n\gamma }^{-\sigma }\right\rangle \leq 2\left( 4+2S(S+1)\right) \hbar ^2
\label{enddigress}
\end{equation}
\bigskip which proves the upper bound (\ref{sfacsum}). \hfill $\blacksquare $

An analogous treatment of the \textit{PAM} anticommutator sum leads to a
similar result, so that one can tabulate the results for $\sum_{\vec{k}%
}\left\langle \left[ A,A^{+}\right] _{+}\right\rangle $ to get...

\begin{itemize}
\item ...for the Heisenberg model: 
\begin{equation}
\sum_{\vec{k}}\left\langle \left[ A,A^{+}\right] _{+}\right\rangle \leq
2S(S+1)\hbar ^2N^2
\end{equation}

\item ...for the Hubbard model: 
\begin{equation}
\sum_{\vec{k}}\left\langle \left[ A,A^{+}\right] _{+}\right\rangle \leq
2\hbar ^2N^2
\end{equation}

\item ...for the s-f/Kondo-lattice model: 
\begin{equation}
\sum_{\vec{k}}\left\langle \left[ A,A^{+}\right] _{+}\right\rangle \leq
\left( 4S(S+1)+10\right) \hbar ^2N^2
\end{equation}

\item ...and for the Periodic Anderson Model: 
\begin{equation}
\sum_{\vec{k}}\left\langle \left[ A,A^{+}\right] _{+}\right\rangle \leq
8\hbar ^2N^2.
\end{equation}
\end{itemize}

\subsection{The double commutator $\left\langle \left[ [C,H]_{-},C^{+}\right]
_{-}\right\rangle $}

The double commutator $\left\langle \left[ [C,H]_{-},C^{+}\right]
_{-}\right\rangle $ still remains to be calculated. For reasons of
simplicity, we shall, in this section, describe the calculation for all
three models step by step. The Hamiltonians all decompose into 
\begin{equation*}
H=H_0+H_b,
\end{equation*}
where $H_b$ denotes the contribution to the Hamiltonian due to the external
magnetic field.

\subsubsection{The Heisenberg case}

The double commutator $\left\langle \left[ [C,H]_{-},C^{+}\right]
_{-}\right\rangle $ in this case is 
\begin{eqnarray}
\left\langle \left[ \lbrack C,H]_{-},C^{+}\right] _{-}\right\rangle
&=&\sum_{\gamma \varepsilon }\left\langle \left[ \left[ S_\gamma ^{+}(\vec{k}%
),H_0+H_b\right] _{-},S_\varepsilon ^{-}(-\vec{k})\right] _{-}\right\rangle 
\notag \\
&=&\sum_{\gamma \varepsilon }\sum_{mp}e^{-i\vec{k}\cdot (\vec{R}_m-\vec{R}%
_p)}\left\langle \left[ [S_{m\gamma }^{+},H]_{-},S_{p\varepsilon }^{-}\right]
_{-}\right\rangle .  \label{Heisen double commut}
\end{eqnarray}
The real-space commutator on the RHS is easily evaluated to give 
\begin{eqnarray}
\left[ \left[ S_{m\gamma }^{+},H\right] _{-},S_{p\varepsilon }^{-}\right]
_{-} &=&-J_{pm}^{\varepsilon \gamma }\hbar ^2\left( 4S_{p\varepsilon
}^zS_{m\gamma }^z+S_{p\varepsilon }^{-}S_{m\gamma }^{+}+S_{m\gamma
}^{+}S_{p\varepsilon }^{-}\right)  \notag \\
&&+2\sum_{\alpha i}\left( J_{im}^{\alpha \gamma }\hbar ^2\delta _{mp}\delta
_{\gamma \varepsilon }\left( 2S_{i\alpha }^zS_{m\gamma }^z+S_{i\alpha
}^{+}S_{m\gamma }^{-}\right) \right)  \notag \\
&&+2b\hbar ^2e^{-i\vec{K}\cdot \vec{R}_m}\delta _{mp}\delta _{\gamma
\varepsilon }S_{m\gamma }^z.
\end{eqnarray}
Inserting this into the full expression (\ref{Heisen double commut}), we
arrive at 
\begin{eqnarray}
\left\langle \left[ \lbrack C,H]_{-},C^{+}\right] _{-}\right\rangle (\vec{k}%
) &=&\sum_{\gamma \varepsilon }\sum_{mp}J_{pm}^{\varepsilon \gamma }\hbar
^2\left( \left( 1-e^{-i\vec{k}\cdot (\vec{R}_m-\vec{R}_p)}\right) \cdot
\right.  \notag \\
&&\left. \cdot \left\langle 2S_{p\varepsilon }^zS_{m\gamma }^z+S_{m\gamma
}^{+}S_{p\varepsilon }^{-}\right\rangle \right) \\
&&+2b\hbar ^2\sum_\varepsilon \sum_me^{-i\vec{K}\cdot \vec{R}_m}\left\langle
S_{m\varepsilon }^z\right\rangle .  \notag
\end{eqnarray}
To this we add the double commutator $\left\langle \left[ [C,H]_{-},C^{+}%
\right] _{-}\right\rangle (-\vec{k})$ which, as discussed above, is a
positive real number. Replacing the spin operator expectation values by the
upper bound $2\hbar ^2S(S+1),$ we find, after some minor algebra, 
\begin{equation}
\left\langle \left[ \lbrack C,H]_{-},C^{+}\right] _{-}\right\rangle \leq
4Nd\hbar ^2\left| B_0M(T,B_0)\right| +8\hbar ^4S(S+1)\sum_{\gamma
\varepsilon }\sum_{mp}\left| J_{pm}^{\varepsilon \gamma }\right| \frac{\vec{k%
}^2(\vec{R}_m-\vec{R}_p)^2}4,
\end{equation}
where we have already used the fact that 
\begin{equation}
1-\cos (\vec{k}\cdot (\vec{R}_m-\vec{R}_p))\leq \frac{\vec{k}^2(\vec{R}_m-%
\vec{R}_p)^2}4.
\end{equation}
With the above definition of the constant $\tilde{Q},$ we arrive at the
final result 
\begin{equation}
\left\langle \left[ \lbrack C,H]_{-},C^{+}\right] _{-}\right\rangle \leq
4Nd\hbar ^2\left( \left| B_0M(T,B_0)\right| +2\hbar ^2S(S+1)\tilde{Q}\vec{k}%
^2\right) .
\end{equation}

\subsubsection{The Hubbard case}

Again, we need to calculate the full double commutator $\left\langle \left[
[C,H]_{-},C^{+}\right] _{-}\right\rangle $. With the Hubbard Hamiltonian
given in eqn. (\ref{Hubbard real-space}), we have 
\begin{equation}
\sum_\gamma \left[ \sigma _\gamma ^{+}(\vec{k}),H\right] _{-}=\hbar
\sum_{ij\alpha \beta }T_{ij}^{\alpha \beta }c_{i\alpha \uparrow
}^{+}c_{j\beta \downarrow }\left( e^{-i\vec{k}\cdot \vec{R}_i}-e^{-i\vec{k}%
\cdot \vec{R}_j}\right) +b\hbar \sum_\alpha \sigma _\alpha ^{+}(\vec{k}+\vec{%
K})
\end{equation}
which still needs to be commuted with $\sum_\varepsilon \sigma _\varepsilon
^{-}(-\vec{k}).$ Replacing the expectation values by their modulus, one gets
the relevant inequality 
\begin{eqnarray}
\left\langle \left[ \lbrack C,H]_{-},C^{+}\right] _{-}\right\rangle (\vec{k}%
) &=&\sum_{\gamma \varepsilon }\left\langle \left[ \left[ \sigma _\gamma
^{+}(\vec{k}),H_0+H_b\right] _{-},\sigma _\varepsilon ^{-}(-\vec{k})\right]
_{-}\right\rangle  \notag \\
&\leq &\hbar ^2\sum_{il\alpha \varepsilon }T_{il}^{\alpha \varepsilon
}\left( e^{-i\vec{k}\cdot (\vec{R}_i-\vec{R}_l)}-1\right) \left( \left|
\left\langle c_{i\alpha \uparrow }^{+}c_{l\varepsilon \uparrow
}\right\rangle \right| \right. +  \notag \\
&&\left. \left| \left\langle c_{l\varepsilon \downarrow }^{+}c_{i\alpha
\downarrow }\right\rangle \right| \right) +2b\hbar ^2\sum_\alpha
\left\langle \sigma _\alpha ^z(\vec{K})\right\rangle .
\end{eqnarray}
For the same reasons as above, we may now add $\left\langle \left[
[C,H]_{-},C^{+}\right] _{-}\right\rangle (-\vec{k})$ to get an upper bound
for the LHS. The expectation values $\left| \left\langle c_{i\alpha \uparrow
}^{+}c_{l\varepsilon \uparrow }\right\rangle \right| +\left| \left\langle
c_{l\varepsilon \downarrow }^{+}c_{i\alpha \downarrow }\right\rangle \right| 
$ can be bounded from above by a procedure analogous to the one carried out
in eqns. (\ref{symccs})-(\ref{enddigress}), leading to $\left| \left\langle
c_{i\alpha \uparrow }^{+}c_{l\varepsilon \uparrow }\right\rangle \right|
+\left| \left\langle c_{l\varepsilon \downarrow }^{+}c_{i\alpha \downarrow
}\right\rangle \right| \leq 4$ (see appendix B, eqn. (\ref{gemischte c}),
for detailed calculation). In total we may thus write, using the notation
above, 
\begin{equation}
\left\langle \left[ \lbrack C,H]_{-},C^{+}\right] _{-}\right\rangle \leq
4Nd\hbar ^2\left( \left| B_0M(T,B_0)\right| +2\tilde{q}\vec{k}^2\right) .
\end{equation}

\subsubsection{The s-f case}

The s-f double commutator 
\begin{equation}
\left\langle \left[ \lbrack C,H]_{-},C^{+}\right] _{-}\right\rangle
=\sum_{\gamma \varepsilon }\left\langle \left[ \left[ S_\gamma ^{+}(\vec{k}%
)+\sigma _\gamma ^{+}(\vec{k}),H_0+H_b\right] _{-},S_\varepsilon ^{-}(-\vec{k%
})+\sigma _\varepsilon ^{-}(-\vec{k})\right] _{-}\right\rangle
\end{equation}
may be computed by considering first the field-independent contribution to
the Hamiltonian 
\begin{equation}
H_0=\sum_{ij\alpha \beta \sigma }T_{ij}^{\alpha \beta }c_{i\alpha \sigma
}^{+}c_{j\beta \sigma }-\frac J2\sum_{i\alpha \sigma }\left( z_\sigma
S_{i\alpha }^zn_{i\alpha \sigma }+S_{i\alpha }^\sigma c_{i\alpha -\sigma
}^{+}c_{i\alpha \sigma }\right)
\end{equation}
and the respective commutators 
\begin{equation}
\left[ \sigma _{m\beta }^{+},H_0\right] _{-}=\sum_{k\gamma }T_{km}^{\gamma
\beta }\hbar \left( c_{m\beta \uparrow }^{+}c_{k\gamma \downarrow
}-c_{k\gamma \uparrow }^{+}c_{m\beta \downarrow }\right) -J\hbar \left(
S_{m\beta }^{+}\sigma _{m\beta }^z-2S_{m\beta }^z\sigma _{m\beta }^{+}\right)
\end{equation}
\begin{equation}
\left[ S_{m\beta }^{+},H_0\right] _{-}=J\hbar \left( S_{m\beta }^{+}\sigma
_{m\beta }^z-2S_{m\beta }^z\sigma _{m\beta }^{+}\right) .
\end{equation}
We then find (neglecting for the moment the external contribution $H_b$) 
\begin{eqnarray}
\left\langle \left[ \lbrack C,H_0]_{-},C^{+}\right] _{-}\right\rangle (\vec{k%
}) &=&\sum_{kmn}\sum_{\beta \gamma \delta }e^{-i\vec{k}\cdot (\vec{R}_m-\vec{%
R}_n)}T_{km}^{\gamma \beta }\hbar ^2\cdot  \notag \\
&&\cdot \left\langle \left[ \left( c_{m\beta \uparrow }^{+}c_{k\gamma
\downarrow }-c_{k\gamma \uparrow }^{+}c_{m\beta \downarrow }\right) ,\left(
S_{n\delta }^{-}+c_{n\delta \downarrow }^{+}c_{n\delta \uparrow }\right) %
\right] _{-}\right\rangle  \notag \\
&=&-\sum_{kn}\sum_{\beta \gamma \delta }\left( 1-e^{i\vec{k}\cdot (\vec{R}_n-%
\vec{R}_m)}\right) \cdot  \notag \\
&&\cdot \left( \delta _{\delta \beta }T_{kn}^{\gamma \beta }c_{n\delta
\downarrow }^{+}c_{k\gamma \downarrow }+\delta _{\delta \gamma
}T_{nk}^{\gamma \beta }c_{k\beta \uparrow }^{+}c_{n\delta \uparrow }\right)
\hbar ^2.
\end{eqnarray}
Following the usual procedure of adding $\left\langle \left[ [C,H]_{-},C^{+}%
\right] _{-}\right\rangle (-\vec{k})$ we arrive at the upper bound 
\begin{eqnarray}
\left\langle \left[ \lbrack C,H_0]_{-},C^{+}\right] _{-}\right\rangle &\leq
&2\hbar ^2\sum_{nk\gamma \beta }\left| T_{nk}^{\gamma \beta }\right| \left(
1-\cos \left( \vec{k}\cdot \left( \vec{R}_n-\vec{R}_k\right) \right) \right)
\cdot  \notag \\
&&\cdot \left( \left| \left\langle c_{n\beta \downarrow }^{+}c_{k\gamma
\downarrow }\right\rangle \right| +\left| \left\langle c_{n\beta \uparrow
}^{+}c_{k\gamma \uparrow }\right\rangle \right| \right)  \notag \\
&\leq &2\hbar ^2\sum_{nk\gamma \beta }\left| T_{nk}^{\gamma \beta }\right| 
\vec{k}^2\left( \vec{R}_n-\vec{R}_k\right) ^2  \notag \\
&=&2Nd\tilde{q}\hbar ^2\vec{k}^2,
\end{eqnarray}
where $\left| \left\langle c_{n\beta \downarrow }^{+}c_{k\gamma \downarrow
}\right\rangle \right| +\left| \left\langle c_{n\beta \uparrow
}^{+}c_{k\gamma \uparrow }\right\rangle \right| \leq 4$ has already been
used.

We still need to add the double commutator $\left\langle \left[
[C,H_b]_{-},C^{+}\right] _{-}\right\rangle $ with the external part $H_b$ of
the Hamiltonian to this preliminary result. Here, too, we need to take into
account the presence of two distinct spin systems. Thus, $H_b$ is given by 
\begin{equation}
H_b=-B_0\frac{\mu _B}\hbar \sum_{i\alpha }(g_JS_{i\alpha }^z+2\sigma
_{i\alpha }^z)e^{-i\vec{K}\cdot \vec{R}_i},
\end{equation}
where the usual definitions of the spin operators apply. As in the previous
cases, one gets 
\begin{equation}
\left\langle \left[ \lbrack C,H_b]_{-},C^{+}\right] _{-}\right\rangle
=-2\hbar ^2NdB_0M(T,B_0),
\end{equation}
where, however, the magnetization is now defined as 
\begin{equation}
M(T,B_0)=\frac 1{Nd}\frac{\mu _B}\hbar \sum_{i\beta }e^{-i\vec{K}\cdot \vec{R%
}_i}\left( \left\langle g_JS_{i\beta }^z\right\rangle +2\left\langle \sigma
_{i\beta }^z\right\rangle \right) .
\end{equation}

Finally, we arrive at 
\begin{equation}
\left\langle \left[ \lbrack C,H]_{-},C^{+}\right] _{-}\right\rangle \leq
2Nd\hbar ^2\left( \left| B_0M(T,B_0)\right| +\tilde{q}\vec{k}^2\right) .
\end{equation}

\subsubsection{The PAM case}

For completeness, we also present the results for the Periodic Anderson
Model. The detailed calculations do not differ much, in spirit, from the
ones carried out above for the Hubbard and the s-f model, except for the
difference in the (anti)commutation relations (\ref{pamacr}) as compared to
the s-f model. The external part $H_{b}$ of the Hamiltonian again leads to a
contribution to the double commutator that is proportional to the total
number of lattice sites $N\cdot d,$ the external field $B_{0}$ and the
magnetization $M(T,B_{0}),$ and its expectation value is bounded from above
by 
\begin{equation}
\left\langle \left[ \lbrack C,H_{b}]_{-},C^{+}\right] _{-}\right\rangle \leq
4Nd\hbar ^{2}\left| B_{0}M(T,B_{0})\right| .
\end{equation}%
What is noteworthy about the double commutator with the unperturbed PAM
Hamiltonian $H_{0}$ is that only the hopping part $H_{s}$ leads to a
non-vanishing contribution, while $H_{f},$ $H_{ff},$ and $H_{hyb}$ do not
even ``survive'' the first of the two commutations,\footnote{%
The fact that in the PAM the anticommutation relations (\ref{pamacr}) hold
(instead of the commutation relations for the s-f model) is irrelevant for
the commutator with $H_{s},$ $H_{f},$ $H_{ff},$ as in these cases only pairs
of $c$- or $f$-operators occur. For the hybridization part $H_{hyb}$ it is
straightforward to show that in this case, too, already the first of the two
commutators vanishes.} 
\begin{equation}
\left[ C,H_{f}\right] _{-}=\left[ C,H_{ff}\right] _{-}=\left[ C,H_{hyb}%
\right] _{-}=0.
\end{equation}%
Thus, the double commutator is essentially of the Hubbard form 
\begin{equation}
\left\langle \left[ \lbrack C,H]_{-},C^{+}\right] _{-}\right\rangle \leq
4Nd\hbar ^{2}\left( \left| B_{0}M(T,B_{0})\right| +2\tilde{q}\vec{k}%
^{2}\right) .
\end{equation}%
The fact that the bound on the double commutator is independent of the
hybridization and other parameters in the Hamiltonian, except for the
hopping, confirms similar results obtained for the purely two-dimensional
case \cite{Pro Lop 1983}.

\subsection{Proving the absence of spontaneous magnetization}

Within all the models discussed (Heisenberg, Hubbard, s-f/Kondo-lattice, and
Periodic Anderson Model), we have found for the double commutator 
\begin{equation}
\left\langle \left[ \lbrack C,H]_{-},C^{+}\right] _{-}\right\rangle \leq \xi
_0^2Nd\left( \left| B_0M(T,B_0)\right| +\xi _1\vec{k}^2\right) ,
\end{equation}
where $\xi _i$ are constants depending, at most, on fixed parameters of the
respective many-body models. We also know that 
\begin{equation}
\left\langle \left[ C,A\right] _{-}\right\rangle =\xi _2NM_\gamma (T,B_0)
\end{equation}
and 
\begin{equation}
\sum_{\vec{k}}\left\langle \left[ A,A^{+}\right] _{+}\right\rangle \leq 2\xi
_3N^2.
\end{equation}

We can now give a generally applicable discussion of the layer
magnetization. With the Bogoliubov inequality (\ref{bog}),

\begin{equation}
\sum_{\vec{k}}\frac{\left| \left\langle \left[ C,A\right] _{-}\right\rangle
\right| ^2}{\left\langle \left[ [C,H]_{-},C^{+}\right] _{-}\right\rangle }%
\leq \frac \beta 2\sum_{\vec{k}}\left\langle \left[ A,A^{+}\right]
_{+}\right\rangle
\end{equation}
we find 
\begin{equation}
\sum_{\vec{k}}\frac{\xi _2^2N^2M_\gamma ^2(T,B_0)}{\xi _0^2Nd\left( \left|
B_0M(T,B_0)\right| +\xi _1\vec{k}^2\right) }\leq \xi _3\beta N^2.
\end{equation}
The LHS of the inequality can be replaced by an integral, using the formula 
\begin{equation}
\sum_{\vec{k}}\hat{=}\frac{L^2}{(2\pi )^2}\int\limits_{\vec{k}}d^2\vec{k},
\end{equation}
where $\frac{L^2}{(2\pi )^2}$ is the area in two-dimensional $\vec{k}$-space
associated with one quantum state. Restricting the support of the integral
to a finite-volume sphere inscribed into the first Brillouin zone only
strengthens the inequality, so 
\begin{equation}
\left( \frac{\xi _2}{\xi _0}\right) ^2\frac 1{2\pi d}\frac{L^2}NM_\gamma
^2(T,B_0)\int\limits_0^{k_0}dk\frac k{\left| B_0M(T,B_0)\right| +\xi
_1k^2}\leq \xi _3\beta ,
\end{equation}
where $k_0$ is the cutoff corresponding to the sphere in $\vec{k}$-space. In
the thermodynamic limit, $L$ and $N$ are taken to infinity $(L,N\rightarrow
\infty )$ in such a way that the two-dimensional specific volume $v_0^{(2)}=%
\frac{L^2}N$ approaches a finite constant. Evaluating the integral and
performing some minor algebra, we have 
\begin{equation}
M_\gamma ^2(T,B_0)\leq \xi \frac{\beta d}{\ln \left( 1+\frac{\xi _1k_0^2}{%
\left| B_0M(T,B_0)\right| }\right) }
\end{equation}
($\xi $ again is an unsignificant constant).

From this formula, it is clear that, for any finite temperature $(\beta
<\infty )$ and finite thickness $(d<\infty ),$ the logarithm in the
denominator will diverge in the limit $B_0\rightarrow 0,$ thus forcing the
layer magnetization $M_\gamma $ to vanish. One should note that the final
result does not depend on the choice of $\vec{K}$ and, thus, excludes both
ferromagnetic and antiferromagnetic ordering. Thus, we have proved that the
Mermin-Wagner theorem which was originally shown to exclude a magnetic phase
transition at finite temperature for one and two-dimensional systems can be
extended to Heisenberg, Hubbard, s-f (Kondo-lattice), and PAM films of 
\textit{any finite thickness}. While this was to be expected from very
general considerations, we have been able to confirm this conjecture through
a detailed microscopic calculation that takes into account specific
Hamiltonians and distinguishes between individual layers.

Our calculation suggests that the parameter $d$, i.e. the number of layers
in a film, plays a similar role as the inverse temperature $\beta .$ If
either one diverges, a phase transition cannot be ruled out. It is, of
course, not \textit{a priori }clear if the behaviour of an upper bound on a
physical quantity has any physical significance itself; however, it is
reasonable to take this result as indicating that in order to describe
(anti)ferromagnetism, one has to take the thermodynamic limit seriously, in
the sense that $N\rightarrow \infty $ does not suffice, but $d\rightarrow
\infty $ is required as well.

\section{Problem of superconducting long-range order}

The disovery of high-temperature superconductivity with its surprising
characteristics has opened up a new field of research within many-body
theory. The details of the origin of superconducting long-range order in the
high-T$_c$ superconductors are still controversially debated, but it seems
certain that the transition to the superconducting state is due to the
strong electron-electron correlations. Thus, the same set of quantum
many-body models used for describing spontaneous magnetic order in
itinerant-electron systems is also appropriate for the superconducting phase
transition. In order to demonstrate that our discussion of the absence of a
magnetic phase transition in film systems can be extended to cover the
superconducting transition as well, we shall carry out the calculations for 
\textit{Hubbard films of finite thickness} which are believed to reflect the
main properties of electrons moving in the set of coupled CuO planes which
is characteristic of many of the high-temperature superconductors under
investigation.

\subsection{$s$-wave Cooper pairing and generalized $\protect\eta $ pairing}

In superconductors the essential new feature is the formation of Cooper
pairs, i.e. of pairs of conduction electrons with opposite spin and opposite
wave vector 
\begin{equation*}
(\vec{k}\uparrow ,-\vec{k}\downarrow )
\end{equation*}
which travel freely through the crystal lattice. Before one attempts to
apply the Bogoliubov inequality to investigate the possibility of a phase
transition to the superconducting state, one must decide on a quantity, or
order parameter, to characterize the phenomenon. For (anti)ferromagnets this
decision was fairly obvious, since the macroscopic quantity measured
experimentally, i.e. the (staggered) magnetization, is straightforwardly
given, on a microscopic level, by the sum over the $z$-component of the
spins. Instead of characterizing superconductors by their vanishing
resistivity on the macroscopic level, we shall focus on the underlying
microscopic process of \textit{pairing} between electrons, which is
characterized by a breakdown of U(1) symmetry of the conduction electrons.

Before giving the Hamiltonian for Hubbard films with an external U(1)
symmetry breaking term, we briefly summarize the local on-site pairing
operators $\eta ^{\pm }$ which describe the creation or annihilation of a
Cooper pair at lattice site $(i,\alpha )$ where $i$ denotes the position in
the 2D Bravais lattice and $\alpha $ as usual is the layer index: 
\begin{eqnarray}
\eta _{i\alpha }^{+} &=&c_{i\alpha \uparrow }^{+}c_{i\alpha \downarrow }^{+}
\label{etaplus} \\
\eta _{i\alpha }^{-} &=&\left( \eta _{i\alpha }^{+}\right) ^{+}  \notag \\
&=&c_{i\alpha \downarrow }c_{i\alpha \uparrow }.  \label{etaminus}
\end{eqnarray}
We also introduce an operator $\eta _{i\alpha }^z$defined as 
\begin{equation}
\eta _{i\alpha }^z=\frac 12\left( n_{i\alpha \uparrow }+n_{i\alpha
\downarrow }-2\right) .
\end{equation}
Going over to $k$-space by Fourier transformation, one can verify the
commutation relations 
\begin{eqnarray}
\left[ \eta _\alpha ^{\pm }(\vec{k}),\eta _\beta ^z(\vec{k}^{\prime })\right]
_{-} &=&\mp \eta ^{\pm }(\vec{k}+\vec{k}^{\prime })\delta _{\alpha \beta } \\
\left[ \eta _\alpha ^{\pm }(\vec{k}),\eta _\beta ^{\mp }(\vec{k}^{\prime })%
\right] _{-} &=&\pm 2\eta ^z(\vec{k}+\vec{k}^{\prime })\delta _{\alpha \beta
}.
\end{eqnarray}

With (\ref{etaplus}) and (\ref{etaminus}), a measure for the breakdown of
U(1) symmetry due to pairing is given by the expectation value\footnote{%
We follow the notation used in similar proofs for the purely one- and
two-dimensional cases (\cite{Su Sch Zit 1997},\cite{Noc Cuo 1999}).} 
\begin{eqnarray}
\mathfrak{F}(\vec{K}) &=&\frac{\left\langle \eta ^{+}(-\vec{K})\right\rangle 
}N \\
&=&\sum_ie^{-i\vec{K}\cdot \vec{R}_i}\frac{\left\langle c_{i\uparrow
}^{+}c_{i\downarrow }^{+}\right\rangle }N.  \notag
\end{eqnarray}
For film systems (and since we are concerned here with local on-site
pairing), we can, as in the magnetic case, write the total order parameter $%
\mathfrak{F}(\vec{K})$ as the sum over intra-layer quantities $\mathfrak{F}%
_\alpha (\vec{K}):$%
\begin{eqnarray}
\mathfrak{F}(\vec{K}) &=&\frac 1{N\cdot d}\sum_\alpha \sum_ie^{-i\vec{K}%
\cdot \vec{R}_i}\left\langle c_{i\alpha \uparrow }^{+}c_{i\alpha \downarrow
}^{+}\right\rangle  \notag \\
&=&\frac 1d\sum_\alpha \mathfrak{F}_\alpha (\vec{K}).
\end{eqnarray}
The ordering wave vector $\vec{K}$ allows one to distinguish between
different types of pairing: for $\vec{K}=0$ one would have $s$-wave pairing,
for $\vec{K}\neq 0$ (generalized) $\eta $-pairing \cite{Yan 1989}, \cite{Su
Sch Zit 1997}.

The Hubbard Hamiltonian with a U(1) symmetry breaking contribution of order $%
\lambda $ is then given by 
\begin{eqnarray}
H &=&\sum_{ij\alpha \beta \sigma }T_{ij}^{\alpha \beta }c_{i\alpha \sigma
}^{+}c_{j\beta \sigma }+\frac U2\sum_{i\alpha \sigma }n_{i\alpha \sigma
}n_{i\alpha -\sigma }  \notag \\
&&-\lambda \sum_{i\alpha }\left( \eta _{i\alpha }^{+}e^{-i\vec{K}\cdot \vec{R%
}_i}+\eta _{i\alpha }^{-}e^{+i\vec{K}\cdot \vec{R}_i}\right) ,
\label{hubbardscfilm}
\end{eqnarray}
where the second line is the symmetry-breaking part $H_{sc}.$

In order to make use of the Bogoliubov inequality 
\begin{equation}
\sum_{\vec{k}}\frac{\left| \left\langle \left[ C,A\right] _{-}\right\rangle
\right| ^2}{\left\langle \left[ \left[ C,H\right] _{-},C^{+}\right]
_{-}\right\rangle }\leq \frac \beta 2\sum_{\vec{k}}\left\langle \left[
A,A^{+}\right] _{+}\right\rangle  \label{boglibubo}
\end{equation}
to derive an upper bound for $\mathfrak{F}_\alpha (\vec{K}),$ the operators $%
C$ and $A$ must be chosen properly. With 
\begin{eqnarray}
C &\equiv &\sum_\gamma \eta _\gamma ^z(\vec{k}) \\
A &\equiv &\eta _\alpha ^{+}(-\vec{k}-\vec{K})
\end{eqnarray}
one finds that indeed the Hamiltonian-independent\textbf{\ commutator}%
\textit{\ }[\textit{\textbf{C,A}}]$_{-}$ reproduces the layer-dependent
quantity $\mathfrak{F}_\alpha (\vec{K}):$%
\begin{eqnarray}
\left\langle \left[ C,A\right] _{-}\right\rangle &=&\left\langle \sum_\gamma %
\left[ \eta _\gamma ^z(\vec{k}),\eta _\alpha ^{+}(-\vec{k}-\vec{K})\right]
_{-}\right\rangle  \notag \\
&=&\left\langle \eta _\alpha ^{+}(-\vec{K})\right\rangle .  \label{commuacsc}
\end{eqnarray}
The \textbf{anticommutator sum} $\sum_{\vec{k}}\left\langle \left[ A,A^{+}%
\right] _{+}\right\rangle $ can be shown to be bounded from above: 
\begin{eqnarray}
\sum_{\vec{k}}\left\langle \left[ A,A^{+}\right] _{+}\right\rangle &=&\sum_{%
\vec{k}}\sum_{ij}e^{i(\vec{k}+\vec{K})\cdot (\vec{R}_i-\vec{R}%
_j)}\left\langle \left[ \eta _{\alpha j}^{+},\eta _{\alpha i}^{-}\right]
_{+}\right\rangle  \notag \\
&=&N\sum_i\left\langle \left[ \eta _{\alpha i}^{+},\eta _{\alpha i}^{-}%
\right] _{+}\right\rangle \leq 4N^2.  \label{antcomsumsc}
\end{eqnarray}

The \textbf{double commutator,} after some algebra, can be reduced to the
familiar structure quadratic in $k$ plus a quantity proportional to the
order parameter: For the contribution due to the unperturbed Hamiltonian $%
H_0 $ (i.e. the first line of (\ref{hubbardscfilm}) only), the double
commutator is 
\begin{equation}
\left\langle \sum_{\gamma \varepsilon }\left[ \left[ \eta _\gamma ^z(\vec{k}%
),H_0\right] _{-},\eta _\varepsilon ^z(-\vec{k})\right] _{-}\right\rangle
\leq Nd\tilde{Q}\vec{k}^2,  \label{dblcomhnullsc}
\end{equation}
where, as usual, $\frac 1{Nd}\sum_{lm\gamma \varepsilon }T_{lm}^{\gamma
\varepsilon }\left( \vec{R}_l-\vec{R}_m\right) ^2=\tilde{Q}<\infty $ is
assumed. The double commutator with the U(1) symmetry breaking term $H_{sc}$
(second line in (\ref{hubbardscfilm})) is bounded by a quantity proportional
to the total number of lattice sites, the total order parameter $\mathfrak{F}%
(\vec{K}),$ and the parameter $\lambda $ with which the U(1)\ symmetry
breaking can be turned on $(\lambda \neq 0)$ or off $(\lambda \rightarrow
0): $%
\begin{equation}
\left\langle \sum_{\gamma \varepsilon }\left[ \left[ \eta _\gamma ^z(\vec{k}%
),H_{sc}\right] _{-},\eta _\varepsilon ^z(-\vec{k})\right] _{-}\right\rangle
\leq 2\lambda Nd\left| \mathfrak{F}(\vec{K})\right| .  \label{dblcomheinssc}
\end{equation}

The absence of $s$-wave (or generalized $\eta )$ pairing in Hubbard films of
finite thickness now follows straightforwardly from the Bogoliubov
inequality (\ref{boglibubo}) in the thermodynamic limit. Formula (\ref%
{boglibubo}) with (\ref{commuacsc}), (\ref{antcomsumsc}) and the sum of (\ref%
{dblcomhnullsc}$)$ and $($\ref{dblcomheinssc}), is equivalent to 
\begin{equation}
\sum_{\vec{k}}\frac{N^2\left| \mathfrak{F}_\alpha (\vec{K})\right| ^2}{Nd%
\tilde{Q}\vec{k}^2+2\lambda Nd\left| \mathfrak{F}(\vec{K})\right| ^2}\leq
4N^2\frac \beta 2
\end{equation}
so that in the thermodynamic limit, with the familiar substitution 
\begin{equation*}
\sum_{\vec{k}}...\mapsto (V^{(d)}/(2\pi )^d)\int d^d\vec{k}...,
\end{equation*}
we get for films consisting of $d$ two-dimensional layers stacked on top of
each other, the result 
\begin{equation}
\frac{v_0^{(2)}}{2\pi }\left| \mathfrak{F}_\alpha (\vec{K})\right| ^2\leq 
\frac{\beta d}{\ln \left( 1+\frac{\tilde{Q}k_0^2}{2\lambda \left| \mathfrak{F%
}(\vec{K})\right| ^2}\right) }
\end{equation}
which in the limit $\lambda \rightarrow 0$ proves conclusively that no
pairing transition of the proposed kind and, thus, no corresponding
superconducting phase transition can occur in the Hubbard model, provided $%
\beta $ and $d$ are finite.

\section{Two remarks on special cases of low-dimensional order}

\subsection{Planar magnetic order}

Ferromagnetic and antiferromagnetic ordering along the $z$-axis are not the
only types of ordering conceivable. Indeed, for the Hubbard model so-called 
\textit{spiral states} have been discussed, which are conjectured to play an
important role in the thermodynamics of the two-dimensional model at low
temperatures \cite{Cha Col 1989}. In particular it has been suggested that
in the zero-temperature case at half filling the 2D Hubbard model with $%
U=\infty $ may display an instability towards the formation of spiral phases %
\cite{Shr Sig 1989}.

In this context, it is of interest to consider order parameters which keep
the spin vectors confined to a plane. It has been shown \cite{Uhr 1992} that
any such order parameter must vanish at finite temperatures in $d\leq 2$
dimensions. By analogy with previous expressions for the magnetization, a
planar magnetization $M_{plan}(T,B_0)$ is defined as 
\begin{equation}
M_{plan}(T,B_0)\equiv \frac 1N\sum_i\left\langle \left( \alpha _i\sigma
_i^y+\beta _i\sigma _i^x\right) \right\rangle ,  \label{planmagn}
\end{equation}
where $\alpha _i,$ $\beta _i$ are arbitrary real numbers except for the
requirement that the sum of their squares be convergent: 
\begin{equation}
\frac 1N\sum_i\left( \alpha _i^2+\beta _i^2\right) \equiv \tilde{Q}<N.
\end{equation}

With the Bogoliubov inequality (\ref{bog}) and by choosing $A,C$ so that 
\begin{equation}
A(\vec{k})=\sum_je^{-i\vec{k}\cdot \vec{R}_j}(\beta _j\sigma _j^y-\alpha
_j\sigma _j^x)
\end{equation}
and 
\begin{equation}
B(\vec{k})=\sum_je^{-i\vec{k}\cdot \vec{R}_j}\sigma _j^z,
\end{equation}
it is easy to verify that for planar magnetic order, too, the order
parameter, $M_{plan}(T,B_0),$ can be bounded from above $(\xi ,\xi
_1=const), $%
\begin{equation}
M_{plan}^2(T,B_0)\leq \xi \frac \beta {\ln \left( 1+\frac{\xi _1k_0^2}{%
\left| B_0M_{plan}(T,B_0)\right| }\right) },
\end{equation}
so that in the limit of vanishing external field at any finite temperature $%
\beta <\infty ,$ the planar magnetization vanishes: 
\begin{equation*}
M_{plan}^2(T,B_0) 
\begin{array}{l}
\begin{array}{l}
\\ 
\overrightarrow{{\footnotesize B}_0{\footnotesize \rightarrow 0}}%
\end{array}%
\end{array}
0.
\end{equation*}
\bigskip

In this remark, we have sketched an extension of the Mermin-Wagner theorem
to a case of magnetic order where ordering within a plane perpendicular to a
fixed direction is considered. The details of the calculation are given in %
\cite{Uhr 1992}; we shall return to this case briefly in a later section of
this thesis where it serves as a simple example for the application of a
more general correlation inequalities approach.

\subsection{Fractal lattices with spectral dimension $\tilde{d}\leq 2$}

We shall briefly outline another interesting special case of the
Mermin-Wagner theorem, namely its derivation for fractal lattices \cite{Cas
1992} whose spectral dimension is $\tilde{d}\leq 2.$ In the previously
discussed cases, extensive use was made of translational invariance of the
underlying 2D Bravais lattice in order to exclude a magnetic (or pairing)
phase transition in film systems. Spatial dimensionality, in these cases,
comes into play in the thermodynamic limit, when the sum over wave vectors
is replaced by a two-dimensional integral in $k$-space. Thus, the
translationally invariant lattice is naturally embedded into a Euclidean
space of integer dimensionality $D$ which determines the continuum limit of
the statistical system under consideration. The question, however, remains
how this embedding into Euclidean space can determine the system's behaviour
when no explicit reference to its dimensionality is contained in the lattice
Hamiltonian. A detailed analysis of the mathematical physics of this problem
would be far beyond the scope of this thesis (see \cite{Hat Hat Wat 1987}
for an elaborate discussion).

It turns out that for irregular lattices, such as fractals, one must adopt a
more sophisticated notion of ``dimension'' which is allowed to take
non-integer values (``spectral dimension'').By use of a mathematical theorem
on divergencies in Gaussian field theories (a proof of which, along with an
extensive discussion, is given in \cite{Hat Hat Wat 1987}) it has been shown
that also for the Heisenberg model on fractal lattices with (spectral)
dimension $\tilde{d}\leq 2$ and nearest-neighbour interactions, the
spontaneous magnetization must vanish at any finite temperatures \cite{Cas
1992}.

The extension to other situations or models, as well as the physical meaning
of this result, are far from obvious and we must, at this point, refer the
reader to refs. \cite{Cas 1992}, \cite{Hat Hat Wat 1987} for a more detailed
discussion.

\chapter{Green's Functions and Correlation Inequalities Approach}

\section{Linear response quantities and their algebraic properties}

\subsection{General formalism of linear response theory}

In examining the reaction of a physical system to an external, or in fact
any, perturbation, one will generally be faced with a Hamiltonian $%
H=H_{0}+H_{ext}$, where $H_{0}$ includes the kinetic (or, more generally,
the one-particle) term \textit{and} the internal interaction due to, say, a
pair-potential. The external contribution $H_{ext}$ is due to an external
scalar field $F_{t}$ which couples to an observable $\hat{B}$ of the system
and which may be explicitly time-dependent $(H_{ext}=F_{t}\hat{B})$. This
perturbation will lead to a change in the system's behaviour affecting other
observables, such as $\hat{A},$ as well.\footnote{%
This section follows the derivation in \cite{Bd7 1997}.} The dynamics of the
process are given by the time-dependent Schr\"{o}dinger equation 
\begin{eqnarray}
i\hbar \frac{\partial }{\partial t}\left| \Psi \right\rangle &=&H\left| \Psi
\right\rangle  \notag \\
&=&He^{-\frac{i}{\hbar }H_{0}t}\left| \Psi _{D}(t)\right\rangle ,
\end{eqnarray}%
where in the second step we have switched to the interaction picture defined
by 
\begin{equation}
i\hbar \frac{\partial }{\partial t}\left| \Psi _{D}(t)\right\rangle
=H_{ext}^{D}(t)\left| \Psi _{D}(t)\right\rangle
\end{equation}%
with 
\begin{eqnarray}
H_{ext}^{D}(t) &=&e^{iH_{0}t/\hbar }H_{ext}e^{-iH_{0}t/\hbar }  \notag \\
&=&e^{i\mathcal{H}_{0}t/\hbar }H_{ext}e^{-i\mathcal{H}_{0}t/\hbar }
\label{diracpicops}
\end{eqnarray}%
(the substitution with $\mathcal{H}_{0}=H_{0}-\mu \hat{N}$ in the last step
is possible if $[H_{0},\hat{N}]_{-}=0).$

Evaluating the change in the expectation value of the observable $\hat{A}$
requires calculating the trace of $\hat{A}$ with the grand canonical density
operator at any given time $t:$%
\begin{equation}
\left\langle \hat{A}\right\rangle _{t}=tr\left( \rho _{t}\hat{A}\right) .
\end{equation}%
The time-dependent density operator within the Schr\"{o}dinger picture
satisfies the von Neumann equation of motion 
\begin{equation}
i\hbar \dot{\rho}_{t}=[\mathcal{H}_{0},\rho _{t}]_{-}+[H_{ext},\rho
_{t}]_{-}.
\end{equation}%
In the interaction picture, $\rho _{t}$ transforms as 
\begin{equation}
\rho _{t}^{D}(t)=\exp \left( \frac{i}{\hbar }\mathcal{H}_{0}t\right) \rho
_{t}\exp \left( -\frac{i}{\hbar }\mathcal{H}_{0}t\right)
\end{equation}%
so that, taking the free initial condition 
\begin{equation}
\lim\limits_{t\rightarrow -\infty }\rho _{t}\equiv \rho _{0}:=\frac{\exp
\left( -\beta \mathcal{H}_{0}\right) }{tr\left( \exp \left( -\beta \mathcal{H%
}_{0}\right) \right) }
\end{equation}%
into account, one has 
\begin{equation}
\rho _{t}^{D}(t)=\rho _{0}-\frac{i}{\hbar }\int\limits_{-\infty
}^{t}dt^{\prime }\left[ H_{ext^{\prime }}^{D}(t^{\prime }),\rho _{t^{\prime
}}^{D}(t^{\prime })\right] _{-}
\end{equation}%
which can be solved by iteration (the somewhat redundant prime in `` $%
ext^{\prime }"$ indicates that the explicit time dependence of $H_{ext}$ is
also to be integrated over). Averages with the density operator at $%
t=-\infty ,$ i.e. with the field-independent density operator $\rho _{0},$
will be denoted by an index ``0'': 
\begin{equation}
\left\langle \hat{A}\right\rangle _{0}\equiv tr\left( \rho _{0}\hat{A}%
\right) .  \label{averagerhonull}
\end{equation}

To first order, the density matrix thus is 
\begin{equation}
\rho _{t}^{D}(t)\approx \rho _{0}-\frac{i}{\hbar }\int\limits_{-\infty
}^{t}dt^{\prime }\left[ H_{ext^{\prime }}^{D}(t^{\prime }),\rho _{0}\right]
_{-}
\end{equation}%
or, in the Schr\"{o}dinger picture (using (\ref{diracpicops})), 
\begin{equation}
\rho _{t}\approx \rho _{0}-\frac{i}{\hbar }\int\limits_{-\infty
}^{t}dt^{\prime }\exp \left( -\frac{i\mathcal{H}_{0}t}{\hbar }\right) \left[
H_{ext^{\prime }}^{D}(t^{\prime }),\rho _{0}\right] _{-}\exp \left( \frac{i%
\mathcal{H}_{0}t}{\hbar }\right) .
\end{equation}%
With this expression, it is now possible to explicitly evaluate the change
in $\hat{A},$%
\begin{eqnarray}
\Delta \hat{A}_{t} &\equiv &\left\langle \hat{A}\right\rangle
_{t}-\left\langle \hat{A}\right\rangle _{0}  \notag \\
&=&-\frac{i}{\hbar }\int\limits_{-\infty }^{t}dt^{\prime }F_{t^{\prime
}}\left\langle \left[ \hat{A}^{D}(t),\hat{B}^{D}(t^{\prime })\right]
_{-}\right\rangle _{0}.  \label{deltaahat}
\end{eqnarray}%
The integrand in (\ref{deltaahat}) depends only on the commutator of the
observables involved and on the ``internal'' dynamics of the interacting
many-body system at zero external field as described by $H_{0},$ which
enters the calculation via the averaging procedure (\ref{averagerhonull}).
To emphasize this fact, one introduces the \textit{retarded Green's function}
\begin{eqnarray}
G_{\hat{A}\hat{B}}^{ret}(t,t^{\prime }) &=&\left\langle \left\langle \hat{A}%
(t);\hat{B}(t^{\prime })\right\rangle \right\rangle ^{ret}  \notag \\
&=&-i\theta (t-t^{\prime })\left\langle \left[ \hat{A}(t),\hat{B}(t^{\prime
})\right] _{-\varepsilon }\right\rangle _{0}.  \label{retgreefu}
\end{eqnarray}%
For $\varepsilon =+,$ the commutator on the RHS is the same as in (\ref%
{deltaahat}), because at zero external field the Heisenberg picture
coincides with the interaction picture. Dividing the expectation value of
the (anti-)commutator in (\ref{retgreefu}) by $2\pi $, we define the
spectral density 
\begin{equation}
S_{\hat{A}\hat{B}}(t,t^{\prime })=\frac{1}{2\pi }\left\langle \left[ \hat{A}%
(t),\hat{B}(t^{\prime })\right] _{-\varepsilon }\right\rangle
\end{equation}%
which is of central importance and which will be discussed later (e.g. in (%
\ref{spectralspectral}) in its spectral representation in Fourier space).

Thus, after Fourier transformation (which requires regularization by use of
an infinitesimal damping factor with $i\eta =i0^{+})$%
\begin{equation}
\Delta \hat{A}_t=\frac 1{2\pi \hbar ^2}\int\limits_{-\infty }^\infty
dEF(E)G_{\hat{A}\hat{B}}^{ret}(E)\exp \left( -\frac i\hbar (E+i\eta )t\right)
\end{equation}
which is known as the\textit{\ Kubo formula.} It can be interpreted as
describing the response of a system to a perturbation in $\hat{B}$ in terms
of the change in the observable $\hat{A}.$

\subsubsection{Equation of motion for the Green's function}

From elementary quantum mechanics it is known that in the Heisenberg picture
a (time-dependent) operator $A_{t}(t)$ evolves according to the equation of
motion 
\begin{equation}
i\hbar \frac{d}{dt}A_{t}(t)=\left[ A_{t},\mathcal{H}\right] _{-}(t)+i\hbar 
\frac{\partial A_{H}}{\partial t}.
\end{equation}%
Calculating the equation of motion for the Green's function $%
G_{AB}(t,t^{\prime })$, which consists of combinations of operators, leads
to 
\begin{equation}
i\hbar \frac{\partial }{\partial t}G_{AB}^{ret}(t,t^{\prime })=\hbar \delta
(t-t^{\prime })\left\langle \left[ A,B\right] _{-\varepsilon }\right\rangle
+\left\langle \left\langle \left[ A,\mathcal{H}\right] _{-}(t);B(t^{\prime
})\right\rangle \right\rangle ^{ret}
\end{equation}%
or, in terms of energy, 
\begin{equation}
E\left\langle \left\langle A;B\right\rangle \right\rangle _{E}^{ret}=\hbar
\left\langle \left[ A,B\right] _{-\varepsilon }\right\rangle +\left\langle
\left\langle \left[ A,\mathcal{H}\right] _{-};B\right\rangle \right\rangle
_{E}^{ret}.
\end{equation}%
By iteration, a hierarchy of higher Green's functions is generated on the
RHS which, if infinite, can be decoupled artificially (by an appropriate
approximation). A procedure for decoupling the higher-order Green's function
in the Heisenberg model will be presented in the next chapter.

\subsection{Transverse spin susceptibility in the Hubbard model}

For a system with an external magnetic field, the response of the
magnetization is the quantity one is interested in and which one can hope to
express in terms of Green's functions. The external contribution $H_{ext}=-%
\vec{m}\cdot \vec{B}_{t}$ to the Hamiltonian $(\vec{m}$: total magnetic
moment; $\vec{B}_{t}$: time-dependent external magnetic field) will lead to
a change in the magnetization $\vec{M}=\frac{1}{V}\left\langle \vec{m}%
\right\rangle .$ Thus, by virtue of the Kubo formula, the response of the $%
\beta $-th component of the magnetization to $\vec{B}_{t}$ is given by 
\begin{equation}
\Delta M_{t}^{\beta }=\frac{-1}{V\hbar }\sum_{\alpha }\int\limits_{-\infty
}^{\infty }dt^{\prime }B_{t^{\prime }}^{\alpha }\left\langle \left\langle
m^{\beta }(t);m^{\alpha }(t^{\prime })\right\rangle \right\rangle .
\end{equation}%
This formula suggest defining a \textit{magnetic susceptibility }by 
\begin{equation}
\chi _{ij}^{\beta \alpha }(t,t^{\prime })=-\frac{\mu _{0}}{V\hbar }\frac{%
g_{J}^{2}\mu _{B}^{2}}{\hbar ^{2}}\left\langle \left\langle \sigma
_{i}^{\beta }(t);\sigma _{j}^{\alpha }(t^{\prime })\right\rangle
\right\rangle  \label{magsus}
\end{equation}%
which generalizes in a straightforward way the more phenomenological
interpretation of the susceptibility as a ``proportionality coefficient''
between external field and magnetization, yielding 
\begin{equation}
\Delta M_{t}^{\beta }=\frac{1}{\mu _{0}}\sum_{ij}\int\limits_{-\infty
}^{\infty }dt^{\prime }\left( \sum_{\alpha }\chi _{ij}^{\beta \alpha
}(t,t^{\prime })B_{t^{\prime }}^{\alpha }\right) .
\end{equation}%
For $(\beta ,\alpha )=(+,-)$ eqn. (\ref{magsus}) describes how the system
reacts to spin flips. With (\ref{sigmpl}) and (\ref{sigmmi}) the \textit{%
transverse spin susceptibility} $\chi ^{+-}(E)$ of the Hubbard model is
given by 
\begin{eqnarray}
\chi _{ij}^{+-}(E) &=&-\frac{\mu _{0}}{V\hbar }\frac{(g_{J}\mu _{B})^{2}}{%
\hbar ^{2}}\left\langle \left\langle \sigma _{i}^{+};\sigma
_{j}^{-}\right\rangle \right\rangle _{E}^{ret} \\
&=&-\frac{\mu _{0}}{V\hbar }(g_{J}\mu _{B})^{2}\left\langle \left\langle
c_{i\uparrow }^{+}c_{i\downarrow };c_{j\downarrow }^{+}c_{j\uparrow
}\right\rangle \right\rangle _{E}^{ret}  \notag
\end{eqnarray}%
or equivalently in $k$-space 
\begin{eqnarray}
\chi _{\vec{q}}^{+-}(E) &=&\frac{1}{N}\sum_{ij}e^{i\vec{q}\cdot \left( \vec{R%
}_{i}-\vec{R}_{j}\right) }\chi _{ij}^{+-}(E)  \notag \\
&=&-\frac{\gamma }{N}\sum_{\vec{k}\vec{p}}\bar{\chi}_{\vec{k}\vec{p}}(\vec{q}%
),
\end{eqnarray}%
where 
\begin{equation}
\gamma :=\frac{\mu _{0}}{\hbar }(g_{J}\mu _{B})^{2}
\end{equation}%
and we have introduced as a shorthand 
\begin{equation}
\bar{\chi}_{\vec{k}\vec{p}}(\vec{q})=\left\langle \left\langle c_{\vec{k}%
\uparrow }^{+}c_{\vec{k}+\vec{q}\downarrow };c_{\vec{p}\downarrow }^{+}c_{%
\vec{p}-\vec{q}\uparrow }\right\rangle \right\rangle _{E}^{ret}.
\end{equation}

\subsection{Dynamical structure factor}

The \textit{dynamical structure factor} $C$ is another important physical
quantity which, as we shall see, provides the same insight into a system's
behaviour as the susceptibility \cite{Bd7 1997}, \cite{Aue 1994}. We shall
derive it from the transverse two-particle spin correlation function 
\begin{equation}
C_{ij}^{+-}(t-t^{\prime })=\left\langle \sigma _{i}^{+}(t)\sigma
_{j}^{-}(t^{\prime })\right\rangle  \label{spincorrelation}
\end{equation}%
by Fourier transformation: 
\begin{eqnarray}
C^{+-}(\vec{q},E) &=&\frac{1}{N}\sum_{ij}\int\limits_{-\infty }^{\infty
}d(t-t^{\prime })\left\langle \sigma _{i}^{+}(t)\sigma _{j}^{-}(t^{\prime
})\right\rangle e^{iE(t-t^{\prime })/\hbar }e^{-i\vec{q}\cdot (\vec{R}_{i}-%
\vec{R}_{j})}  \notag \\
&=&\frac{1}{N}\int\limits_{-\infty }^{\infty }d(t-t^{\prime })\left\langle
\sigma ^{+}(\vec{q},t)\sigma ^{-}(-\vec{q},t^{\prime })\right\rangle
e^{iE(t-t^{\prime })/\hbar }.  \label{spinspincorrfct}
\end{eqnarray}

\subsection{Connection between susceptibility and dynamical structure factor}

Along with (\ref{spinspincorrfct}) we also know, by definition, that 
\begin{equation}
G_{+-}^{ret}(\vec{R}_{i}-\vec{R}_{j};t-t^{\prime })=-i\theta (t-t^{\prime
})\left( \left\langle \sigma _{i}^{+}(t)\sigma _{j}^{-}(t^{\prime
})\right\rangle -\left\langle \sigma _{j}^{-}(t^{\prime })\sigma
_{i}^{+}(t)\right\rangle \right)
\end{equation}%
for the retarded Green's function and%
\begin{eqnarray}
\chi _{\vec{k}}^{+-}(E) &=&-\frac{\mu _{0}(g_{J}\mu _{B})^{2}}{NV\hbar }%
\int\limits_{-\infty }^{\infty }d\tau e^{iE\tau /\hbar }(-i)\theta (\tau
)\sum_{ij}\left\{ \left\langle \sigma _{i}^{+}(\tau )\sigma
_{j}^{-}(0)\right\rangle \right.  \notag \\
&&\left. -\left\langle \sigma _{j}^{-}(0)\sigma _{i}^{+}(\tau )\right\rangle
\right\} e^{i\vec{k}\cdot (\vec{R}_{i}-\vec{R}_{j})} \\
&=&-\frac{\mu _{0}(g_{J}\mu _{B})^{2}}{NV\hbar }\int\limits_{-\infty
}^{\infty }d\tau e^{iE\tau /\hbar }\theta (\tau )(-i)\left\{ \left\langle
\sigma ^{+}(-\vec{k},\tau )\sigma ^{-}(\vec{k},0)\right\rangle \right. 
\notag \\
&&\left. -\left\langle \sigma ^{-}(\vec{k},0)\sigma ^{+}(-\vec{k},\tau
)\right\rangle \right\} .
\end{eqnarray}

By use of the integral representation for the Heaviside step function, 
\begin{equation}
\theta (\tau )=\frac 1{2\pi i}\int\limits_{-\infty }^\infty dE^{\prime }%
\frac{e^{iE^{\prime }\tau /\hbar }}{E^{\prime }-i\eta },
\end{equation}
the susceptibility can indeed be expressed completely in terms of the
dynamical structure factor: 
\begin{eqnarray}
&&\chi _{\vec{k}}^{+-}(E;i\eta )  \notag \\
&=&\frac{\mu _0(g_J\mu _B)^2}{NV\hbar }\frac 1{2\pi }\int\limits_{-\infty
}^\infty d\tau \int\limits_{-\infty }^\infty dE^{\prime \prime }\frac{%
e^{iE^{\prime \prime }\tau /\hbar }\left( \left\langle \sigma ^{+}(-\vec{k}%
,\tau )\sigma ^{-}(\vec{k},0)\right\rangle -\left\langle \sigma ^{-}(\vec{k}%
,0)\sigma ^{+}(-\vec{k},\tau )\right\rangle \right) }{(E^{\prime \prime
}-E)-i\eta }  \notag \\
&=&\left( -\frac{\mu _0(g_J\mu _B)^2}{V\hbar }\right) \frac 1{2\pi
}\int\limits_{-\infty }^\infty dE^{\prime \prime }\frac{C^{+-}(\vec{k}%
,E^{\prime \prime })-C^{-+}(\vec{k},-E^{\prime \prime })}{(E-E^{\prime
\prime })+i\eta }  \notag \\
&=&\left( -\frac{\mu _0(g_J\mu _B)^2}{V\hbar }\right) \frac 1{2\pi
}\int\limits_{-\infty }^\infty dE^{\prime \prime }\frac{\left( 1-e^{-\beta
E^{\prime \prime }}\right) C^{+-}(\vec{k},E^{\prime \prime })}{(E-E^{\prime
\prime })+i\eta },  \label{susceptthrustrucfac}
\end{eqnarray}
where use has been made of the quite general relation 
\begin{equation}
C_{AB}(-E)=e^{-\beta E}C_{BA}(E)
\end{equation}
which can be verified by inspecting the spectral representation of the
dynamical structure factor, 
\begin{eqnarray}
C^{+-}(\vec{q},E) &=&\frac 1N\int\limits_{-\infty }^\infty dte^{iEt/\hbar
}\left\langle \sigma ^{+}(\vec{q},t)\sigma ^{-}(-\vec{q},0)\right\rangle 
\notag \\
&=&\frac 1N\int\limits_{-\infty }^\infty dte^{iEt/\hbar }\sum_n\frac{%
e^{-\beta E_n}}\Xi \left\langle n\left| e^{i\mathcal{H}t/\hbar }\sigma ^{+}(%
\vec{q},0)e^{-i\mathcal{H}t/\hbar }\sigma ^{-}(-\vec{q},0)\right|
n\right\rangle  \notag \\
&=&\sum_{nm}\frac{e^{-\beta E_n}}{N\Xi }\int\limits_{-\infty }^\infty
dte^{\frac i\hbar (E+(E_n-E_m))t}\left\langle n\left| \sigma ^{+}(\vec{q}%
,0)\right| m\right\rangle \left\langle m\left| \sigma ^{-}(-\vec{q}%
,0)\right| n\right\rangle  \notag \\
&=&2\pi \hbar \sum_{nm}\frac{e^{-\beta E_n}}{N\Xi }\left| \left\langle
n\left| \sigma ^{+}(\vec{q},0)\right| m\right\rangle \right| ^2\delta
(E+(E_n-E_m)).  \label{spectralrepstrucfac}
\end{eqnarray}
From this expression one also concludes $C^{+-}(\vec{q},E)\geq 0$, which is
true for any dynamical structure factor $C_{A^{+}A}(\vec{q},E)$
corresponding to a pair of adjunct operators $A,A^{+}.$ Thus, the dynamical
structure factor is\textit{\ positive semi-definite.} The equation for the
susceptibility (\ref{susceptthrustrucfac}) can easily be generalized for
arbitrary operators $A,B:$%
\begin{eqnarray}
\chi _{AB}(E;i\eta ) &=&\frac 1{2\pi }\int\limits_{-\infty }^\infty
dE^{\prime }\frac{C_{AB}(E^{\prime })-C_{BA}(-E^{\prime })}{E-E^{\prime
}+i\eta }  \notag \\
&=&\frac 1{2\pi }\int\limits_{-\infty }^\infty dE^{\prime }\frac{\left(
1-e^{-\beta E^{\prime }}\right) C_{BA}(E^{\prime })}{E-E^{\prime }+i\eta }.
\label{gensusceptthrustrucfac}
\end{eqnarray}
The \textit{generalized susceptibility} still depends on the energy $E$ and,
on a more formal level, the regularization $i\eta .$ Often, one is
interested in the\textit{\ static limit,} where $E\rightarrow 0$. The\textit{%
\ static susceptibility} is then defined as 
\begin{equation}
\chi _{AB}\equiv \mathfrak{R}\left( \lim_{E,\eta \rightarrow 0}\chi
_{AB}(E;i\eta )\right)
\end{equation}
and thus, for $k$-dependent operators $A(\vec{k}),$ $B(\vec{k}^{\prime }),$%
\begin{equation}
\chi _{AB}(\vec{k},\vec{k}^{\prime })=\mathfrak{R}\left( \lim_{E,\eta
\rightarrow 0}\frac 1{2\pi }\int\limits_{.-\infty }^\infty dE^{\prime }\frac{%
\left( 1-e^{-\beta E^{\prime }}\right) C_{BA}(\vec{k}^{\prime },\vec{k}%
;E^{\prime })}{E-E^{\prime }+i\eta }\right) .
\end{equation}

The latter formula will later serve as the starting point for the discussion
of certain algebraic properties of the static susceptibility that will allow
us to regard $\chi _{AB}$ as defining a scalar product between $A$ and $B$
in operator space.

At this point, however, it appears useful to connect the static
susceptibility with yet another ubiquitous quantity, i.e. the spectral
density, which in its spectral representation is $(\varepsilon =\pm 1)$%
\begin{equation}
S_{AB}^{(\varepsilon )}(\vec{k},\vec{k}^{\prime };E^{\prime })=\frac \hbar
\Xi \sum_{mn}\left\langle n\left| A(\vec{k},0)\right| m\right\rangle
\left\langle m\left| B(\vec{k}^{\prime },0)\right| n\right\rangle e^{-\beta
E_m}\left( e^{\beta E^{\prime }}-\varepsilon \right) \delta (E^{\prime
}-(E_m-E_n)).  \label{spectralspectral}
\end{equation}
Comparing this with the spectral representation of the dynamical structure
factor 
\begin{eqnarray}
&&C_{BA}(\vec{k}^{\prime },\vec{k};E^{\prime })  \notag \\
&=&\int\limits_{-\infty }^\infty dte^{iE^{\prime }t/\hbar }\sum_n\frac{%
e^{-\beta E_n}}{N\Xi }\left\langle n\left| e^{i\mathcal{H}t/\hbar }B(\vec{k}%
^{\prime },0)e^{-i\mathcal{H}t/\hbar }A(\vec{k},0)\right| n\right\rangle \\
&=&\int\limits_{-\infty }^\infty dt\sum_{mn}\frac{e^{-\beta E_n}}{N\Xi }%
\left\langle m\left| A(\vec{k},0)\right| n\right\rangle \left\langle n\left|
B(\vec{k}^{\prime },0)\right| m\right\rangle e^{i(E^{\prime
}-(E_m-E_n))t/\hbar }  \notag \\
&=&2\pi \hbar \sum_{mn}\frac{e^{-\beta E_n}}{N\Xi }\left\langle m\left| A(%
\vec{k},0)\right| n\right\rangle \left\langle n\left| B(\vec{k}^{\prime
},0)\right| m\right\rangle \delta (E^{\prime }-(E_m-E_n))  \notag
\end{eqnarray}
and reminding oneself of (\ref{gensusceptthrustrucfac}), it is easy to see
that in the static limit 
\begin{eqnarray}
\mathfrak{R}\left( \lim_{E,\eta \rightarrow 0}\chi _{AB}(\vec{k},\vec{k}%
^{\prime };E,i\eta )\right) &\equiv &\chi _{AB}(\vec{k},\vec{k}^{\prime }) 
\notag \\
&=&\mathcal{P}\int\limits_{-\infty }^\infty dE^{\prime }\frac{S_{AB}^{(+)}(%
\vec{k},\vec{k}^{\prime};E^{\prime })}{E^{\prime }}.
\label{integralstatsuscept}
\end{eqnarray}

\subsubsection{Algebraic properties of the static susceptibility}

Earlier it was mentioned that the algebraic properties of the static
susceptibility, notably the fact that it defines a scalar product in the
space of operators $A,B$ lead to interesting exact relations between
different physical quantities.\footnote{%
This section partly follows ref. \cite{Wag 1966}.} Three properties need to
be verified in order to prove that 
\begin{equation}
\left\langle A(\vec{k});B(\vec{k}^{\prime })\right\rangle :=\chi _{AB}(\vec{k%
},\vec{k}^{\prime })
\end{equation}%
is a valid scalar product:

\begin{enumerate}
\item $\chi _{AB}(\vec{k},\vec{k}^{\prime })$ is linear in $A(\vec{k})$ and $%
B(\vec{k}^{\prime })$

\item $\left( \chi _{AB}(\vec{k},\vec{k}^{\prime })\right) ^{*}=\chi
_{B^{+}A^{+}}(\vec{k}^{\prime },\vec{k})$

\item $\chi _{AA^{+}}(\vec{k})\equiv \chi _{AA^{+}}(\vec{k},\vec{k})=%
\mathcal{P}\int\limits_{-\infty }^\infty dE^{\prime }\frac{S_{AA^{+}}(\vec{k}%
,\vec{k};E^{\prime })}{E^{\prime }}\geq 0.$
\end{enumerate}

The linearity is trivial, as can be seen directly from the definition of $%
\chi _{AB}.$ The last two properties can be demonstrated by use of
convenient representations in terms of either the spectral density or the
dynamical structure factor. For 2.) we start from eqn. (\ref%
{gensusceptthrustrucfac}): 
\begin{eqnarray}
&&\left( \chi _{AB}(\vec{k},\vec{k}^{\prime };E,i\eta )\right) ^{*}  \notag
\\
&=&\frac 1{2\pi }\mathcal{P}\int\limits_{-\infty }^\infty dE^{\prime }\frac{%
S_{AB}^{*}(\vec{k},\vec{k}^{\prime };E^{\prime })-S_{BA}^{*}(\vec{k}^{\prime
},\vec{k};-E^{\prime })}{(E-E^{\prime }+i\eta )^{*}}  \notag \\
&=&\frac 1{2\pi }\mathcal{P}\int\limits_{-\infty }^\infty dE^{\prime }\frac{%
\left( 1-e^{-\beta E^{\prime }}\right) \left( \int\limits_{-\infty }^\infty
dt\left\langle \left( A_{\vec{k}}(t)B_{\vec{k}^{\prime }}(0)\right)
^{+}\right\rangle e^{-iE^{\prime }t/\hbar }\right) }{E-E^{\prime }-i\eta } 
\notag \\
&=&\frac 1{2\pi }\mathcal{P}\int\limits_{-\infty }^\infty dE^{\prime }\frac{%
\left( 1-e^{-\beta E^{\prime }}\right) \left( \int\limits_{-\infty }^\infty
dt\left\langle B_{\vec{k}^{\prime }}^{+}(t)A_{\vec{k}}^{+}(0)\right\rangle
e^{iE^{\prime }t/\hbar }\right) }{E-E^{\prime }-i\eta }  \notag \\
&=&\chi _{B^{+}A^{+}}(\vec{k}^{\prime },\vec{k};E,-i\eta )
\end{eqnarray}
so that in the static limit, where $\mathfrak{R}\left( \lim_{E,\eta
\rightarrow 0}...\right) $ is to be calculated, one indeed has 
\begin{equation}
\left( \chi _{AB}(\vec{k},\vec{k}^{\prime })\right) ^{*}=\chi _{B^{+}A^{+}}(%
\vec{k}^{\prime },\vec{k}).
\end{equation}

For property 3.) it is easiest to start directly from the integral
representation of the static susceptibility (\ref{integralstatsuscept}): 
\begin{eqnarray}
&&\chi _{AA^{+}}(\vec{k},\vec{k})  \notag \\
&\equiv &\chi _{AA^{+}}(\vec{k})  \notag \\
&=&\mathcal{P}\int\limits_{-\infty }^\infty dE\frac{S_{AA^{+}}(\vec{k},\vec{k%
};E)}E \\
&=&\mathcal{P}\int\limits_{-\infty }^\infty dE\frac 1E\frac \hbar \Xi
\sum_{mn}\left| \left\langle n\left| A^{+}\right| m\right\rangle \right|
^2e^{-\beta E_n}\delta (E-(E_n-E_m))\left( e^{\beta E}-1\right) \geq 0 
\notag
\end{eqnarray}
by use of the spectral representation for the spectral density in the last
step. Thus, the static susceptibility indeed defines a positive
semi-definite scalar product in the operator space.

\section{Correlation inequalities approach}

Having shown in the previous section that the static susceptibility defines
a scalar product between $A$ and $B$, we shall discuss, on quite general
grounds, an inequality for the static susceptibility. The inequality, in
fact, will be a special case of the Schwarz inequality which holds for any
scalar product and, thus, in the case of the susceptibility reads 
\begin{equation}
\left| \chi _{AB}(\vec{k},\vec{k}^{\prime })\right| ^{2}\leq \chi _{AA^{+}}(%
\vec{k})\chi _{B^{+}B}(\vec{k}^{\prime }).  \label{schwarzsuscept}
\end{equation}%
For the specific example, we require $A_{\vec{k}}(t)$ to be of the form $A_{%
\vec{k}}(t):=i\hbar \frac{\partial }{\partial t}Q_{\vec{k}}(t).$\footnote{%
See e.g. ref. \cite{Wag 1966}.} The spectral density for the ``original''
operators $A$ and $B$ can then be expressed in terms of $B$ and the newly
introduced operator $Q:$%
\begin{eqnarray}
S_{AB}(\vec{k},\vec{k}^{\prime };E) &=&\int\limits_{-\infty }^{\infty
}d(t-t^{\prime })e^{\frac{i}{\hbar }E(t-t^{\prime })}S_{AB}(\vec{k},\vec{k}%
^{\prime };t,t^{\prime })  \notag \\
&=&\frac{1}{2\pi }\int\limits_{-\infty }^{\infty }d(t-t^{\prime })e^{\frac{i%
}{\hbar }E(t-t^{\prime })}\left\langle \left[ i\hbar \frac{\partial }{%
\partial t}Q_{\vec{k}}(t),B_{\vec{k}^{\prime }}(t^{\prime })\right]
_{-}\right\rangle  \notag \\
&=&\left\langle \frac{1}{2\pi }\int\limits_{-\infty }^{\infty }d(t-t^{\prime
})e^{\frac{i}{\hbar }E(t-t^{\prime })}i\hbar \left( \frac{\partial Q_{\vec{k}%
}(t-t^{\prime })}{\partial (t-t^{\prime })}B_{\vec{k}^{\prime }}(0)\right.
\right.  \notag \\
&&\left. \left. -B_{\vec{k}^{\prime }}(0)\frac{\partial Q_{\vec{k}%
}(t-t^{\prime })}{\partial (t-t^{\prime })}\right) \right\rangle
\end{eqnarray}%
The next step is to integrate out the RHS by partial integration. For any
physical problem, one can multiply the integrand with a factor $e^{-\alpha
\left| t-t^{\prime }\right| },$ $\alpha >0,$ and take the limit $%
\lim_{\alpha \rightarrow 0}$ after the integration has been carried out.
Thus, contributions from the infinitely distant past or future are
neglected. In this case, one arrives at 
\begin{eqnarray}
S_{AB}(\vec{k},\vec{k}^{\prime };E) &=&\left\langle \frac{1}{2\pi }%
\int\limits_{-\infty }^{\infty }d\tau e^{\frac{i}{\hbar }E\tau }E\left( Q_{%
\vec{k}}(\tau )B_{\vec{k}^{\prime }}(0)-B_{\vec{k}^{\prime }}(0)Q_{\vec{k}%
}(\tau )\right) \right\rangle  \notag \\
&=&\frac{1}{2\pi }E\int\limits_{-\infty }^{\infty }d(t-t^{\prime })e^{\frac{i%
}{\hbar }E(t-t^{\prime })}\left\langle \left[ Q_{\vec{k}}(t-t^{\prime }),B_{%
\vec{k}^{\prime }}(0)\right] _{-}\right\rangle  \notag \\
&=&ES_{QB}(\vec{k},\vec{k}^{\prime };E).
\end{eqnarray}

With an analogous calculation it can be shown that 
\begin{equation}
S_{AA^{+}}(\vec{k};E)=E^2S_{QQ^{+}}(\vec{k};E).
\end{equation}

With these formulas one can now make use of Schwarz's inequality for the
susceptibilities (\ref{schwarzsuscept}). Rearranging the latter as 
\begin{equation}
\chi _{B^{+}B}(\vec{k}^{\prime })\geq \frac{\left| \chi _{AB}(\vec{k},\vec{k}%
^{\prime })\right| ^{2}}{\chi _{AA^{+}}(\vec{k})}
\end{equation}%
and deriving from (\ref{integralstatsuscept}) the expressions 
\begin{eqnarray}
\chi _{AB}(\vec{k},\vec{k}^{\prime };t,t) &=&\left\langle \left[ Q_{\vec{k}%
}(t),B_{\vec{k}^{\prime }}(t)\right] _{-}\right\rangle \\
\chi _{AA^{+}}(\vec{k}) &=&\int\limits_{-\infty }^{\infty }dEES_{QQ^{+}}(%
\vec{k},E)  \notag \\
&=&\left\langle \left[ \left[ Q_{\vec{k}},H\right] _{-},Q_{\vec{k}}^{+}%
\right] _{-}\right\rangle ,
\end{eqnarray}%
one has 
\begin{eqnarray}
\mathcal{P}\int\limits_{-\infty }^{\infty }dE\frac{S_{B^{+}B}(\vec{k}%
^{\prime };E)}{E} &\geq &\frac{\left| \left\langle \left[ Q_{\vec{k}},B_{%
\vec{k}^{\prime }}\right] _{-}\right\rangle \right| ^{2}}{%
\int\limits_{-\infty }^{\infty }dEES_{QQ^{+}}(\vec{k},E)}  \notag \\
&=&\frac{\left| \left\langle \left[ Q_{\vec{k}},B_{\vec{k}^{\prime }}\right]
_{-}\right\rangle \right| ^{2}}{\left\langle \left[ \left[ Q_{\vec{k}},H%
\right] _{-},Q_{\vec{k}}^{+}\right] _{-}\right\rangle }  \label{geninequ}
\end{eqnarray}

Since the LHS of this inequality is essentially identical with $\chi
_{B^{+}B}(\vec{k}^{\prime })$, eqn. (\ref{geninequ}) defines a relation
between the response function for the observable $B$ and commutators which
can, in principle, be calculated directly from our knowledge of $Q,$ $B,$
and $H.$

\section{Finite-temperature case and Bogoliubov inequality}

So far in this chapter, quite general relationships between different
physical quantities have been stated. In order to show that the formalism
developed up to this point is also capable of reproducing the substantially
stronger results obtained in Ch. III by use of Bogoliubov's inequality, it
is useful to verify that the latter can indeed be obtained within the
framework of this chapter.

Pitaevskii and Stringari \cite{Pit Str 1991} have suggested to define a
scalar product simply through the anticommutator: 
\begin{equation}
\left\langle A;B\right\rangle :=\left\langle [A^{+},B]_{+}\right\rangle
\end{equation}
which gives the Schwarz inequality 
\begin{equation}
\left\langle \lbrack A^{+},A]_{+}\right\rangle \left\langle
[B^{+},B]_{+}\right\rangle \geq \left| \left\langle
[A^{+},B]_{+}\right\rangle \right| ^2.  \label{schwarzpitstring}
\end{equation}
It is possible \cite{Pit Str 1991} to define auxiliary operators, denoted by
a tilde, in a such a way that 
\begin{eqnarray}
&\left\langle n\left| \tilde{C}\right| 0\right\rangle :=&\left\langle
n\left| C\right| 0\right\rangle  \notag \\
&\left\langle 0\left| \tilde{C}\right| n\right\rangle :=&-\left\langle
0\left| C\right| n\right\rangle
\end{eqnarray}
from which it follows that 
\begin{eqnarray}
\left\langle n\left| \tilde{C}^{+}\right| 0\right\rangle &=&-\left\langle
n\left| C^{+}\right| 0\right\rangle  \notag \\
\left\langle 0\left| \tilde{C}^{+}\right| n\right\rangle &=&\left\langle
0\left| C^{+}\right| n\right\rangle .
\end{eqnarray}
If we assume, for the moment, that the expectation values in (\ref%
{schwarzpitstring}) are evaluated in the ground state $(T=0)$ and instead of 
$B$ the newly defined operator $\tilde{B}$ is used, one deduces from 
\begin{eqnarray}
\left\langle 0\left| [A^{+},\tilde{B}]_{+}\right| 0\right\rangle
&=&\left\langle 0\left| [A^{+},B]_{-}\right| 0\right\rangle  \notag \\
\left\langle 0\left| [\tilde{B}^{+},\tilde{B}]_{+}\right| 0\right\rangle
&=&\left\langle 0\left| [B^{+},B]_{+}\right| 0\right\rangle
\label{elimtilde}
\end{eqnarray}
that the Schwarz inequality holds not only for the anticommutator on the RHS
but also for the commutator, so that (writing down (\ref{schwarzpitstring})
for the operators $A,\tilde{B},$ and making use of (\ref{elimtilde}) to
express $\tilde{B}$ in terms of $B)$ one finds 
\begin{equation}
\left\langle \lbrack A^{+},A]_{+}\right\rangle \left\langle
[B^{+},B]_{+}\right\rangle \geq \left| \left\langle
[A^{+},B]_{-}\right\rangle \right| ^2.  \label{commuschwarz}
\end{equation}

The extension of this method to finite temperatures is simple. The modified
operators required for this purpose (denoted by a hat) are given by 
\begin{equation}
\left\langle n\left| \hat{C}\right| m\right\rangle :=\frac{W_{m}-W_{n}}{%
W_{m}+W_{n}}\left\langle n\left| C\right| m\right\rangle ,
\label{finitetempauxils}
\end{equation}%
where $W_{m}=\frac{1}{\Xi }\exp \left( -\beta E_{m}\right) $ is the
statistical weight in the grand canonical ensemble. The expectation value
can be calculated by expanding into eigenstates: 
\begin{eqnarray}
\left\langle \lbrack A^{+},\hat{B}]_{+}\right\rangle &=&\frac{1}{\Xi }%
\sum_{mn}\left( e^{-\beta E_{m}}+e^{-\beta E_{n}}\right) \left\langle
m\left| A^{+}\right| n\right\rangle \frac{W_{m}-W_{n}}{W_{m}+W_{n}}%
\left\langle n\left| B\right| m\right\rangle  \notag \\
&=&\frac{1}{\Xi }\sum_{mn}\left( e^{-\beta E_{m}}-e^{-\beta E_{n}}\right)
\left\langle m\left| A^{+}\right| n\right\rangle \left\langle n\left|
B\right| m\right\rangle  \notag \\
&=&\left\langle [A^{+},B]_{-}\right\rangle .
\end{eqnarray}%
If both commutators are constructed according to (\ref{finitetempauxils}),
one finds 
\begin{eqnarray}
\left\langle \lbrack \hat{A}^{+},\hat{B}]_{+}\right\rangle &=&\frac{1}{\Xi }%
\left( e^{-\beta E_{m}}-e^{-\beta E_{n}}\right) \frac{e^{-\beta
E_{n}}-e^{-\beta E_{m}}}{e^{-\beta E_{n}}+e^{-\beta E_{m}}}\left\langle
m\left| A^{+}\right| n\right\rangle \left\langle n\left| B\right|
m\right\rangle  \notag \\
&=&\frac{1}{\Xi }\sum_{mn}\left( e^{-\beta E_{m}}-e^{-\beta E_{n}}\right)
\int dE\tanh \left( \frac{\beta E}{2}\right) \delta (E-(E_{n}-E_{m}))\cdot 
\notag \\
&&\cdot \left\langle m\left| A^{+}\right| n\right\rangle \left\langle
n\left| B\right| m\right\rangle .
\end{eqnarray}%
Once again making use of the spectral representation of the spectral density
(\ref{spectralspectral}), one sees 
\begin{equation}
\left\langle \lbrack \hat{A}^{+},\hat{B}]_{+}\right\rangle =\frac{1}{\hbar }%
\int dE\tanh \left( \frac{\beta E}{2}\right) S_{A^{+}B}(E)
\end{equation}%
or, if specifically $\hat{A}=\hat{B},$%
\begin{equation}
\left\langle \lbrack \hat{A}^{+},\hat{A}]_{+}\right\rangle =\frac{1}{\hbar }%
\int dE\tanh \left( \frac{\beta E}{2}\right) S_{A^{+}A}(E).
\end{equation}%
This compares with the anticommutator for ``normal'' operators $(A,B$
instead of $\hat{A},\hat{B}),$ for which 
\begin{equation}
\left\langle \lbrack A^{+},A]_{+}\right\rangle =\frac{1}{\hbar }\int dE\coth
\left( \frac{\beta E}{2}\right) S_{A^{+}A}(E).
\end{equation}%
Furthermore, from the definition of the spectral density it is clear that
for the \textit{commutator} 
\begin{equation}
\left\langle \lbrack A^{+},B]_{-}\right\rangle =\frac{1}{\hbar }\int
dES_{A^{+}B}(E).
\end{equation}%
The Schwarz inequality (\ref{schwarzpitstring}) may then be written as 
\begin{equation}
\int dES_{A^{+}A}(E)\coth \left( \frac{\beta E}{2}\right) \int
dES_{B^{+}B}(E)\tanh \left( \frac{\beta E}{2}\right) \geq \left| \int
dES_{A^{+}B}(E)\right| ^{2}.
\end{equation}%
Earlier (Ch. II) we had encountered the definition of the Bogoliubov scalar
product between two operators: 
\begin{eqnarray}
\mathfrak{B}(A;B) &=&\sum_{mn}^{E_{n}\neq E_{m}}\left\langle n\left|
A^{+}\right| m\right\rangle \left\langle m\left| B\right| n\right\rangle 
\frac{W_{m}-W_{n}}{E_{n}-E_{m}}  \notag \\
&=&\frac{1}{\Xi }\int dE\sum_{mn}^{E_{n}\neq E_{m}}\left\langle n\left|
A^{+}\right| m\right\rangle \left\langle m\left| B\right| n\right\rangle
\left( e^{-\beta E_{m}}\right.  \notag \\
&&\left. -e^{-\beta E_{n}}\right) \frac{1}{E}\delta (E-(E_{n}-E_{m}))  \notag
\\
&=&\frac{1}{\hbar }\int dE\frac{S_{A^{+}B}(E)}{E}.
\end{eqnarray}%
Hence, the Schwarz inequality for the Bogoliubov scalar product can be
expressed in terms of the spectral density alone, 
\begin{equation}
\left( \int \frac{dE}{E}S_{AA^{+}}(E)\right) \left( \int \frac{dE}{E}%
S_{BB^{+}}(E)\right) \geq \left| \int \frac{dE}{E}S_{BA^{+}}(E)\right| ^{2}.
\end{equation}%
From the series expansion of $\coth $ the simple relation 
\begin{equation}
\coth \left( \frac{\beta E}{2}\right) \geq \frac{2}{\beta E}
\end{equation}%
may be used to strengthen inequalities with $\frac{1}{E}$ appearing under
the integral sign, 
\begin{equation}
\left( \frac{1}{\hbar }\int \frac{dE}{E}\frac{\beta E}{2}S_{AA^{+}}(E)\coth
\left( \frac{\beta E}{2}\right) \right) \left( \frac{1}{\hbar }\int \frac{dE%
}{E}S_{BB^{+}}(E)\right) \geq \left| \frac{1}{\hbar }\int \frac{dE}{E}%
S_{BA^{+}}(E)\right| ^{2}.
\end{equation}

With the energy-weighted sum rule, which relates the first spectral moment
to the double commutator 
\begin{equation}
\left\langle \lbrack A^{+},[H,C]_{-}]_{-}\right\rangle =\frac 1\hbar \int
dEES_{A^{+}C}(E)
\end{equation}
and by choosing $B:=[H,C]_{-},$ we get the following result: 
\begin{equation}
\frac \beta 2\left\langle [A,A^{+}]+\right\rangle \left( \frac 1\hbar \int 
\frac{dE}ES_{[H,C]_{-};[H,C]_{-}^{+}}(E)\right) \geq \left| \frac 1\hbar
\int \frac{dE}ES_{[H,C]_{-};A^{+}}(E)\right| ^2.  \label{einsvorbog}
\end{equation}
The two remaining integrals in (\ref{einsvorbog}) are calculated in appendix
C and shown to lead to the Bogoliubov inequality: 
\begin{equation}
\frac \beta 2\left\langle [A,A^{+}]_{+}\right\rangle \left\langle
[C,[H,C^{+}]_{-}]_{-}\right\rangle \geq \left| \left\langle
[C,A^{+}]_{-}\right\rangle \right| ^2.  \label{bogthecia}
\end{equation}

\section{Low-lying excitations}

Having seen how certain algebraic properties impose constraints on the
behaviour of the many-body quantities discussed in this chapter and having
demonstrated that the Bogoliubov inequality can be reproduced within this
framework, we point out, in this section, some additional applications of
susceptibility and correlation inequalities. We do not intend to give a
complete survey, as this would be far beyond the scope of this work, but
rather outline some aspects of low-lying excitations of a many-body system
and the Hubbard model in particular.

\subsection{Relationship between magnetization and transverse susceptibility
at zero temperature}

The definition of the dynamical structure factor $C^{+-}(\vec{q},E)$ as the
Fourier transform of the transverse spin-spin correlation function still
holds at zero temperature; the correlation function, however, must now be
evaluated with the $T=0$ ground state: 
\begin{eqnarray}
\left. C^{+-}(\vec{q},E)\right| _{T=0} &\equiv &\frac 1N\int\limits_{-\infty
}^\infty dt\left. e^{iEt/\hbar }\left\langle \sigma ^{+}(\vec{q},t)\sigma
^{-}(-\vec{q},0)\right\rangle \right| _{T=0}  \notag \\
&=&\frac 1N\int\limits_{-\infty }^\infty dte^{iEt/\hbar }\sum_m\left\langle
0\left| e^{i\mathcal{H}t/\hbar }\sigma ^{+}(\vec{q},0)e^{-i\mathcal{H}%
t/\hbar }\right| m\right\rangle \left\langle m\left| \sigma ^{-}(-\vec{q}%
,0)\right| 0\right\rangle  \notag \\
&=&\frac{2\pi \hbar }{N} \sum_m\delta (E-(E_m-E_0))\left| \left\langle
0\left| \sigma ^{+}(\vec{q},0)\right| m\right\rangle \right| ^2.
\end{eqnarray}
Since $E_0$ is the ground state energy, we have $E_m\geq E_0$, so the $%
\delta $-function only gives a contribution if $E\geq 0.$ As a corollary 
\begin{equation}
EC^{+-}(\vec{q},E)\left\{ 
\begin{array}{l}
\geq 0 \\ 
=0%
\end{array}
\right. \text{if} 
\begin{array}{l}
E>0 \\ 
E\leq 0.%
\end{array}
\label{positivity}
\end{equation}

If one defines 
\begin{eqnarray}
C_{\perp }(\vec{q},E) &=&\frac 1{2}\left( C^{+-}(\vec{q},E)+C^{-+}(-\vec{q}%
,E)\right)  \notag \\
&=&\frac 1{2}\left\langle \left[ \sigma ^{+}(\vec{q},E),\sigma ^{-}(-\vec{q}%
,E)\right] _{+}\right\rangle
\end{eqnarray}
and 
\begin{equation}
C_{\perp }(\vec{q})=\int dEC_{\perp }(\vec{q},E),
\end{equation}
one can make use of (\ref{commuschwarz}) for operators $\sigma ^{+}(\vec{q}+%
\vec{Q}),$ $\sigma ^{-}(-\vec{q})$ to get 
\begin{equation}
C_{\perp }(\vec{q}+\vec{Q})C_{\perp }(\vec{q})\geq \left| M(\vec{Q})\right|
^2,
\end{equation}
where the magnetization 
\begin{equation}
M(\vec{Q})=\frac{2\hbar }N\left\langle \sum_i\sigma _i^ze^{-i\vec{Q}\cdot 
\vec{R}_i}\right\rangle
\end{equation}
has been introduced (the reciprocal lattice vector $\vec{Q}$ as usual
accounts for possible antiferromagnetic order). Eqn. (\ref{positivity})
tells us that $C_{\perp }(\vec{q},E)$ is nonvanishing only for $E>0$, where
it is positive, so the integral quantity $C_{\perp }(\vec{q})$ may be
formally written as 
\begin{equation}
C_{\perp }(\vec{q})=\int dE\sqrt{EC_{\perp }(\vec{q},E)}\cdot \sqrt{\frac{%
C_{\perp }(\vec{q},E)}E}.  \label{hoeldereins}
\end{equation}

In this form, we can apply \textit{H\"{o}lder's inequality} (\ref{hoeldapp})
which states that for real-valued function $f$ and $g$ for which $\left|
f(x)\right| ^{p}$ and $\left| g(x)\right| ^{q}$ are integrable (where $p$
and $q$ are numbers that satisfy the relations $p^{-1}+q^{-1}=1$ and $p>1)$
the following inequality holds: 
\begin{equation}
\int_{a}^{b}dxf(x)g(x)\leq \left( \int_{a}^{b}dx\left| f(x)\right|
^{p}\right) ^{1/p}\left( \int_{a}^{b}\left| g(x)\right| ^{q}\right) ^{1/q}.
\label{hoeldmath}
\end{equation}

For $p=q=2$ and with $f,g$ defined as 
\begin{eqnarray}
f(x) &=&\sqrt{EC_{\perp }(\vec{q},E)}  \notag \\
g(x) &=&\sqrt{\frac{C_{\perp }(\vec{q},E)}{E}}
\end{eqnarray}%
H\"{o}lder's inequality (\ref{hoeldmath}) applied to (\ref{hoeldereins})
gives 
\begin{equation}
C_{\perp }(\vec{q})\leq \sqrt{\int dEEC_{\perp }(\vec{q},E)}\cdot \sqrt{\int
dE\frac{C_{\perp }(\vec{q},E)}{E}}.  \label{hoelderzwei}
\end{equation}

Earlier (Ch.III) we had calculated 
\begin{eqnarray}
\int dEEC_{\perp }(\vec{q},E) &=&\left\langle \left[ \sigma ^{+}(\vec{q},E),%
\left[ \mathcal{H},\sigma ^{-}(-\vec{q},E)\right] _{-}\right]
_{-}\right\rangle  \notag \\
&\equiv &f(\vec{q})Nq^2\leq const\cdot Nq^2
\end{eqnarray}
which suggests writing 
\begin{equation}
\frac{\int dEEC_{\perp }(\vec{q},E)}{\int dE\frac{C_{\perp }(\vec{q},E)}E}%
\equiv \omega ^2(\vec{q})\cdot q^2,  \label{omega}
\end{equation}
thus defining the new quantity $\omega (\vec{q}).$ Inequality (\ref%
{hoeldereins}) can be strengthened using (\ref{hoelderzwei}) and rearranged
by use of (\ref{omega}) to give 
\begin{equation}
C_{\perp }(\vec{q}+\vec{Q})\geq \frac{\left| M(\vec{Q})\right| ^2}{\omega (%
\vec{q})q\cdot \chi ^{+-}(\vec{q})}.  \label{strucfacscale}
\end{equation}

This equation relates the order parameter $M,$ the transverse susceptibility 
$\chi ^{+-}$ and the structure factor $C^{+-}$ (via $C_{\perp })$. The
relation between these quantities can be made even more pronounced by
summing both sides of (\ref{strucfacscale}) over the first Brillouin zone.
Drawing on our previous results for the Hubbard model we have 
\begin{equation}
\xi _0\cdot N\geq \sum_{\vec{q}}C_{\perp }(\vec{q}+\vec{Q})\geq \sum_{\vec{q}%
}\frac{\left| M(\vec{Q})\right| ^2}{\omega (\vec{q})q\cdot \chi ^{+-}(\vec{q}%
)}
\end{equation}
or, in the thermodynamic limit, 
\begin{equation}
\xi _0\geq \frac{v^{(d)}}{(2\pi )^d}\int\limits_{1.B.Z.}d^d\vec{q}\frac 1q%
\frac{\left| M(\vec{Q})\right| ^2}{\omega (\vec{q})\chi ^{+-}(\vec{q})}.
\label{xinull}
\end{equation}
Provided that $\omega (\vec{q})$ and $\chi ^{+-}(\vec{q})$ remain finite in
the limit $q\rightarrow 0,$ this means that in 1D not even at zero
temperature a transition to a state with nonvanishing $M(\vec{Q})$ can
occur, as this would lead to a logarithmic divergence of the RHS and thus
would contradict $\xi _0$ being a constant. For $\omega (\vec{q}),$ our
previous results indicate that it is bounded from above. That $\chi ^{+-}(%
\vec{q})$ will also remain finite under quite general conditions can be made
plausible starting from the zero-temperature formulation of (\ref%
{integralstatsuscept}): 
\begin{equation}
\chi ^{+-}(\vec{q})=\mathcal{P}\int\limits_{-\infty }^\infty dE^{\prime }%
\frac{S_{(\varepsilon =-)}^{+-}(\vec{k},E^{\prime })}{E^{\prime }},
\end{equation}
where $(\varepsilon =-)$ indicates that we are now dealing with the
anticommutator spectral density. Making use of the spectral representation
one shows that at $T=0:$%
\begin{eqnarray}
\left. \chi ^{+-}(\vec{q})\right| _{T=0} &=&2\pi \hbar \mathcal{P}%
\int\limits_{-\infty }^\infty dE^{\prime }\frac 1{E^{\prime }}\sum_m\delta
(E^{\prime }-(E_m-E_0))\cdot  \notag \\
&\cdot &\left( \left| \left\langle 0\left| \sigma ^{+}(\vec{q},0)\right|
m\right\rangle \right| ^2+\left| \left\langle 0\left| \sigma ^{-}(-\vec{q}%
,0)\right| m\right\rangle \right| ^2\right) .  \label{suscisumm}
\end{eqnarray}

Integrating the $\delta $-function gives a factor $1/(E_{m}-E_{0})$
appearing in the sum over the complete set of states $\left| m\right\rangle
. $ Thus, for $\left| m\right\rangle =\left| 0\right\rangle $, i.e. $%
E_{m}=E_{0},$ one will get a divergent term, so that at first sight this
could be thought of as indicating, via the divergence near $q=0$ of the RHS
of (\ref{xinull}), a phase transition to a state of finite $M$. This
divergence, however, is spurious and does not usually indicate a phase
transition. The ``excitations'' which the divergent term in (\ref{suscisumm}%
) seems to suggest, are just other ground states and thus should not be
included in our description of a possibly ordered phase. Note that this
argument should merely be taken as giving a hint as to the peculiarities of
zero-temperature phase transitions. A detailed and mathematically rigorous
discussion would require more elaborate limiting procedures as $q\rightarrow
0$ and $E\rightarrow 0.$ (For related problems see a series of papers by
Lange \cite{Lan 1966},\cite{Lan 1967}.)

\subsection{Goldstone's theorem}

Returning to the finite-temperature case, we want to conclude this chapter
by determining heuristically, whether in the limit $k\rightarrow 0,$
assuming a \textit{ferromagnetically ordered }phase, there exist elementary
excitations with vanishing energy $\overline{E(k)}%
\begin{array}{l}
\\ 
\overrightarrow{k\rightarrow 0}%
\end{array}%
0$ as this would mean that Goldstone's theorem is satisfied. We start from
the inequality 
\begin{eqnarray}
\chi ^{+-}(\vec{k}) &\leq &\frac{\beta }{2V}\left\langle \left[ \sigma ^{+}(%
\vec{k})-\left\langle \sigma ^{+}(\vec{k})\right\rangle ,\sigma ^{-}(-\vec{k}%
)-\left\langle \sigma ^{-}(-\vec{k})\right\rangle \right] _{+}\right\rangle 
\notag \\
&=&\frac{\beta }{2V}\sum_{mn}e^{i\vec{k}\cdot (\vec{R}_{m}-\vec{R}%
_{n})}\left\langle c_{m\uparrow }^{+}c_{m\downarrow }c_{n\downarrow
}^{+}c_{n\uparrow }+c_{n\downarrow }^{+}c_{n\uparrow }c_{m\uparrow
}^{+}c_{m\downarrow }\right\rangle
\end{eqnarray}%
which holds for any $\vec{k}.$ The inequality becomes only stronger when
summing the LHS over all $\vec{k}$ with $\left| \vec{k}\right| \geq \left| 
\vec{k}_{0}\right| >0$ (i.e. excluding only the null vector) while summing
the RHS over all $\vec{k}$ in the first Brillouin zone. One then has 
\begin{eqnarray}
\frac{1}{V}\sum\nolimits_{\vec{k}}^{\prime }\chi ^{+-}(\vec{k}) &\leq &\frac{%
\beta }{2V^{2}}\sum_{mn}\left( \sum\nolimits_{\vec{k}}e^{i\vec{k}\cdot
\left( \vec{R}_{m}-\vec{R}_{n}\right) }\right) \left\langle c_{m\uparrow
}^{+}c_{m\downarrow }c_{n\downarrow }^{+}c_{n\uparrow }+c_{n\downarrow
}^{+}c_{n\uparrow }c_{m\uparrow }^{+}c_{m\downarrow }\right\rangle  \notag \\
&\leq &\frac{2\beta }{\hbar ^{2}}\frac{N}{V^{2}}\sum_{n}\left\langle \left(
\sigma _{n}^{z}\right) ^{2}\right\rangle =const\cdot \beta \frac{N^{2}}{V^{2}%
},
\end{eqnarray}%
where $\sigma _{i}^{z}=\frac{\hbar }{2}\left( c_{i\uparrow }^{+}c_{i\uparrow
}-c_{i\downarrow }^{+}c_{i\downarrow }\right) $ has been employed.

In the thermodynamic limit, the specific volume approaches a finite limit, $%
V/N\rightarrow v^{(d)},$ and the sum can be replaced by an integral. The
lower cutoff $k_{0}$ becomes infinitesimal, while for any finite $\beta $
the integral remains bounded from above by a finite constant: 
\begin{equation}
\frac{1}{\left( 2\pi \right) ^{3}}\int\limits_{\left| \vec{k}\right|
>k_{0}\rightarrow 0^{+}}d^{3}\vec{k}\chi ^{+-}(\vec{k})\leq const\cdot \beta
.
\end{equation}%
From this we easily read off that as $k\rightarrow 0$, $\chi ^{+-}$ behaves
as 
\begin{equation}
\chi ^{+-}\sim \frac{f\left( \frac{\vec{k}}{\left| \vec{k}\right| }\right) }{%
k^{2+\varepsilon }}  \label{chiasym}
\end{equation}%
with $0\leq \varepsilon <1$ and $f$ not necessarily isotropic in $k$-space.

With this asymptotic behaviour of $\chi ^{+-}$ in mind, we shall now discuss
Goldstone's theorem. As a measure for the mean energy of elementary
excitations, we make use of the following modified definition of spectral
moments (see also \cite{Bre 1962}, \cite{Wag 1966}, \cite{Hoh Bri 1974}, %
\cite{Str 1992}): 
\begin{equation}
\overline{E_{B}^{n}(\vec{k})}:=\lim_{B_{0}\rightarrow 0}\lim_{N,V\rightarrow
\infty }\frac{1}{N}\frac{\int\limits_{0}^{\infty }dEE^{n-1}S_{B^{+}B}(\vec{k}%
,E)}{\mathcal{P}\int\limits_{0}^{\infty }dE\frac{1}{E}S_{B^{+}B}(\vec{k},E)}
\end{equation}%
(Note that this definition fulfills all the intuitive requirements for a
mean energy, such as $\overline{E_{B}^{2}(\vec{k})}-\left( \overline{E_{B}(%
\vec{k})}\right) ^{2}\geq 0$ and $\overline{E_{B}^{n}(\vec{k})}\geq 0.$)

The mean square value of the energy (i.e. the second modified spectral
moment) is bounded from above, via eqn. (\ref{geninequ}) and the spectral
moments, by 
\begin{equation}
\overline{E_{B}^{2}(\vec{k})}\leq \frac{\xi _{0}(v_{0}^{(d)})^{2}k^{4}}{%
M^{2}(T,B_{0})}
\end{equation}%
(where $\xi _{0}$ is a constant and the limit $B_{0}\rightarrow 0$ has
already been taken). We also know that 
\begin{equation}
M^{2}(T,B_{0})\leq 2v_{0}^{(d)}\tilde{q}\cdot \chi ^{+-}(\vec{k})\cdot k^{2}.
\end{equation}%
Multiplying both inequalities and taking the limit $k\rightarrow 0$, which
allows us to make use of eqn. (\ref{chiasym}), we arrive at 
\begin{equation}
\overline{E_{B}^{2}(\vec{k})}\leq const\cdot \frac{f\left( \vec{k}/\left| 
\vec{k}\right| \right) }{M^{4}(T,B_{0})}\frac{k^{6}}{k^{2+\varepsilon }}%
=g\left( \frac{\vec{k}}{\left| \vec{k}\right| }\right) k^{4-\varepsilon }.
\end{equation}%
(Note that in this section we implicitly assume that a phase transition to a
state with finite magnetization has already occurred.) From $\overline{E_{B}(%
\vec{k})}\leq \sqrt{\overline{E_{B}^{2}(\vec{k})}}$ one immediately
concludes that gapless excitations exist, with the energy spectrum vanishing
as $\sim k^{2-\frac{\varepsilon }{2}}$ as $k\rightarrow 0$. This appears to
indicate that, indeed, Goldstone's theorem holds for the Hubbard model.%
\footnote{%
Note that our discussion is not meant to be a rigorous proof, as this would
require further refinement of the argument.}

\chapter{Relevance of the Mermin-Wagner Theorem and its Validity in
Approximative Theories}

The statement that ``magnetic long-range order in low dimensions is
destroyed at any finite temperature due to collective long-wavelength
excitations'' has become common wisdom amongst theoretical physicists. While
the absence of long-range magnetic order is proved conclusively via the
Bogoliubov inequality, it does not provide a detailed picture of how the
destruction of order comes about. With respect to approximative treatments
of many-body models such as the Hubbard model, it will be helpful to improve
one's understanding of the microscopic processes that determine the
existence, or absence, of a phase transition. In general, this will be a
rather difficult task, due to the diversity of possible approximations. In
the following, however, we shall show that simple improvements of mean-field
theory, though physically appealing, generally do not suffice to enforce the
Mermin-Wagner theorem, whereas in the Heisenberg case the Tyablikov method
already seems to incorporate the effect of dimensionality on long-range
order in all the relevant limiting cases.

\section{Onsager reaction-field theory}

Reaction-field theory as an improvement of ordinary mean-field theory goes
back to Onsager's treatment of a system of dipoles \cite{Ons 1936}. Onsager
suggested that the orienting part of the mean-field acting on a given dipole
must not include that part of the contribution from the dipoles in the
vicinity which is due to their correlation with the given dipole under
consideration. We shall not give a detailed review of Onsager's original
work for a system of dipoles, but instead introduce the main concepts at
first for the Ising model and then for the Heisenberg model. We do not claim
any originality for the technical calculations which for the Ising and the
Heisenberg model have been carried out by \cite{Bro Tho 1967} and \cite{Sin
2000}, respectively.

\subsection{The Ising model}

As the purpose of this section is to introduce the main ideas, we shall
restrict our attention for the moment to the unmagnetized phase of the Ising
model. The total mean field acting on a spin (in the sense of \textbf{Weiss
mean-field theory}) is given by 
\begin{equation}
B_i^{Weiss}=B_i^{ext}+\sum_{j\neq i}V_{ij}\left\langle \mu _j\right\rangle ,
\label{Weiss eqn}
\end{equation}
where $B_i^{ext}$ is an infinitesimal external field and $-V_{ij}\mu _i\mu
_j $ is the spin-spin interaction between sites $i$ and $j$, which must be
summed over all $j\neq i$ to yield the total mean field. The mean field has
an orienting field on the magnetic moment of the spin $i$ under
consideration, thus giving 
\begin{eqnarray}
\left\langle \mu _i\right\rangle &=&c_0\beta B_i^{Weiss}  \notag \\
&=&c_0\beta \left( B_i^{ext}+\sum_jV_{ij}\left\langle \mu _j\right\rangle
\right) ,
\end{eqnarray}
where $c_0$ is a constant accounting for proper physical dimension and $%
V_{ii}\equiv 0$ as usual. By Fourier transformation one arrives at 
\begin{equation}
\left\langle \mu _{\vec{q}}\right\rangle =c_0\beta \left( B_{\vec{q}%
}^{ext}+V(\vec{q})\left\langle \mu _{\vec{q}}\right\rangle \right)
\end{equation}
which can be solved for the mean-field susceptibility $\chi =\frac{\partial M%
}{\partial B}$ at vanishing external field: 
\begin{equation}
\chi ^{Weiss}(\vec{q})=\lim_{B_{\vec{q}}\rightarrow 0}\frac{\left\langle \mu
_{\vec{q}}\right\rangle }{B_{\vec{q}}^{ext}}=\frac{c_0\beta }{1-c_0\beta V(%
\vec{q})}.
\end{equation}
The fluctuation-dissipation theorem (see e.g. \cite{Whi 1983}, p.116) gives
an expression for the mean fluctuation of the order parameter: 
\begin{equation}
\left\langle \left| \mu _{\vec{q}}\right| ^2\right\rangle =\frac 1{1-\beta
c_0V(\vec{q})}.
\end{equation}

So far, no distinction between the correlated and uncorrelated contributions
to the mean-field has been made; this we will now attempt, within the
framework of \textbf{Onsager's reaction-field theory.} First, in the
language of infinitesimal variations of both the external and, thus, also
the full mean field (starting from zero), the Weiss equation (\ref{Weiss eqn}%
) can be written as 
\begin{equation}
\delta B_{i}^{Weiss}=\delta B_{i}^{ext}+\sum_{j}V_{ij}\delta \left\langle
\mu _{j}\right\rangle .
\end{equation}%
The Onsager prescription is to assume that the full mean field separates
into two independent contributions, the correlated and the uncorrelated one,
called\textit{\ reaction field} and \textit{cavity field,} respectively. The
latter can be formally derived by subtracting the reaction field from the
full mean field: 
\begin{equation}
\delta B_{i}^{cavity}=\delta B_{i}^{ext}+\sum_{j}V_{ij}\delta \left\langle
\mu _{j}\right\rangle -\sum_{j}V_{ij}\delta \left\langle \mu
_{j}\right\rangle _{\mu _{i}},  \label{cavity}
\end{equation}%
where the index ``$\mu _{i}"$ symbolizes the assumption that it is this part
of the contribution to the field acting on $\mu _{i}$ which is itself
determined by the correlation between spin $i$ and the surrounding spins $j$
that are summed over. In a heuristic way, the value of $\delta \left\langle
\mu _{j}\right\rangle _{\mu _{i}}$ can, in principle, be determined from a
kind of \textit{master equation} written in terms of conditional averages: 
\begin{equation}
\delta \left\langle \mu _{j}\right\rangle _{\mu _{i}}=\delta \left(
p_{i}(+1)\left\langle \mu _{j}\right\rangle _{i\uparrow
}+p_{i}(-1)\left\langle \mu _{j}\right\rangle _{i\downarrow }\right) ,
\label{master}
\end{equation}%
where $\left\langle \mu _{j}\right\rangle _{i\uparrow ,\downarrow }$ are the
conditional average values of $\mu _{j}$ for spin $j$ when $\mu _{i}=\pm 1,$
and $p_{i}(\mu _{i})$ is the probability for spin $i$ to have the magnetic
moment $\mu _{i}.$ Thus, since we are still dealing with the unmagnetized
phase, we have 
\begin{equation}
p_{i}(\mu _{i})=\frac{1}{2}\left( 1\pm \left\langle \mu _{i}\right\rangle
\right) .
\end{equation}%
The conditional averages are fixed values and so in (\ref{master}) only the
probabilities $p_{i}(\pm 1)$ can have a finite variation: 
\begin{eqnarray}
\delta \left\langle \mu _{j}\right\rangle _{\mu _{i}} &=&\frac{1}{2}\left(
\delta \left\langle \mu _{i}\right\rangle \right) \left( \left\langle \mu
_{j}\right\rangle _{i\uparrow }-\left\langle \mu _{j}\right\rangle
_{i\downarrow }\right)  \notag \\
&=&\left( \delta \left\langle \mu _{i}\right\rangle \right) \left\langle \mu
_{i}\mu _{j}\right\rangle ,
\end{eqnarray}%
where in the last step we have again made use of the fact that we are
dealing with the unmagnetized phase. Substituting this expression in the
formula for the cavity field (\ref{cavity}) gives 
\begin{equation}
\delta B_{i}^{cavity}=\delta B_{i}^{ext}+\left( \sum_{j}V_{ij}\left( \delta
\left\langle \mu _{j}\right\rangle \right) \right) -\lambda \delta
\left\langle \mu _{i}\right\rangle ,  \label{cavity2}
\end{equation}%
where we have introduced the parameter $\lambda $ as a measure of the
interaction energy per spin: 
\begin{equation}
\lambda =\sum_{j}V_{ij}\left\langle \mu _{i}\mu _{j}\right\rangle .
\end{equation}

By comparison with the earlier Weiss result, we see that the reaction field
has been isolated from the other contributions to the mean field, i.e. $%
\delta B_{i}^{reaction}=+\lambda \,\delta \left\langle \mu _{i}\right\rangle
,$which must be subtracted as in (\ref{cavity2}) when calculating the
response of spin $i$ to the Onsager cavity field. By way of Fourier
transformation one gets 
\begin{equation}
\delta \left\langle \mu _{\vec{q}}\right\rangle =c_{0}\beta \left( \delta B_{%
\vec{q}}^{ext}+V_{\vec{q}}\delta \left\langle \mu _{\vec{q}}\right\rangle
-\lambda \delta \left\langle \mu _{\vec{q}}\right\rangle \right)
\end{equation}%
and from this the susceptibility 
\begin{equation}
\chi (\vec{q})=\frac{\delta \left\langle \mu _{\vec{q}}\right\rangle }{%
\delta B_{\vec{q}}^{ext}}=\frac{c_{0}\beta }{1-c_{0}\beta (V_{\vec{q}%
}-\lambda )}.
\end{equation}%
In the light of the Weiss calculation this result could easily have been
guessed; at this point, however, the calculation served the simple purpose
of introducing the basic concepts of Onsager's reaction-field theory.

\subsection{The Heisenberg model}

We shall now apply the Onsager method to the Heisenberg model. While this
generalization is quite straightforward, we want to point out one of the
main shortcomings of the method, namely the fact that Onsager reaction-field
theory is incapable of incorporating the effect of dimensionality on
long-range order in a consistent way, contrary to what has been conjectured
e.g. in \cite{Bro Tho 1967} or \cite{Tus Szc Log 1995}. We shall follow \cite%
{Sin 2000} in treating the spin-1/2 anisotropic Heisenberg model given by 
\begin{equation}
H=-\sum_{n.n.ij}\left( J_{ij}\left(
S_{i}^{x}S_{j}^{x}+S_{i}^{y}S_{j}^{y}\right)
+J_{ij}^{z}S_{i}^{z}S_{j}^{z}\right) ,  \label{Heisanis}
\end{equation}%
where $J_{ij}^{z}=\varepsilon J_{ij}>J_{ij}$ is assumed. This kind of
anisotropy in spin space favouring uniaxial ordering is widely used (for a
Green's function approach to be discussed later within the Tyablikov
approximation, see e.g. \cite{Hau Bro Cor Cos 1972}) and should, even in 2D,
certainly break the symmetry one encounters for the corresponding isotropic
model. Once we have established the Onsager reaction-field theory for the
Heisenberg model (\ref{Heisanis}), we will be able to check the validity of
the Onsager method by comparing its predictions with this expectation.

Again, we start with a simple \textbf{Weiss mean-field theory} of an
infinitesimal external field giving rise to an total mean field 
\begin{equation}
B_{iz}^{Weiss}=B_{iz}^{ext}+\sum_jJ_{ij}^z\left\langle S_j^z\right\rangle ,
\label{WeissHeis}
\end{equation}
where the expectation value $\left\langle S_j^z\right\rangle $ is to be
calculated self-consistently from 
\begin{equation}
\left\langle S_i^z\right\rangle =\frac{\sum\limits_{\{S\}}m_ie^{\beta
m_iB_{iz}^{Weiss}}}{\sum\limits_{\{S\}}e^{\beta m_iB_{iz}^{Weiss}}}
\label{weissszett}
\end{equation}
and possible proportionality constants have been dropped for clarity ($m$:
quantum number characterizing the $z$-component of the spin). Taking the
condition of infinitesimal external field seriously, we can expand the
numerator and denominator of (\ref{weissszett}) to first order, 
\begin{eqnarray}
\left\langle S_i^z\right\rangle &\approx &\frac{\sum\limits_{\{S\}}m_i\left(
1+\beta m_iB_{iz}^{Weiss}\right) }{\sum\limits_{\{S\}}\left( 1+\beta
m_iB_{iz}^{Weiss}\right) }  \notag \\
&\approx &\left( \sum\limits_{\{S\}}\frac{\left( m_i\right) ^2}{2S+1}\right)
\beta B_{iz}^{Weiss},
\end{eqnarray}
where in the second step the denominator has been expanded to first order,
i.e. 
\begin{equation*}
\left( \sum\limits_{\{S\}}\left( 1+\beta m_iB_{iz}^{Weiss}\right) \right)
^{-1}\approx \left( \sum\limits_{\{S\}}1\right) ^{-1}=\frac 1{2S+1},
\end{equation*}
and we have also made use of $\sum\limits_{\{S\}}m_i=0.$ The sum over spin
configurations, symbolized by $\sum_{\{S\}},$ can be written explicitly as
the sum over integer steps $m_i=-S,-S+1,...,+S:$%
\begin{eqnarray}
\left\langle S_i^z\right\rangle &=&\left( \sum_{m=-S}^{+S}\frac{m^2}{2S+1}%
\right) \beta B_{iz}^{Weiss}=\frac 13S(S+1)\beta B_{iz}^{Weiss}  \notag \\
&=&\left\langle \left( m_i\right) ^2\right\rangle ^{free}\beta B_{iz}^{Weiss}
\label{sizchi0}
\end{eqnarray}
making use of the familiar partial sum formula $\sum_{k=1}^nk^2=\frac{%
n(n+1)(2n+1)}6$ and the isotropic distribution of spin orientations in the
free case. Defining 
\begin{equation}
\chi ^0\equiv \beta \left\langle (m_i)^2\right\rangle =\beta \frac{S(S+1)}3,
\label{chi0}
\end{equation}
and using (\ref{sizchi0}) and (\ref{WeissHeis}), we get for the Fourier
transform of the spin $z$-component the result 
\begin{eqnarray}
\left\langle S_z(\vec{q})\right\rangle &=&\sum_i\chi ^0e^{i\vec{q}\cdot \vec{%
R}_i}\left( B_{iz}^{ext}+\sum_jJ_{ij}^z\left\langle S_j^z\right\rangle
\right)  \notag \\
&\equiv &\chi ^0\left( B_z^{ext}(\vec{q})+J^z(\vec{q})\left\langle S^z(\vec{q%
})\right\rangle \right)
\end{eqnarray}
leading to the Weiss mean-field susceptibility for the anisotropic
Heisenberg model 
\begin{equation}
\chi _{MF}^z(\vec{q})=\frac{\chi ^0}{1-\chi ^0J^z(\vec{q})}
\label{chizettqmf}
\end{equation}
from which we can deduce an expression for the transition temperature by
setting the denominator of (\ref{chizettqmf}) to zero so that the
longitudinal susceptibility diverges, indicating a phase transition at the
ordering wave vector $\vec{q}=\vec{Q}$: 
\begin{equation}
T_c^{MF}=\frac{S(S+1)}{3k_B}J^z(\vec{Q}).  \label{crittempmf}
\end{equation}

Now, going over to \textbf{Onsager reaction-field theory} we write, by
analogy with the Ising model, 
\begin{equation}
B_{iz}^{cavity}=\sum_{j}J_{ij}^{z}\left\langle S_{j}^{z}\right\rangle
-\sum_{j}\lambda _{ij}J_{ij}^{z}\left\langle S_{i}^{z}\right\rangle ,
\end{equation}%
the second sum being again the reaction field which must be subtracted from
the total mean field. The parameters $\lambda _{ij}$ will in general be
temperature-dependent and are determined by the correlation of the
surrounding spins $j$ with the ``frozen'' spin $i.$ One easily derives the
Onsager susceptibility 
\begin{equation}
\chi _{Ons}^{z}(\vec{q})=\frac{\chi ^{0}}{1-\chi ^{0}\left( J^{z}(\vec{q}%
)-\lambda \right) },
\end{equation}%
where $\lambda =\sum_{j}\lambda _{ij}J_{ij}^{z}.$ Assuming, as before, a
phase transition at an ordering wave vector $\vec{q}=\vec{Q}$ one gets for
the critical temperature 
\begin{equation}
T_{c}^{Ons}=\frac{S(S+1)}{3k_{B}}(J^{z}(\vec{Q})-\lambda )
\end{equation}%
which, dividing by (\ref{crittempmf}), gives 
\begin{equation}
\frac{T_{c}^{Ons}}{T_{c}^{MF}}=\left( 1-\frac{\lambda }{J^{z}(\vec{Q})}%
\right) .  \label{tconsovertcmf}
\end{equation}

Singh \cite{Sin 2000} has noted that a simple procedure to determine the
parameter $\lambda $ appearing in (\ref{tconsovertcmf}) is provided by the
fluctuation-dissipation theorem \cite{Whi 1983} which, for zero external
field, is 
\begin{equation}
\beta \left\langle S_i^zS_j^z\right\rangle ^0=\sum_{\vec{q}}\chi ^z(\vec{q}%
)e^{i\vec{q}\cdot (\vec{R}_i-\vec{R}_j)}.  \label{fdt}
\end{equation}
In the isotropic case, one would have $\left\langle \left( m_i\right)
^2\right\rangle =\frac{S(S+1)}3$ and all values of $m_i$ would have equal
probability. In the anisotropic case larger values of $\left| \left\langle
S_i^z\right\rangle \right| $ are favoured, while the corresponding pairs $%
\left\langle S_i^z\right\rangle =\pm m$ remain equally probable, so that $%
\left\langle \left( m_i\right) ^2\right\rangle ^0=\alpha \frac{S(S+1)}3.$
Theorem (\ref{fdt}) gives 
\begin{eqnarray}
\beta \left\langle \left( m_i\right) ^2\right\rangle ^0 &=&\sum_{\vec{q}%
}\chi ^z(\vec{q})=\sum_{\vec{q}}\frac{\chi ^0}{1-\chi ^0(J^z(\vec{q}%
)-\lambda )}  \notag \\
&=&\alpha \cdot \beta \frac{S(S+1)}3
\end{eqnarray}

With (\ref{chi0}) and (\ref{crittempmf}) one finds $\chi
^{0}(T_{c}^{Ons})\cdot J^{z}(\vec{Q})=\frac{T_{c}^{MF}}{T_{c}^{Ons}}$ which
together with $\lambda =J^{z}(\vec{Q})-J^{z}(\vec{Q})\frac{T_{c}^{Ons}}{%
T_{c}^{MF}}$ gives the final result for the Onsager susceptibility: 
\begin{eqnarray}
\beta \left\langle \left( m_{i}\right) ^{2}\right\rangle ^{0} &=&\alpha \chi
^{0}=\sum_{\vec{q}}\chi ^{z}(\vec{q})  \notag \\
&=&\sum_{\vec{q}}\frac{\chi ^{0}}{J^{z}(\vec{Q})\chi ^{0}-\frac{T_{c}^{MF}}{%
T_{c}^{Ons}}\frac{J^{z}(\vec{q})}{J^{z}(\vec{Q})}}  \notag \\
\alpha \chi ^{0} &=&\sum_{\vec{q}}\frac{\chi ^{0}}{\frac{T_{c}^{MF}}{%
T_{c}^{Ons}}\left( 1-\frac{J^{z}(\vec{q})}{J^{z}(\vec{Q})}\right) }.
\end{eqnarray}%
The constant $\chi ^{0}$ cancels and, since we are in the spin-1/2 case, the
constant $\alpha ,$ which in the anisotropic case would normally ``shift
probability'' to pairs $\left\langle S_{i}^{z}\right\rangle =\pm m$ with
higher $\left| m\right| $, is unity, $\alpha _{S=1/2}\equiv 1,$ because
\thinspace $\pm 1/2$ are the only orientations possible. Thus, we get the
final result 
\begin{equation*}
\frac{T_{c}^{Ons}}{T_{c}^{MF}}=\frac{1}{\sum\limits_{\vec{q}}\frac{1}{\left(
1-\frac{J^{z}(\vec{q})}{J^{z}(\vec{Q})}\right) }}.
\end{equation*}%
From this, we see that while Onsager reaction-field theory does modify the
ordinary mean-field result, it is independent of the anisotropy assumed
above by setting $J_{ij}^{z}=\varepsilon J_{ij}$, since the anisotropy
parameter $\varepsilon $ will drop out of the fraction $J^{z}(\vec{q})/J^{z}(%
\vec{Q})$ in the denominator. Effects of dimensionality that stem from the
summation $\sum_{\vec{q}}$ are, however, retained, e.g. its logarithmic
divergence in two dimensions.

In the next section, we shall discuss the same Hamiltonian using the
Tyablikov method and find that it correctly treats this and other limiting
cases. The attempt to enforce correct Mermin-Wagner type behaviour within
mean-field theory obviously fails, at least as far as the Onsager
improvement of mean-field theory is concerned. We note that this observation %
\cite{Sin 2000} refutes contrary assertions in the literature (such as \cite%
{Tus Szc Log 1995}).

\section{Mermin-Wagner Theorem and Tyablikov Method}

Tyablikov \cite{Bog Tya 1959} proposed a method of approximation for the
Heisenberg model by suggesting a decoupling in the higher-order Green's
functions derived within the equation of motion method. Let us remind
ourselves of the Heisenberg Hamiltonian, 
\begin{equation}
H=-\sum_{ij}J_{ij}(S_{i}^{+}S_{j}^{-}+S_{i}^{z}S_{j}^{z})-g_{J}\frac{\mu _{B}%
}{\hbar }B_{0}\sum_{i}S_{i}^{z},  \label{Heisenberg}
\end{equation}%
where we have added a symmetry-breaking term $H_{b}=g_{j}\frac{\mu _{B}}{%
\hbar }B_{0}\sum_{i}S_{i}^{z}:=b\sum_{i}S_{i}^{z}$ to the unperturbed
Hamiltonian $H_{0}.$ The coupling constants $J_{ij}$ fulfill the symmetry
conditions ($J_{ij}=J_{ji}$; $J_{ii}=0)$ and are taken to give the
ferromagnetic case: $J_{ij}>0.$

In order to evaluate the problem, we employ the equation of motion method by
use of Green's functions. Since we are interested in the question whether or
not the system described by the many-body model Hamiltonian exhibits
spontaneous ferromagnetic order, we choose as the appropriate Green's
function 
\begin{eqnarray}
G_{ij}^{ret}(t-t^{\prime }) &=&\langle \langle
S_{i}^{+}(t);S_{j}^{-}(t^{\prime })\rangle \rangle ^{ret}  \notag \\
&=&-i\theta (t-t^{\prime })\left\langle \left[ S_{i}^{+}(t),S_{j}^{-}(t^{%
\prime })\right] _{-}\right\rangle
\end{eqnarray}%
which, when plugged into the first-order equation of motion gives 
\begin{eqnarray}
EG_{ij}^{ret}(E) &=&\hbar \left\langle \left[ S_{i}^{+},S_{j}^{-}\right]
_{-}\right\rangle +\left\langle \left\langle \left[ S_{i}^{+},H\right]
_{-};S_{j}^{-}\right\rangle \right\rangle _{E}^{ret}  \notag \\
&=&2\hbar ^{2}\delta _{ij}\left\langle S_{i}^{z}\right\rangle +\hbar
bG_{ij}^{ret}(E)-2\hbar \sum_{m}J_{im}\left( \left\langle \left\langle
S_{m}^{+}S_{i}^{z};S_{j}^{-}\right\rangle \right\rangle _{E}^{ret}\right. 
\notag \\
&&-\left. \left\langle \left\langle
S_{m}^{z}S_{i}^{+};S_{j}^{-}\right\rangle \right\rangle _{E}^{ret}\right) .
\label{eom}
\end{eqnarray}%
\linebreak

The higher-order Green's functions must be decoupled in order to make the
problem analytically and numerically tractable. In the Tyablikov decoupling
scheme, this is done by separating $S^{z}$ from the spin-flip operators $%
S^{\pm }$ in the higher-order Green's function \textit{ad hoc} via the
substitution 
\begin{equation}
\left\langle \left\langle S_{m}^{z}S_{i}^{+};S_{j}^{-}\right\rangle
\right\rangle \longrightarrow \left\langle S_{m}^{z}\right\rangle
\left\langle \left\langle S_{i}^{+};S_{j}^{-}\right\rangle \right\rangle .
\label{decoupling}
\end{equation}

One then arrives at the equation 
\begin{equation}
\left( E-g_{J}\mu _{B}B_{0}-2\hbar \left\langle S^{z}\right\rangle
\sum_{m}J_{im}\right) G_{ij}^{ret}(E)=2\hbar ^{2}\delta _{ij}\left\langle
S_{i}^{z}\right\rangle -2\hbar \sum_{m}\left\langle S_{i}^{z}\right\rangle
G_{mj}^{ret}(E)J_{im}  \label{ortseom}
\end{equation}%
which can be solved for the Green's function explicitly in momentum space
with the standard Fourier transforms $(\vec{q}$ is chosen from the first
Brillouin zone)

\begin{eqnarray}
G_{\vec{q}}^{ret}(E) &=&\frac{1}{N}\sum_{ij}G_{ij}^{ret}(E)e^{i\vec{q}\cdot (%
\vec{R}_{i}-\vec{R}_{j})} \\
\delta _{ij} &=&\frac{1}{N}\sum_{\vec{q}}e^{i\vec{q}\cdot (\vec{R}_{i}-\vec{R%
}_{j})},
\end{eqnarray}%
thus arriving at the solution 
\begin{equation}
G_{\vec{q}}^{ret}(E)=\frac{2\hbar ^{2}\left\langle S^{z}\right\rangle }{E-E(%
\vec{q})+i0^{+}},  \label{impulsgf}
\end{equation}%
where the energy 
\begin{equation}
E(\vec{q})=2\hbar \left\langle S^{z}\right\rangle (J_{0}-J(\vec{q}%
))+g_{J}\mu _{B}B_{0}
\end{equation}%
has been introduced $(J_{0}:=\sum_{i}J_{ij})$. Note that in the transition
from eqn. (\ref{ortseom}) to eqn. (\ref{impulsgf}) we have made use of
translational symmetry 
\begin{equation}
\left\langle S_{i}^{z}\right\rangle =\left\langle S_{m}^{z}\right\rangle
\equiv \left\langle S^{z}\right\rangle ,  \label{translatsym}
\end{equation}%
thus assuming the ferromagnetic case where all spins point into the same
direction. The Green's function, by virtue of the relation $S_{\vec{q}}(E)=-%
\frac{1}{\pi }\mathfrak{I}G_{\vec{q}}^{ret}(E)$ gives the spectral density 
\begin{equation}
S_{\vec{q}}(E)=2\hbar ^{2}\left\langle S^{z}\right\rangle \delta \left(
E-2\hbar \left\langle S^{z}\right\rangle \left( J_{0}-J(\vec{q})\right)
-g_{J}\mu _{B}B_{0}\right)
\end{equation}%
which allows one to make use of the spectral theorem for calculating the
expectation value 
\begin{equation}
\left\langle S_{j}^{-}S_{i}^{+}\right\rangle =\frac{1}{N}\sum_{\vec{q}}e^{-i%
\vec{q}\cdot (\vec{R}_{i}-\vec{R}_{j})}\frac{1}{\hbar }\int_{-\infty
}^{\infty }dE\frac{S_{\vec{q}}(E)}{e^{\beta E}-1}.
\end{equation}%
In the diagonal case $(i=j)$, spin algebra yields (restricting our
attention, for the moment, to the $S=\frac{1}{2}$ case) the formula $%
\left\langle S_{i}^{-}S_{i}^{+}\right\rangle =\hbar ^{2}S-\hbar \left\langle
S_{(i)}^{z}\right\rangle .$ (Because of translational symmetry (\ref%
{translatsym}) we shall drop the site index $i$.) Combining the two
formulas, a compact though implicit equation for the expectation value of
the $z-$component of the spin is derived: 
\begin{equation}
\left\langle S^{z}\right\rangle =\hbar S-2\frac{\left\langle
S^{z}\right\rangle }{N}\sum_{\vec{q}}\frac{1}{e^{\beta E(\vec{q})}-1}.
\label{spontmag}
\end{equation}%
Up to this point, we have only summarized the well-known Tyablikov procedure
for calculating the magnetization. In the following section, we shall
examine whether or not the Tyablikov decoupling conserves the Mermin-Wagner
theorem, which states that there can be no finite-temperature $(\beta
<\infty )$ magnetic phase transition in one or two dimensions.

\subsection{Tyablikov decoupling and Mermin-Wagner theorem}

The standard proof of the Mermin-Wagner theorem for a\textit{\ given model
Hamiltonian} starts with the Bogoliubov inequality, from which an upper
bound for the magnetization is derived which is then shown to vanish in the
limit $B_{0}\rightarrow 0$ at any finite temperature. In our case,
discussing an approximative method, the Bogoliubov inequality cannot be used
for testing the Mermin-Wagner theorem since the method does not approximate
the Hamiltonian but the higher-order Green's functions instead. However,
since with eqn. (\ref{spontmag}) an expression for the magnetization $%
\left\langle S^{z}\right\rangle $ is given, it is only natural to examine
this equation in order to check whether the Mermin-Wagner theorem remains
indeed valid within the Tyablikov method.

First we reformulate (\ref{spontmag}) writing it as 
\begin{equation}
\frac{\hbar S}{\left\langle S^z\right\rangle }=\frac 1N\sum_{\vec{q}}\frac{%
e^{\beta E(\vec{q})/2}+e^{-\beta E(\vec{q})/2}}{e^{\beta E(\vec{q}%
)/2}-e^{-\beta E(\vec{q})/2}}=\frac 1N\sum_{\vec{q}}\coth \left( \frac \beta
2E(\vec{q})\right)
\end{equation}
which, in the continuum notation appropriate for the thermodynamic limit,
can be written as 
\begin{equation}
\frac{\hbar S}{\left\langle S^z\right\rangle }=v^{(d)}\cdot \frac 1{(2\pi
)^d}\int\limits_{1.\text{B.Z}.}d^d\vec{q}\coth \left( \frac \beta 2\left(
2\hbar \left\langle S^z\right\rangle \left( J_0-J(\vec{q})\right) +g_J\mu
_BB_0\right) \right) ,
\end{equation}
where the full expression for the energy $E(\vec{q})$ has been inserted and $%
v^{(d)}=\lim\limits_{TDL}\frac{V^{(d)}}N$ is the specific $(d-$dimensional)
volume. A possible phase transition can reasonably be expected only for
temperatures below a certain critical value $T_c,$ for which the limit of a
vanishing external field must then be considered: 
\begin{equation}
B_0=0^{+};T-T_c=0^{-}\Leftrightarrow \left\langle S^z\right\rangle =0^{+}.
\label{szettnullplus}
\end{equation}
Using the series expansion of the hyperbolic cotangent in the assumed region
of finite spontaneous magnetization, 
\begin{equation}
\coth x=1+2\sum_{m=1}^\infty \exp \left( -2mx\right) ,
\end{equation}
one finally arrives at 
\begin{eqnarray}
\frac{\left\langle S^z\right\rangle }{\hbar S} &=&\frac
1{1+2\sum\limits_{m=1}^\infty e^{-\beta mg_J\mu _BB_0}\frac{v^{(d)}}{(2\pi
)^d}\int\limits_{1.\text{B.Z.}}d^d\vec{q}\exp \left( -2m\beta \hbar
\left\langle S^z\right\rangle \left( J_0-J(\vec{q})\right) \right) }  \notag
\\
&=&\frac 1{1+2\sum\limits_{m=1}^\infty e^{-\beta mg_J\mu _BB_0}\cdot
I^{(d)}}=\frac 1{1+2\phi ^{(d)}}.  \label{szettandphi}
\end{eqnarray}
(which at the same time defines the quantities $I^{(d)}$ and $\phi ^{(d)}).$

The choice of dimension $D$ will turn out to be crucial for evaluating the
behaviour of $\left\langle S^z\right\rangle $ in the limit $B_0\rightarrow
0. $

\subsection{Proof of Mermin-Wagner theorem in two dimensions}

To prove the Mermin-Wagner theorem, it must be shown that in two dimensions
there exists a bound to $\left\langle S^z\right\rangle $ that vanishes in
the limit $B_0\rightarrow 0$ (see eqn. (\ref{szettnullplus})). This is
certainly the case if the quantity 
\begin{equation}
\phi ^{(2)}=\sum\limits_{m=1}^\infty e^{-\beta mg_J\mu _BB_0}\frac{v^{(2)}}{%
(2\pi )^2}\int\limits_{1.\text{B.Z.}}d^2\vec{q}\exp \left( -2m\beta \hbar
\left\langle S^z\right\rangle \left( J_0-J(\vec{q})\right) \right)
\end{equation}
diverges.

In a first step, the support of the integral is restricted to a sphere
inscribed into the first Brillouin zone with radius $q_0$. Secondly, the
integrand itself can be bounded from below by making use of the fact that $%
J_0-J(\vec{q})\leq \tilde{Q}q^2,$ as long as the interaction decays rapidly
enough for the quantity $\tilde{Q}=\frac 1{4N}\sum_{ij}\left| \vec{R}_i-\vec{%
R}_j\right| ^2\left| J_{ij}\right| $ to converge. Now spherical coordinates
can be used, and since the integrand has been replaced by a function of $%
\left| \vec{q}\right| $ only, one finds $(\xi _i=const.)$%
\begin{eqnarray}
\phi ^{(2)} &\geq &\sum_{m=1}^\infty e^{-\beta mg_J\mu _BB_0}\frac{v^{(2)}}{%
2\pi }\int\limits_0^{q_0}dqqe^{-2m\beta \hbar \left\langle S^z\right\rangle 
\tilde{Q}q^2}  \notag \\
&=&\sum_{m=1}^\infty e^{-\beta mg_J\mu _BB_0}\frac{v^{(2)}}{2\pi }%
\int\limits_0^{q_0}dq\left( \frac d{dq}\exp \left( -2m\beta \hbar \tilde{Q}%
q^2\left\langle S^z\right\rangle \right) \right) \frac{(-1)}{4m\beta \hbar 
\tilde{Q}\left\langle S^z\right\rangle }  \notag \\
&=&\frac{v^{(2)}}{8\pi \beta \hbar \tilde{Q}\left\langle S^z\right\rangle }%
\sum_{m=1}^\infty \frac{\exp \left( -\beta mg_J\mu _BB_0\right) }m\left(
1-\exp \left( -2m\beta \hbar \tilde{Q}\left\langle S^z\right\rangle
q_0^2\right) \right)  \notag \\
&=&\xi _0\sum_{m=1}^\infty \left( \frac{e^{-\xi _1mB_0}}m-e^{-\xi
_1mB_0}\cdot \frac{e^{-2m\beta \hbar \tilde{Q}\left\langle S^z\right\rangle
q_0^2}}m\right) .  \label{lowboundphi}
\end{eqnarray}

Assuming for the moment a finite value of $\left\langle S^z\right\rangle ,$
the second sum converges in the limit $B_0\rightarrow 0,$ since subsequent
terms vanish exponentially with $m.$ The first sum, however, is the harmonic
series and, thus, diverges. But as $\phi ^{(2)}$ diverges, eqn. (\ref%
{szettandphi}) states that in this case $\left\langle S^z\right\rangle $
must be zero. Starting from a finite value of $\left\langle S^z\right\rangle 
$ one arrives at a contradiction, so the assumption must be wrong.

At first sight, it may appear obvious from (\ref{lowboundphi}) and (\ref%
{szettandphi}) that $\left\langle S^{z}\right\rangle =0$ is indeed
consistent with both equations, since the two series appearing in the lower
bound of $\phi ^{(2)}$ cancel exactly. However, in the derivation it has
been tacitly assumed that $\left\langle S^{z}\right\rangle \neq 0.$ This can
be seen by explicitly inserting (\ref{lowboundphi}) into (\ref{szettandphi})
to get the full relation for the magnetization: 
\begin{equation}
\frac{\left\langle S^{z}\right\rangle }{\hbar S}\leq \frac{1}{1+\frac{v^{(2)}%
}{4\pi \beta \hbar \tilde{Q}\left\langle S^{z}\right\rangle }%
\sum\limits_{m=1}^{\infty }\frac{1}{m}\exp \left( -\beta mg_{J}\mu
_{B}B_{0}\right) \left[ 1-\exp \left( -2m\beta \hbar \tilde{Q}\left\langle
S^{z}\right\rangle q_{0}^{2}\right) \right] },
\end{equation}%
where the assumption $\left\langle S^{z}\right\rangle =0$ would lead to an
expression ``$0/0"$ in the denominator. Thus, this case can only be analyzed
by means of l'Hospital's rule, where one considers the limit $%
\lim_{\left\langle S^{z}\right\rangle \rightarrow 0}\lim_{B_{0}\rightarrow
0}\left( \cdots \right) $: 
\begin{eqnarray}
&&\lim_{\left\langle S^{z}\right\rangle \rightarrow 0}\left(
\lim_{B_{0}\rightarrow 0}\frac{\left\langle S^{z}\right\rangle }{\hbar S}%
\right)  \notag \\
&\leq &\frac{1}{1+\frac{v^{(2)}}{4\pi \beta \hbar \tilde{Q}}%
\lim\limits_{\left\langle S^{z}\right\rangle \rightarrow 0}\frac{1}{%
\left\langle S^{z}\right\rangle }\sum\limits_{m=1}^{\infty }\frac{1}{m}%
\left( 1-\exp \left( -2m\beta \hbar \tilde{Q}\left\langle S^{z}\right\rangle
q_{0}^{2}\right) \right) }  \notag \\
&=&\frac{1}{1+\frac{v^{(2)}}{4\pi \beta \hbar \tilde{Q}}\sum\limits_{m=1}^{%
\infty }\frac{2m\beta \hbar \tilde{Q}q_{0}^{2}}{m}\lim\limits_{\left\langle
S^{z}\right\rangle \rightarrow 0}\exp \left( -2m\beta \hbar \tilde{Q}%
\left\langle S^{z}\right\rangle q_{0}^{2}\right) },
\end{eqnarray}%
where the last equality is due to l'Hospital's rule. Reformulating the
initial equation in this way, the denominator now is explicitly divergent,
so indeed $\left\langle S^{z}\right\rangle =0$.

The situation changes when an external field $B_0\neq 0$ is applied: As one
can easily see from (\ref{lowboundphi}), both series then converge; thus $%
\phi $ is also convergent and one arrives at a non-vanishing upper bound to
the magnetization. The upper bound is still a function of $\left\langle
S^z\right\rangle ,$ so it will be difficult to extract more detailed
information about $\left\langle S^z\right\rangle $ from this relation except
for the mere possibility of a finite value of the magnetization.

\subsection{Three-dimensional case}

Testing the above procedure in the 3-dimensional case provides another
valuable check. We shall, at this point, show that the lower bound of $\phi $
derived by analogy with the two-dimensional case does \textit{not} diverge
in three dimensions. This is, of course, only a negative statement about the
particular lower bound. Typically, for tests of the Mermin-Wagner theorem in
three dimensions, this is all one can expect, since a positive statement
about a possible lower bound of the magnetization (i.e. a converging \textit{%
upper} bound of $\phi $) will rarely be possible. Later, we will present a
different treatment of the three-dimensional case which will also allow us
to discuss the dimensional crossover from two to three dimensions (or vice
versa) in much greater detail. For the moment we note that, indeed, the
finite upper bound to the magnetization survives the limit $B_0\rightarrow 0$
in three dimensions:

\begin{eqnarray}
&&\frac{\left\langle S^z\right\rangle }{\hbar S}  \notag \\
&=&\frac 1{1+2\sum\limits_{m=1}^\infty e^{-\beta mg_J\mu _BB_0}\frac{v^{(3)}%
}{2\pi ^2}\int\limits_0^{q_0}dqq^2\exp \left( -2m\beta \hbar \left\langle
S^z\right\rangle \frac 1N\sum_{ij}\left( 1-e^{i\vec{q}\cdot \left( \vec{R}_i-%
\vec{R}_j\right) }\right) J_{ij}\right) }  \notag \\
&\leq &\left( 1+2\sum\limits_{m=1}^\infty e^{-\beta mg_J\mu _BB_0}\frac{%
v^{(3)}}{2\pi ^2}\int\limits_0^{q_0}dqq^2\exp \left( -2m\beta \hbar
\left\langle S^z\right\rangle \tilde{Q}q^2\right) \right) ^{-1}  \notag \\
&=&\left( 1+2\sum\limits_{m=1}^\infty e^{-\beta mg_J\mu _BB_0}\frac{v^{(3)}}{%
2\pi ^2}\frac d{d\left\langle S^z\right\rangle }\frac{(-1)}{2\beta m\hbar 
\tilde{Q}}\int\limits_0^{q_0}dq\exp \left( -2m\beta \hbar \left\langle
S^z\right\rangle \tilde{Q}q^2\right) \right) ^{-1}  \notag \\
&=&\left( 1+2\sum\limits_{m=1}^\infty e^{-\beta mg_J\mu _BB_0}\frac{v^{(3)}}{%
2\pi ^2}\frac d{d\left\langle S^z\right\rangle }\frac{(-1)}{\left( 2\beta
m\hbar \tilde{Q}\right) ^{3/2}\sqrt{\left\langle S^z\right\rangle }}%
\int\limits_0^{\sqrt{2\beta m\hbar \left\langle S^z\right\rangle \tilde{Q}}%
q_0}due^{-u^2}\right) ^{-1}  \notag \\
&=&\left( 1+2\sum\limits_{m=1}^\infty e^{-\beta mg_J\mu _BB_0}\frac{v^{(3)}}{%
2\pi ^2}\Delta _m\right) ^{-1}  \label{dreideszett}
\end{eqnarray}
with the quantity $\Delta _m$ defined as 
\begin{eqnarray}
\Delta _m &=&\frac d{d\left\langle S^z\right\rangle }\frac{-1}{\left( 2\beta
m\hbar \tilde{Q}\right) ^{3/2}\sqrt{\left\langle S^z\right\rangle }}%
\int\limits_0^{\sqrt{2m\beta \hbar \left\langle S^z\right\rangle \tilde{Q}}%
q_0}due^{-u^2}  \notag \\
&=&\frac{-1}{\left( 2\beta m\hbar \tilde{Q}\right) ^{3/2}\sqrt{\left\langle
S^z\right\rangle }}\left( \frac{d\xi }{d\left\langle S^z\right\rangle }%
\right) \frac d{d\xi }\int\limits_0^\xi due^{-u^2}  \notag \\
&&-\left( \frac d{d\left\langle S^z\right\rangle }\frac 1{\left( 2\beta
m\hbar \tilde{Q}\right) ^{3/2}\sqrt{\left\langle S^z\right\rangle }}\right)
\int\limits_0^{\sqrt{2\beta m\hbar \tilde{Q}\left\langle S^z\right\rangle }%
q_0}due^{-u^2},
\end{eqnarray}
where we have made use of the chain rule for the new variable 
\begin{equation*}
\xi =\sqrt{2\beta m\hbar \tilde{Q}\left\langle S^z\right\rangle }q_0.
\end{equation*}
The first integral is of the form $\frac d{dx}\int\limits_0^xf(y)dy=f(x)$,
so that 
\begin{equation}
\Delta _m=\frac{-q_0}{4\beta \hbar \tilde{Q}\left\langle S^z\right\rangle }%
\frac{\left( e^{-2\beta \hbar \tilde{Q}q_0^2\left\langle S^z\right\rangle
}\right) ^m}m+\frac 1{2\left( 2\beta m\hbar \tilde{Q}\right)
^{3/2}\left\langle S^z\right\rangle ^{3/2}}\int\limits_0^{\sqrt{2\beta \hbar 
\tilde{Q}\left\langle S^z\right\rangle m}q_0}due^{-u^2}.  \label{deltaem}
\end{equation}
Let us discuss this equation in relation to the inequality (\ref{dreideszett}%
). Assuming a finite non-vanishing value of the magnetization, we choose a
fixed non-zero number $\varepsilon $ such that $\left\langle
S^z\right\rangle >\varepsilon >0.$ We can then approximate the second
integral in (\ref{deltaem}) from below by setting the integrand equal to the
value of $e^{-u^2}$ at its upper bound, thus strengthening the inequality in
(\ref{dreideszett}): 
\begin{equation}
\Delta _m\geq \frac{q_0}{4\beta \hbar \tilde{Q}\left\langle S^z\right\rangle 
}\left( \varepsilon -1\right) \frac{\left( \exp \left( -2\beta \hbar \tilde{Q%
}q_0^2\varepsilon \right) \right) ^m}m.
\end{equation}
The series in the denominator of the R.H.S. in (\ref{dreideszett}) then
always converges, even in the limit $B_0\rightarrow 0,$ so that a magnetic
phase transition is not ruled out. This is, of course, no proof that a
magnetic phase transition must occur; as one can easily see from (\ref%
{deltaem}) by taking the limit $\left\langle S^z\right\rangle \rightarrow 0$%
, the individual terms of the series vanish: $\Delta _m 
\begin{array}{l}
\\ 
\overrightarrow{\left\langle S^z\right\rangle \rightarrow 0}%
\end{array}
0,$ which is not in contradiction with (\ref{dreideszett}).

\subsection{Absence of spontaneous sublattice magnetization in ABAB-type
antiferromagnets}

While the Tyablikov method has been very popular for treating ferromagnets,
where it has led to convincing results over a broad temperature range, it
can, in fact, be generalized to cover antiferromagnets as well \cite{Fuc
1960}, \cite{Hew Ter 1964}. Recently, there has been renewed interest in the
antiferromagnetic Heisenberg model with relation to high-temperature
superconductors such as YBa$_2$Cu$_3$O$_{6-x}$ which, for low doping,
displays antiferromagnetic behaviour.

Quite generally, in the Heisenberg antiferromagnet, the respective
sublattices can be treated using the Tyablikov formalism essentially in the
way discussed above for ferromagnets. As to antiferromagnets, we want to
sketch briefly how to proceed. Following Hewson and ter Haar \cite{Hew Ter
1964} we consider, for simplicity, the case of the lattice being divided
into two sublattices A and B so that the nearest neighbours of every atom of
one sublattice all belong to the other sublattice (ABAB-type ordering).
(More complicated types of antiferromagnetic order can be discussed in a
similar way; for layered antiferromagnets see e.g. ref. \cite{Lin 1963}.) If
we restrict our attention to the case of nearest-neighbour interaction, thus
keeping only the essential features of the interaction, the Hamiltonian (\ref%
{Heisenberg}) can be written as 
\begin{equation}
H=\sum_f\sum_gJ_{fg}\left( S_f^{+}S_g^{-}+S_f^zS_g^z\right) -\frac{g_J\mu
_BB_0}\hbar \left( \sum_gS_g^z-\sum_fS_f^z\right)
\end{equation}
with $f$ and $g$ referring to the sublattice with predominantly spin-down or
spin-up orientation, respectively. (The sign change for $\downarrow $-spins
coupling to the external field is a matter of convention.)

Since we are now dealing with antiferromagnetic order, we can no longer
assume translational symmetry as in eqn. (\ref{translatsym}). Instead, for
purely geometric reasons, we must distinguish between ``same-lattice''
Green's functions $\left\langle \left\langle S_{f}^{+},S_{f^{\prime
}}^{-}\right\rangle \right\rangle $ or $\left\langle \left\langle
S_{g}^{+};S_{g^{\prime }}^{-}\right\rangle \right\rangle $ (where $f,$ $%
f^{\prime }$ and $g,g^{\prime }$ denote sites of the same sublattice $A$ or $%
B)$ on the one hand and ``mixed'' Green's functions $\left\langle
\left\langle S_{g}^{+};S_{f}^{-}\right\rangle \right\rangle $ or $%
\left\langle \left\langle S_{f}^{+};S_{g}^{-}\right\rangle \right\rangle $
on the other hand. The equation of motion (\ref{eom}) must be modified in
order to accommodate the distinction between the two kinds of Green's
functions. Thus, one arrives at a system of two equations of motion (one for
the ``same-lattice'' Green's function, the other for the ``mixed'' Green's
function) which can be written in matrix form. In the ``mixed'' case there
will be no inhomogenity as spin operators at different lattice sites commute
(which will always be the case when dealing with different sublattices). The
Fourier transforms of ``same-lattice'' and ``mixed'' Green's functions
(denoted by $G_{1\vec{k}}$ and $G_{2\vec{k}}$) are taken over the respective
reciprocal sublattices in $k$-space: 
\begin{eqnarray}
G_{1\vec{k}} &=&\sum_{g-g^{\prime }}\left\langle \left\langle
S_{g}^{+};S_{g^{\prime }}^{-}\right\rangle \right\rangle e^{-i\vec{k}\cdot
\left( \vec{R}_{g}-\vec{R}_{g^{\prime }}\right) } \\
G_{2\vec{k}} &=&\sum_{g-f}\left\langle \left\langle
S_{g}^{+};S_{f}^{-}\right\rangle \right\rangle e^{-i\vec{k}\cdot \left( \vec{%
R}_{g}-\vec{R}_{f}\right) },
\end{eqnarray}%
where the summation is over sites of the same sublattice \textit{or} over
different sublattices, respectively. The higher-order Green's functions are
decoupled, as in the ferromagnetic case, through the Tyablikov procedure (%
\ref{decoupling}). Inverting the 2$\times $2 Green's function matrix, one
can solve for the same-lattice Green's function $G_{1\vec{k}},$ which, via
the spectral theorem and the identity $\left\langle
S_{i}^{-}S_{i}^{+}\right\rangle =\hbar ^{2}S-\hbar \left\langle
S_{(i)}^{z}\right\rangle $ for the $S=\frac{1}{2}$ case, gives an expression
for the sublattice magnetization: 
\begin{equation}
\frac{\left\langle S_{A/B}^{z}\right\rangle }{\hbar S}=\frac{1}{1+\frac{2}{N}%
\sum_{\vec{k}}^{\prime }\left( \frac{J_{0}}{\sqrt{J_{0}^{2}-J(\vec{q})^{2}}}%
\coth \left( \beta \left\langle S_{A/B}^{z}\right\rangle \sqrt{J_{0}^{2}-J(%
\vec{q})^{2}}\right) -1\right) }.  \label{szundphi}
\end{equation}%
The formula resembles closely that for the ferromagnet, except for the
typical square-root dependence on the exchange integrals which is related to
the spin-wave dispersion in simple antiferromagnetic spin-wave theory \cite%
{QdM2 1986}. (For more details on the derivation see also refs. \cite{Lin
1964} and \cite{Lee Liu 1967}.) Earlier, for reasons of simplicity, we had
restricted the interaction to nearest neighbours. Thus, 
\begin{eqnarray}
J_{0} &=&\sum_{\substack{ n.n.  \\ i=fix}}J_{ij} \\
J(\vec{q}) &=&\sum_{\substack{ n.n.  \\ i=fix}}J_{ij}e^{i\vec{q}\cdot \left( 
\vec{R}_{i}-\vec{R}_{j}\right) },
\end{eqnarray}%
where ``$n.n."$ stands for ``nearest neighbours only'' and $i=$\textit{fix}
reflects the fact that only the dummy variable $j$ is summed over.

From (\ref{szundphi}) it is possible to derive an explicit formula for the
RPA N\'{e}el temperature $T_N^{RPA}$, by making use of the series expansion $%
\coth (x)\approx \frac 1x+...$ for small values of $x,$ i.e. close to the
transition point where $\left\langle S^z\right\rangle (T=T_N)=0:$%
\begin{equation}
T_N^{RPA}=\frac{J_0}{2k_B}\frac 1{\frac 2N\sum\limits_{\vec{k}}\frac
1{\left( 1-\frac{J^2(\vec{k})}{J_0^2}\right) }}.  \label{tnrpa}
\end{equation}
In our case, where $J_{ij}=J\neq 0$ for nearest neighbours $i,j$ only and
where the lattice is of ABAB structure, it is straightforward to show that
in the thermodynamic limit the sum in the denominator diverges in two
dimensions while remaining finite in three dimensions. Due to a $1/k^2$
divergence of the integrand close to $k=0$, which is not cancelled by the 2D
volume element $kdk,$ the sum diverges logarithmically in two dimensions. In
three dimensions, the sum remains finite. Thus, one recovers the
Mermin-Wagner theorem which states that there can be no finite-temperature
phase transition in the two-dimensional isotropic Heisenberg model.

\subsection{Transition from two to three dimensions via 3D anisotropic
interlayer coupling}

Previously, we saw (eqn. (\ref{szettandphi}),(\ref{szundphi})) that the
dimension determines the (sublattice) magnetization via summation over $D$%
-dimensional $k$-space. In those cases where one starts from an \textit{a
priori} restriction in geometry, the choice of dimension is fixed from the
very beginning. These cases were discussed at length for the ferromagnet
and, in the previous section, for the antiferromagnet as well. There is,
however, another method for changing effective dimensionality. Instead of
restricting the geometry of the system \textit{a priori}, one can change the
interaction in such a way as to mimic the change in dimension.

As an example, we shall return to our discussion of the antiferromagnet. The
interaction had been restricted to nearest neighbours. If we choose the
interaction parameters $J_{ij}$ so that (for a fixed lattice site $i)$ $%
J_{ij}=J_{||}$ if sites $i$ and $j$ lie within a plane and $J_{ij}=J_{\bot }$
if $j$ is located in the direction orthogonal to this plane, then, by
varying the anisotropy parameter 
\begin{equation}
\varepsilon :=\frac{J_{\bot }}{J_{||}}  \label{epsilon}
\end{equation}
we can change the coupling continuously from quasi two-dimensional $%
(\varepsilon =0)$ to (isotropic) three-dimensional $(\varepsilon =1).$

Of course we have to distinguish between the coordination numbers within the
plane and those along the orthogonal direction, $z_{||}$ and $z_{\perp }.$
Modifying (\ref{tnrpa}), the formula for the RPA N\'{e}el temperature, to
take into account the anisotropy gives (see also ref. \cite{Sin Sin 1992}
for comparison)%
\begin{equation}
T_{N}^{RPA}=\frac{J_{||}}{2k_{B}}\frac{z_{||}+\varepsilon z_{\perp }}{\frac{2%
}{N}\sum_{\vec{q}}\frac{1}{1-\left( \frac{z_{||}\gamma _{||}\left( \vec{q}%
)\right) +\varepsilon z_{\perp }\gamma _{\perp }(\vec{q})}{%
z_{||}+\varepsilon z_{\perp }}\right) ^{2}}}  \label{rpatemp}
\end{equation}%
where the substitution $J_{0}=\sum\limits_{\substack{ n,m  \\ i=fix}}%
J_{ij}=z_{||}J_{||}+\varepsilon z_{\perp }J_{||}$ has been carried out and
where we have used the common abbreviations%
\begin{eqnarray}
\gamma _{||}(\vec{q}) &=&\frac{1}{z_{||}}\sum_{n.n.||}e^{i\vec{q}\cdot \vec{%
\Delta}_{||}} \\
\gamma _{\perp }(\vec{q}) &=&\frac{1}{z_{\perp }}\sum_{n.n.\perp }e^{i\vec{q}%
\cdot \vec{\Delta}_{\perp }}.
\end{eqnarray}

Note that in the strictly two-dimensional case (where $z_{\bot }=0$and $\vec{%
q}=(q_x,q_y)$) the sum in the denominator (\ref{rpatemp}) diverges as
expected. In the following, however, we shall focus on the limit with $D=3$,
i.e. $\vec{q}=(q_x,q_y,q_z)=(\vec{q}_{||},q_{\bot }),$but with quasi
two-dimensional interaction: $\varepsilon \rightarrow 0.$

Apart from prefactors, in the thermodynamic limit the sum can be replaced by
an integral: 
\begin{equation}
\int d^{3}\vec{q}\frac{1}{1-\left( \frac{z_{||}\gamma _{||}(\vec{q}%
)+\varepsilon z_{\bot }\gamma _{\bot }(\vec{q})}{z_{||}+\varepsilon z_{\bot }%
}\right) ^{2}}.
\end{equation}%
Evidently, the main contribution to the integral will come from near $q=0,$
since there $\gamma _{||}(\vec{q})$ and $\gamma _{\bot }(\vec{q})$ can be
approximated by $1-\alpha _{||/\bot }q^{2}.$ Note that by a proper choice of 
$\alpha _{||}$and $\alpha _{\bot }$one can isolate a finite sphere around
the origin throughout which the approximation (as far as the total value of
the integral is concerned) holds \textit{exactly}. With a suitable choice of 
$\tilde{z}_{||}$, it is also possible to write 
\begin{equation}
\gamma _{||}(\vec{q})=\frac{1}{\tilde{z}_{||}}\sum_{n.n.\vec{R}}e^{i\vec{q}%
_{||}\cdot \vec{R}}
\end{equation}%
instead of the original definition, while preserving the qualitative
dependence on $\left| \vec{q}\right| $ (at least in the thermodynamic limit.)

With the above modifications, we find for the integral $I$over a suitable
sphere around the origin 
\begin{eqnarray}
I &=&\int_{S}d^{3}\vec{q}\frac{1}{1-\left( \frac{z_{||}}{z||+\varepsilon
z_{\bot }}\right) ^{2}(1-\alpha _{||}q_{||}^{2})^{2}-\left( \frac{%
\varepsilon z_{\bot }}{z_{||}+\varepsilon z_{\bot }}\right) ^{2}(1-\alpha
_{\bot }q_{\bot }^{2})^{2}+...}  \notag \\
&&\frac{{}}{...+4\left( \frac{\varepsilon z_{\bot }z_{||}}{%
z_{||}+\varepsilon z_{\bot }}\right) ^{2}(1-\alpha _{||}q_{||}^{2})(1-\alpha
_{\bot }q_{\bot }^{2})^{2}}  \notag \\
&\geq &\int\limits_{S}d^{3}\vec{q}\frac{1}{A_{\varepsilon }+B_{\varepsilon
}(1-\beta \varepsilon ^{2})q_{||}^{2}+\varepsilon ^{2}C_{\varepsilon }\gamma
q_{\bot }^{2}}=:I^{\prime }
\end{eqnarray}%
where only quadratic contributions have been retained in the last line.
Quantities with $\varepsilon $ as an index are smooth functions of $%
\varepsilon $ remaining finite as $\varepsilon \rightarrow 0.$ We have also
made extensive use of our previous notation $\vec{q}=(\vec{q}_{||},q_{\bot
}) $ where $\vec{q}_{||}$ is a vector of the basal plane and $q_{\bot }$ the
coordinate orthogonal to the plane.

The evaluation of the last integral is carried out in cylinder coordinates,
first integrating in $z$-direction from a lower cutoff $-\Lambda _{\bot }$
to the upper cutoff $+\Lambda _{\bot }.$ Afterwards, we integrate in the
basal plane using polar coordinates: 
\begin{eqnarray}
I^{\prime } &=&\int d^{2}\vec{q}\int\limits_{-\Lambda _{\bot }}^{+\Lambda
_{\bot }}dq_{\bot }\frac{1}{\left( A_{\varepsilon }+B_{\varepsilon }(1-\beta
\varepsilon ^{2})q_{||}^{2}\right) +\varepsilon ^{2}C_{\varepsilon }\gamma
q_{\bot }^{2}}  \notag \\
&=&2\pi \int\limits_{0}^{\Lambda _{||}}dq_{||}\frac{2q_{||}}{\sqrt{%
A_{\varepsilon }+B_{\varepsilon }\left( 1-\beta \varepsilon ^{2}\right)
q_{||}^{2}}}\arctan \left( \frac{\varepsilon \sqrt{C_{\varepsilon }\gamma }%
\Lambda _{\bot }}{\sqrt{A_{\varepsilon }+B_{\varepsilon }\left( 1-\beta
\varepsilon ^{2}\right) q_{||}^{2}}}\right)  \notag \\
&=&-\frac{4\pi /\left( \varepsilon \sqrt{C_{\varepsilon }\gamma }\Lambda
_{\bot }\right) }{B_{\varepsilon }(1-\beta \varepsilon ^{2})}%
\int\limits_{u_{0}}^{u_{1}}du\frac{\arctan u}{u^{2}},
\end{eqnarray}%
where in the last integral the substitution 
\begin{equation}
u:=\frac{\varepsilon \sqrt{C_{\varepsilon }\gamma }\Lambda _{\bot }}{\sqrt{%
A_{\varepsilon }+B_{\varepsilon }\left( 1-\beta \varepsilon ^{2}\right)
q_{||}^{2}}}
\end{equation}%
has been used, with $u_{0}=\infty $ and $u_{1}=\frac{\varepsilon \sqrt{%
C_{\varepsilon }\gamma }\Lambda _{\bot }}{\sqrt{A_{\varepsilon
}+B_{\varepsilon }(1-\beta \varepsilon ^{2})\Lambda _{||}^{2}}}$ as the
integration limits. The integral vanishes at infinity and, thus, we find 
\begin{eqnarray}
I^{\prime } &=&\frac{4\pi /\left( \varepsilon \sqrt{C_{\varepsilon }\gamma }%
\Lambda _{\bot }\right) }{B_{\varepsilon }(1-\beta \varepsilon ^{2})}\cdot %
\left[ \frac{\sqrt{A_{\varepsilon }+B_{\varepsilon }(1-\beta \varepsilon
^{2})\Lambda _{||}^{2}}}{\varepsilon \sqrt{C_{\varepsilon }\gamma }\Lambda
_{\bot }}\arctan \left( \frac{\varepsilon \sqrt{C_{\varepsilon }\gamma }%
\Lambda _{\bot }}{\sqrt{A_{\varepsilon }+B_{\varepsilon }(1-\beta
\varepsilon ^{2})\Lambda _{||}^{2}}}\right) \right.  \notag \\
&&\left. +\frac{1}{2}\ln \left( 1+\frac{A_{\varepsilon }+B_{\varepsilon
}(1-\beta \varepsilon ^{2})\Lambda _{||}^{2}}{\varepsilon ^{2}C_{\varepsilon
}\gamma \Lambda _{\bot }}\right) \right] .
\end{eqnarray}%
From this, one can read off the following result: While the arctan
approaches zero for $\varepsilon \rightarrow 0$ and thus may cancel the $%
1/\varepsilon ^{2}$ divergency due to the prefactors, there will always
remain a divergency in the logarithmic contribution. Since the integral $%
I^{\prime }$ determines the N\'{e}el temperature via eqn. (\ref{rpatemp}) 
\begin{equation}
T_{N}^{RPA}=\frac{J_{||}(z_{||}+\varepsilon z_{\bot })}{\tilde{\xi}k_{B}}%
\cdot \frac{1}{I^{\prime }}
\end{equation}%
we see that in the quasi-2D limit of vanishing interlayer coupling $%
(\varepsilon \rightarrow 0)$ the N\'{e}el temperature (as calculated from
Tyablikov RPA) vanishes as 
\begin{equation}
T_{N}^{RPA}%
\begin{array}{l}
\\ 
\widetilde{\varepsilon {\small \rightarrow }0}%
\end{array}%
\frac{\varepsilon }{\left| \ln \left( \varepsilon \right) \right| },
\end{equation}%
in complete accordance with the Mermin-Wagner theorem.

\subsection{Extending the results to higher spins $(S>1/2)$}

The results so far were derived for the $S=1/2$ case only. They can,
however, be extended to arbitrary spin by a rather simple procedure. The
definitions of $\phi ^{(d)}$ and $E(\vec{q})$ remain unchanged in relation
to the previous discussion for ferromagnets \cite{Pra 1963} as well as for
antiferromagnets \cite{Hew Ter 1964}. Only eqn. (\ref{szettandphi}) must be
modified to run 
\begin{equation}
\frac{\left\langle S^z\right\rangle }\hbar =\frac{\left[ S-\phi (S)\right] %
\left[ 1+\phi (S)\right] ^{2S+1}+\left( 1+S+\phi (S)\right) \left( \phi
(S)\right) ^{2S+1}}{\left( 1+\phi (S)\right) ^{2S+1}-\left( \phi (S)\right)
^{2S+1}}.  \label{high}
\end{equation}
The easiest way to reproduce the above $S=1/2$ results for higher spins is
to show that the convergence, or divergence, of $\phi ^{(d)}$ will lead to a
finite or vanishing $\left\langle S^z\right\rangle $ value, respectively.
This suffices because $\phi ^{(d)}$ is the only quantity that depends
explicitly on the dimension.

For large values of $\phi ^{(d)},$ one can expand (\ref{high}) in powers of $%
1/\phi ^{(d)}$to get \cite{QdM2 1986} 
\begin{equation}
\frac{\left\langle S^z\right\rangle }\hbar \approx \frac{S(S+1)}{3\phi (S)},
\end{equation}
thus establishing to order $\mathcal{O}(1/\phi ^2)$ that the results
obtained above carry over to this section.

Alternatively, writing (\ref{high}) in powers of $\phi ,$ one finds 
\begin{eqnarray}
\frac{\left\langle S^z\right\rangle }\hbar &=&\frac{\left[ S-\phi \right] %
\left[ 1+\phi \right] ^{2S+1}+\left( 1+S+\phi \right) \phi ^{2S+1}}{\left(
1+\phi \right) ^{2S+1}-\phi ^{2S+1}}  \notag \\
&=&\frac{\sum\limits_{n=0}^{2S+1}S\binom{2S+1}n\phi
^n-\sum\limits_{n=0}^{2S+1}\binom{2S+1}n\phi ^{n+1}+\phi ^{2S+1}(1+S)+\phi
^{2S+2}}{\phi ^{2S+1}-\sum\limits_{n=0}^{2S+1}\binom{2S+1}n\phi ^n}  \notag
\\
&=&\frac{\sum\limits_{n=1}^{2S}\left( S\binom{2S+1}n-\binom{2S+1}{n-1}%
\right) \phi ^n-S}{\sum\limits_{n=0}^{2S}\binom{2S+1}n\phi ^n}  \notag \\
&=&\frac{\sum\limits_{n=1}^{2S-1}\left( S-\frac n{2S-n+2}\right) \binom{2S+1}%
n\phi ^n-S}{\sum\limits_{n=0}^{2S}\binom{2S+1}n\phi ^n},
\end{eqnarray}
where in the last step we have used the recursion relation for binomial
coefficients. For $S\geq 1$ the coefficients in the numerator and the
denominator are finite, so from the general form 
\begin{equation}
\frac{\left\langle S^z\right\rangle }\hbar =\frac{\sum\limits_{n=1}^{2S-1}%
\alpha _n\phi ^n-S}{\sum\limits_{n=0}^{2S}\beta _n\phi ^n}  \label{genform}
\end{equation}
we can see that the polynomial in the denominator is of higher order than
the numerator since $\beta _{2S}=\binom{2S+1}{2S}\neq 0$. If, as in 2D, $%
\phi $ diverges assuming finite $\left\langle S^z\right\rangle ,$ eqn. (\ref%
{genform}) immediately contradicts the assumption. Thus, in two dimensions, $%
\left\langle S^z\right\rangle $ must vanish. On the other hand, if $\phi $
remains finite (as expected in 3D), a finite value of $\left\langle
S^z\right\rangle $ is consistent with (\ref{genform}), thus allowing for a
non-zero spontaneous magnetization.\vspace{1.5cm}

To summarize our discussion of the Tyablikov method, it has been shown that
the Tyablikov method for the Heisenberg model conserves the Mermin-Wagner
theorem: In two dimensions, a finite value of $\left\langle S^z\right\rangle 
$ can only be achieved by applying an external field; a magnetic phase
transition at finite temperature is ruled out. Spatial dimensionality enters
the calculation explicitly through the $k$-space integral appearing in the
definition of $\phi .$ The Tyablikov method, in this respect, differs
markedly from mean-field theory, where the coordination number of a given
lattice is the only variable accounting for the geometry of the system, or
from simple improvements thereof, such as Onsager reaction-field theory
discussed in the previous section.

\chapter{Summary}

In this paper, we have investigated several aspects of dimensionality in
many-body theories of long-range order. The Bogoliubov inequality, together
with the concept of quasi-averages, which we have discussed in Chapter II,
provides the mathematical tools for examining rigorously whether or not a
phase transition occurs. For film systems, i.e. systems of $d$ identical
two-dimensional layers stacked on top of each other, we have shown that a
magnetic phase transition at finite temperatures can be rigorously ruled
out, unless $d\rightarrow \infty .$ The proof holds for a wide class of
many-body Hamiltonians, such as the Heisenberg, Hubbard, Kondo-lattice, and
Periodic Anderson Model. The general nature of the proof suggests that
analogous results would be obtained for other Hamiltonians and other kinds
of phase transition as well. For $s$-wave pairing in Hubbard films this
conjecture has been confirmed.

The Bogoliubov inequality has been adapted to fit in with a more general
framework of correlation inequalities as presented in Chapter IV. From the
algebraic properties of some of the main statistical quantities
characterizing a phase transition (most notably the static susceptibility),
rigorous relations have been derived which may be used to derive relations
between different types of order parameters or phase transitions. The
general applicability of the formalism presented, particularly the fact that
the Bogoliubov inequality and the Goldstone theorem fit in well with it,
suggest that further research in this direction might be able to clarify
some conceptual issues in the theory of (magnetic) phase transitions.

Finally (Chapter V), we have contrasted some approximative methods of
many-body physics with respect to whether or not they conserve the
Mermin-Wagner theorem. As it turns out, simple extensions of mean-field
theory (which is known to violate the Mermin-Wagner theorem) satisfy the
Mermin-Wagner theorem at best on a superficial level. By including
fluctuations as in the Tyablikov method, it is possible to construct an
approximative theory that reproduces the Mermin-Wagner theorem in all the
relevant limiting cases.

As of today, the Mermin-Wagner theorem is perhaps the strongest rigorous
statement available, as it covers a variety of different models, is
independent of the microscopic parameters and is applicable in the
thermodynamic limit. It is, however, a \textit{negative} statement about
phase transitions. Therefore, it is unclear whether or not future
investigations of this problem will eventually be able to satisfactorily
resolve the question of precisely \textit{how} phase transitions do occur in
the real world. With these considerations in mind, it is perhaps appropriate
to close our discussion of the Mermin-Wagner theorem and related exact
results on a philosopher's remark, namely that

\begin{quote}
``It is the mark of an educated man to look for exactness in each class of
things just so far as the nature of the subject admits.''\footnote{%
Aristotle, \textit{Nicomachean Ethics.}}
\end{quote}

\appendix 

\chapter{Positive semi-definiteness of the Bogoliubov scalar product}

With (\ref{bogscalprod}) the \textit{Bogoliubov scalar product} $\mathfrak{B}
$ has been defined as 
\begin{equation}
\mathfrak{B}(A;B)=\sum_{\substack{ n.m  \\ E_{n}\neq E_{m}}}\left\langle
n\left| A^{+}\right| m\right\rangle \left\langle m\left| B\right|
n\right\rangle \frac{w_{m}-w_{n}}{E_{n}-E_{m}}  \label{bogapp}
\end{equation}%
with 
\begin{equation}
w_{n}=\frac{\exp (-\beta E_{n})}{tr\left( \exp (-\beta H)\right) },
\end{equation}%
where $H$ is the Hamiltonian and $A,B$ are arbitrary operators.

In order to verify that $\mathfrak{B}$ indeed has the properties one expects
from the mathematical definition of a scalar product, we check the axioms:

\begin{itemize}
\item $\mathfrak{B}(A;B)$ is a \textit{complex number with} $\mathfrak{B}%
(A;B)=\left( \mathfrak{B}(B;A)\right) ^{*}.$ This follows from (\ref{bogapp}%
) by noting that $(w_m-w_n)/(E_n-E_m)$ is a real number and 
\begin{equation}
\left( \left\langle n\left| B^{+}\right| m\right\rangle \left\langle m\left|
A\right| n\right\rangle \right) ^{*}=\left\langle n\left| A^{+}\right|
m\right\rangle \left\langle m\left| B\right| n\right\rangle .
\end{equation}

\item $\mathfrak{B}(A;B)$ is \textit{linear with respect to its arguments:} $%
\mathfrak{B}(A;\alpha _1B_1+\alpha _2B_2)=\alpha _1\mathfrak{B}%
(A;B_1)+\alpha _2\mathfrak{B}(A;B_2).$ This follows from the properties of
the matrix element $\left\langle m\left| B\right| n\right\rangle .$ By use
of the first axiom, one can derive an analogous relation for the first
argument of $\mathfrak{B}.$

\item $\mathfrak{B}$ induces a \textit{positive semi-definite} norm in the
space of operators, i.e. 
\begin{equation}
\mathfrak{B}(A;A)\geq 0,  \label{seminrm}
\end{equation}
because of $(w_m-w_n)/(E_n-E_m)\geq 0.$

\item $\mathfrak{B}(A;A)=0$ does \textit{not}, in general, imply $A=0$ (for
which (\ref{seminrm}) indeed vanishes); e.g. for the Hamiltonian, $\mathfrak{%
B}(H;H)=0$ , while $H=0$ is no physically meaningful Hamiltonian.
\end{itemize}

\chapter{Upper bound for $|\langle c_{n\protect\beta \protect\sigma }^{+}c_{k%
\protect\gamma \protect\sigma }\rangle |$}

In order to derive an upper bound for the expectation value $\left|
\left\langle c_{n\beta \sigma }^{+}c_{k\gamma \sigma }\right\rangle \right| $
one uses the procedure that has been discussed in the main text (see eqns. (%
\ref{symccs}) ff.) for the correlation function $\left\langle c_{n\gamma
\sigma }^{+}c_{n\gamma -\sigma }S_{n\gamma }^{-\sigma }\right\rangle $ which
we encountered in the s-f case.

For $(n,\beta )=(k,\gamma )$ one obviously has 
\begin{equation}
\left\langle c_{n\beta \sigma }^{+}c_{n\beta \sigma }\right\rangle \leq 1.
\label{gleichindiz}
\end{equation}
Starting, in the general case, with the decomposition 
\begin{eqnarray}
\left\langle c_{i\alpha \sigma }^{+}c_{l\nu \sigma }\right\rangle &=&\frac
14\left\langle \left( c_{i\alpha \sigma }^{+}+c_{l\nu \sigma }^{+}\right)
\left( c_{i\alpha \sigma }+c_{l\nu \sigma }\right) \right.  \notag \\
&&-\left( c_{i\alpha \sigma }^{+}-c_{l\nu \sigma }^{+}\right) \left(
c_{i\alpha \sigma }-c_{l\nu \sigma }\right)  \notag \\
&&+i\left( c_{i\alpha \sigma }^{+}+ic_{l\nu \sigma }^{+}\right) \left(
c_{i\alpha \sigma }-ic_{l\nu \sigma }\right)  \notag \\
&&\left. -i\left( c_{i\alpha \sigma }^{+}-ic_{l\nu \sigma }^{+}\right)
\left( c_{i\alpha \sigma }+ic_{l\nu \sigma }\right) \right\rangle
\end{eqnarray}
one can apply the spectral theorem, 
\begin{equation}
\sum_\sigma \left\langle c_{i\alpha \sigma }^{+}c_{l\nu \sigma
}\right\rangle =\frac 1{4\hbar }\sum_{j=1}^4\int\limits_{-\infty }^\infty
dE\frac 1{e^{\beta E}+1}\phi (j)\sum_\sigma S_{A_{j\sigma }^{+}A_{j\sigma
}}^{(-)}(E),
\end{equation}
where $\phi :(j=1,2,3,4)\rightarrow (\phi (j)=+1,-1,+i,-i)$ defines a phase
factor for every term as in (\ref{symccs}), and $S_{A_{j\sigma
}^{+}A_{j\sigma }}^{(-)}(E)=\frac 1{2\pi }\left\langle \left[ A_{j\sigma
}^{+},A_{j\sigma }\right] _{+}\right\rangle (E)$ is the spectral density in
its energy representation. Hence 
\begin{eqnarray}
\left| \left\langle c_{n\beta \sigma }^{+}c_{k\gamma \sigma }\right\rangle
\right| &\leq &\frac 14\sum_{j=1}^4\frac 1\hbar \int\limits_{-\infty
}^\infty dE\left| \phi (j)\right| \frac 1{e^{\beta
E}+1}S_{A_jA_j^{+}}^{(-)}(E)  \notag \\
&\leq &\frac 14\sum_{j=1}^4M_{A_jA_j^{+}}^{(0)},  \label{noughthmom}
\end{eqnarray}
where each of the $0$-th spectral moments in (\ref{noughthmom}) can be
expressed by the expectation value of the anticommutator: 
\begin{equation}
\frac 1\hbar \int\limits_{-\infty }^\infty dES_{A_{j\sigma }^{+}A_{j\sigma
}}^{(-)}(E)\equiv M_{A_{j\sigma }^{+}A_{j\sigma }}^{(0)}=\left\langle \left[
A_{j\sigma }^{+},A_{j\sigma }\right] _{+}\right\rangle .
\end{equation}
In (\ref{noughthmom}), use has already been made of the triangle inequality 
\begin{equation*}
\left| \sum_j\phi (j)\cdot ...\right| \leq\sum_j\left| \phi (j)\cdot
...\right|
\end{equation*}
and of the fact that for pairs of adjunct operators, the spectral density is
positive definite.

Since we have discussed the case $(n,\beta )=(k,\gamma )$ in (\ref%
{gleichindiz}) separately, we can now assume $(n,\beta )\neq (k,\gamma ).$
The $0$-th spectral moments can be easily calculated; they each give $%
M_{A_jA_j^{+}}^{(0)}=2.$ Thus for all $n,k,\beta ,\gamma $ we have as an
upper bound 
\begin{equation}
\left| \left\langle c_{n\beta \sigma }^{+}c_{k\gamma \sigma }\right\rangle
\right| \leq 2.  \label{gemischte c}
\end{equation}

\chapter{From correlation inequalities to the Bogoliubov inequality}

For completeness, we derive in this section the remaining integrals in eqn. (%
\ref{einsvorbog}), thus completing the derivation of the Bogoliubov
inequality within the correlation inequalities approach:

\begin{itemize}
\item For the integral on the LHS of (\ref{einsvorbog}) it must be shown
that 
\begin{equation}
\frac 1\hbar \int \frac{dE}ES_{\left[ H,C\right] _{-};\left[ H,C\right]
_{-}^{+}}(E)=\left\langle \left[ C,\left[ H,C^{+}\right] _{-}\right]
_{-}\right\rangle .
\end{equation}
Since the double commutator is related to the spectral density $S_{CC^{+}}$
by 
\begin{equation}
\left\langle \left[ C,\left[ H,C^{+}\right] _{-}\right] _{-}\right\rangle
=\int dEES_{CC^{+}}(E),
\end{equation}
due to the general commutator representation of the spectral moments, it
suffices to show directly that 
\begin{equation}
S_{\left[ H,C\right] _{-};\left[ H,C\right] _{-}^{+}}(E)=E^2S_{CC^{+}}(E).
\end{equation}
This is indeed the case: 
\begin{eqnarray}
S_{\left[ H,C\right] _{-};\left[ H,C\right] _{-}^{+}}(E) &=&\frac 1\Xi
\sum_{mn}\left( e^{-\beta E_m}-e^{-\beta E_n}\right) \left\langle m\left|
HC-CH\right| \right\rangle \cdot  \notag \\
&&\left\langle n\left| (HC)^{+}-(CH)^{+}\right| m\right\rangle \delta
(E-(E_n-E_m))  \notag \\
&=&\frac 1\Xi \sum_{mn}(e^{-\beta E_m}-e^{-\beta E_n})\left(
-(E_n-E_m)\right) ^2\left\langle m\left| C\right| n\right\rangle \cdot 
\notag \\
&&\cdot \left\langle n\left| C^{+}\right| m\right\rangle \delta (E-(E_n-E_m))
\notag \\
&=&\frac 1\Xi \sum_{mn}\left( e^{-\beta E_n}-e^{-\beta E_m}\right)
E^2\left\langle m\left| C\right| n\right\rangle \left\langle n\left|
C^{+}\right| m\right\rangle \cdot  \notag \\
&&\cdot \delta (E-(E_n-E_m))  \notag \\
&=&E^2S_{CC^{+}}(E),
\end{eqnarray}
where we have twice made use of the spectral representation (\ref%
{spectralspectral}) of the spectral density.

\item The integral on the LHS of (\ref{einsvorbog}) can be treated
similarly: 
\begin{eqnarray}
&&\left| \int \frac{dE}ES_{\left[ C,H\right] _{-};A^{+}}(E)\right| ^2  \notag
\\
&=&\left| \frac 1\Xi \int \frac{dE}E\sum_{mn}\left( e^{-\beta E_m}-e^{-\beta
E_n}\right) \left\langle m\left| CH-HC\right| n\right\rangle \cdot \right. 
\notag \\
&&\left. \left\langle n\left| A^{+}\right| m\right\rangle \delta
(E-(E_n-E_m))\right| ^2  \notag \\
&=&\left| \int dE\left[ \frac 1\Xi \sum_{mn}\left( e^{-\beta E_m}-e^{-\beta
E_n}\right) \left\langle m\left| C\right| n\right\rangle \left\langle
n\left| A^{+}\right| m\right\rangle \delta \left( E-(E_n-E_m)\right) \right]
\right| ^2  \notag \\
&=&\left| \int dES_{CA^{+}}(E)\right| ^2=\left| \left\langle \left[ C,A^{+}%
\right] _{-}\right\rangle \right| ^2.
\end{eqnarray}
\end{itemize}

Thus the transition from the correlation inequality (\ref{einsvorbog}) to
the Bogoliubov inequality (\ref{bogthecia}) can indeed be made.

\chapter{Integral inequalities}

We give here, for reference, a list of important mathematical inequalities
that hold for real-valued functions $f(x),$ $g(x)$ under the additional
conditions given with each inequality.\footnote{%
The list follows the section on integral inequalities in ref. \cite{Gra Ryz
1994}. The standard reference for inequalities from a mathematical viewpoint
is \cite{Har Lit Pol 1934}.} Not all of these inequalities are used in this
thesis, but it appears worthwhile to contrast those we do use with related
ones.

\subsubsection{Cauchy-Schwarz inequality}

Let $f(x)$ and $g(x)$ be any two integrable functions on $\left[ a,b\right]
. $ Then, 
\begin{equation}
\left( \int_a^bdxf(x)g(x)\right) ^2\leq \left( \int_a^bdxf^2(x)\right)
\left( \int_a^bdxg^2(x)\right) ,
\end{equation}
and the equality will hold if, and only if, $f(x)=kg(x),$ with $k$ real.

\subsubsection{H\"{o}lder's inequality}

Let $f(x)$ and $g(x)$ be any two functions for which $\left| f(x)\right| ^p$
and $\left| g(x)\right| ^q$ are integrable on $\left[ a,b\right] $ with $p>1$
and $p^{-1}+q^{-1}=1;$ then, 
\begin{equation}
\int_a^bdxf(x)g(x)\leq \left( \int_a^bdx\left| f(x)\right| ^p\right)
^{1/p}+\left( \int_a^bdx\left| g(x)\right| ^q\right) ^{1/q}.
\label{hoeldapp}
\end{equation}
The equality holds if, and only if, $\alpha \left| f(x)\right| ^p=\beta
\left| g(x)\right| ^q,$ where $\alpha $ and $\beta $ are positive constants.

\subsubsection{Minkowski's inequality}

Let $f(x)$ and $g(x)$ be any two functions for which $\left| f(x)\right| ^p$
and $\left| g(x)\right| ^p$ are integrable on $\left[ a,b\right] $ with $%
p>0; $ then, 
\begin{equation}
\left( \int_a^bdx\left| f(x)+g(x)\right| ^p\right) ^{1/p}\leq \left(
\int_a^bdx\left| f(x)\right| ^p\right) ^{1/p}+\left( \int_a^bdx\left|
g(x)\right| ^p\right) ^{1/p}.
\end{equation}
The equality holds if, and only if, $f(x)=kg(x)$ for some non-negative $k.$

\subsubsection{Jensen's inequality}

Let $f(x)$ and $g(x)$ be defined for $a\leq x\leq b$ such that $\alpha \leq
f(x)\leq \beta $ and $g(x)\geq 0,$ with $g(x)\neq 0.$ If $\phi (u)$ is a
convex function on the interval $\alpha \leq u\leq \beta ,$ i.e. if for any
two points $u_1,u_2$ in $\left[ \alpha ,\beta \right] $%
\begin{equation*}
\phi \left( \frac{u_1+u_2}2\right) \leq \frac{\phi (u_1)+\phi (u_2)}2,
\end{equation*}
then the following relation between $f(x),$ $g(x),$ and $\phi (f(x))$ holds: 
\begin{equation}
\phi \left( \frac{\int_a^bdxf(x)g(x)}{\int_a^bdxg(x)}\right) \leq \frac{%
\int_a^bdx\phi (f(x))g(x)}{\int_a^bdxg(x)}.
\end{equation}
Due to the importance of convex functions in statistical mechanics, Jensen's
inequality is one of the more widely used integral inequalities. For the use
of Jensen's inequality in statistical mechanics see the paper by Luttinger,
ref. \cite{Lut 1966}.

\chapter*{Acknowledgments}

I would like to thank Prof. Wolfgang Nolting and Dr. Michael Potthoff for
helpful discussions, and Prof. Robert Keiper for stimulating conversations.

I gratefully acknowledge financial support from the \textit{Studienstiftung,
Bonn,} during my time at the Humboldt University of Berlin. I also wish to
thank the \textit{Arnold Gerstenberg Trust} and the\textit{\ European Trust, 
}both of the University of Cambridge, for financial support while revising
the original draft.

\end{document}